\documentclass[journal]{IEEEtran}

\usepackage{amsmath}
\usepackage{amssymb}
\usepackage{amsthm}
\usepackage{graphicx}
\usepackage{color}
\usepackage{mathrsfs}
 
\usepackage{epsfig}
\usepackage[numbers]{natbib}
\usepackage[linesnumbered,ruled]{algorithm2e}
\usepackage{url}
\usepackage{tabularx}

\begin{document}

 


\newcommand{\be}{\begin{equation}}
\newcommand{\ee}{\end{equation}}
\newcommand{\bes}{\begin{equation*}}
\newcommand{\ees}{\end{equation*}}
\newcommand{\beqn}{\begin{eqnarray}}
\newcommand{\eeqn}{\end{eqnarray}}
\newcommand{\beqns}{\begin{eqnarray*}}
\newcommand{\eeqns}{\end{eqnarray*}}
\newcommand{\begal}{\begin{align}}
\newcommand{\egal}{\end{align}}
\newcommand{\beals}{\begin{align*}}
\newcommand{\eeals}{\end{align*}}
\newcommand{\besp}{\begin{split}}
\newcommand{\eesp}{\end{split}}

\newcommand{\lkr}{\left(}
\newcommand{\lkv}{\left[}
\newcommand{\rkv}{\right]}
\newcommand{\rkr}{\right)}
\newcommand{\lfi}{\left\{}
\newcommand{\rfi}{\right\}}
\newcommand{\lnor}{\left\|}
\newcommand{\rnor}{\right\|}

\newcommand{\fr}[1]{(\ref{#1})}

\newcommand{\vart}{\vartheta}
\newcommand{\ro}{\varrho}
\newcommand{\ph}{\varphi}
\newcommand{\del}{\delta}
\newcommand{\Del}{\Delta}
\newcommand{\dn}{\delta_n}
\newcommand{\al}{\alpha}
\newcommand{\af}{\alpha}
\newcommand{\eps}{\epsilon}
\newcommand{\Ga}{\Gamma}
\newcommand{\ga}{\gamma}
\newcommand{\te}{\theta}
\newcommand{\om}{\omega}
\newcommand{\lam}{\lambda}
\newcommand{\Up}{\Upsilon}
\newcommand{\up}{\upsilon}
\newcommand{\dz}{\zeta}
\newcommand{\sig}{\sigma}
\newcommand{\sgmd}{\sigma^2}
\newcommand{\Lam}{\Lambda}
\newcommand{\Om}{\Omega}
\newcommand{\Sig}{\Sigma}
\newcommand{\Te}{\Theta}

\newcommand{\EE}{\ensuremath{{\mathbb E}}}
\newcommand{\JJ}{\ensuremath{{\mathbb J}}}
\newcommand{\II}{\ensuremath{{\mathbb I}}}
\newcommand{\ZZ}{\ensuremath{{\mathbb Z}}}
\newcommand{\PP}{\ensuremath{{\mathbb P}}}
\newcommand{\QQ}{\ensuremath{{\mathbb Q}}}
\newcommand{\KK}{\ensuremath{{\mathbb K}}}
\newcommand{\RR}{{\mathbb R}}
\newcommand{\real}{{\mathbb R}}

\newcommand{\iter}{\mathrm{iter}}
\newcommand{\low}{\mbox{low}}
\newcommand{\vect}{\mbox{vec}}
\newcommand{\Pen}{\mbox{Pen}}
\newcommand{\Span}{\mbox{Span}}
\newcommand{\intg}{\mbox{int}}
\newcommand{\card}{\mbox{card}}
\newcommand{\Range}{\mbox{Range}}
\newcommand{\Var}{\mbox{Var}}
\newcommand{\Cov}{\mbox{Cov}}
\newcommand{\diag}{{\rm diag}}
\newcommand{\supp}{\mbox{supp}}
\newcommand{\etal}{{\it et  al. }}
\newcommand{\std}{\mbox{std}}
\newcommand{\SNR}{\mbox{SNR}}
\newcommand{\Tr}{\mbox{Tr}}
\newcommand{\proj}{\mbox{proj}}
\newcommand{\Expect}{\mathrm{Expect}}
\newcommand{\Skew}{\mathrm{Skew}}
\newcommand{\Kmax}{K_{\max}}
\newcommand{\Kmin}{K_{\min}}
\newcommand{\argmin}{\mbox{argmin}}
\newcommand{\argmax}{\mbox{argmax}}
\newcommand{\mat}{\mbox{mat}}
\newcommand{\tr}{\mbox{tr}}
\newcommand{\rank}{\mbox{rank}}
\newcommand{\Reg}{\mbox{Reg}}
\newcommand{\SVD}{\mbox{SVD}}
\newcommand{\EVD}{\mbox{EVD}}
\newcommand{\sgn}{\mbox{sign}}
\newcommand{\iid}{\mbox{i.i.d.}}
\newcommand{\nnz}{\mbox{nnz}}
\newcommand{\polylog}{\mbox{polylog}}

\newcommand{\rhon}{\rho_{n}} 
\newcommand{\rhonl}{\rho_{n,l}} 

\newcommand{\sinTe}{\sin\Te}

\theoremstyle{plain}
\newtheorem{theorem}{Theorem}
\newtheorem{lem}{Lemma}
\newtheorem{prop}{Proposition}
\newtheorem{cor}{Corollary}
\newtheorem{rem}{Remark}

\newcommand{\ba}{\mathbf{a}}
\newcommand{\bb}{\mathbf{b}}
\newcommand{\bc}{\mathbf{c}}
\newcommand{\bd}{\mathbf{d}}
\newcommand{\boe}{\mathbf{e}}
\newcommand{\bof}{\mathbf{f}}
\newcommand{\bg}{\mathbf{g}}
\newcommand{\bi}{\mathbf{i}}
\newcommand{\bj}{\mathbf{j}}
\newcommand{\bk}{\mathbf{k}}
\newcommand{\bh}{\mathbf{h}}
\newcommand{\bp}{\mathbf{p}}
\newcommand{\bq}{\mathbf{q}}
\newcommand{\bt}{\mathbf{t}}
\newcommand{\bu}{\mathbf{u}}
\newcommand{\bv}{\mathbf{v}}
\newcommand{\bw}{\mathbf{w}}
\newcommand{\bx}{\mathbf{x}}
\newcommand{\by}{\mathbf{y}}
\newcommand{\bz}{\mathbf{z}}

\newcommand{\bA}{\mathbf{A}}
\newcommand{\boB}{\mathbf{B}}
\newcommand{\bC}{\mathbf{C}}
\newcommand{\bD}{\mathbf{D}}
\newcommand{\bE}{\mathbf{E}}
\newcommand{\bF}{\mathbf{F}}
\newcommand{\bG}{\mathbf{G}}
\newcommand{\bH}{\mathbf{H}}
\newcommand{\bI}{\mathbf{I}}
\newcommand{\bJ}{\mathbf{J}}
\newcommand{\bK}{\mathbf{K}}
\newcommand{\bM}{\mathbf{M}}
\newcommand{\bO}{\mathbf{O}}
\newcommand{\bP}{\mathbf{P}}
\newcommand{\bQ}{\mathbf{Q}}
\newcommand{\bR}{\mathbf{R}}
\newcommand{\bS}{\mathbf{S}}
\newcommand{\bU}{\mathbf{U}}
\newcommand{\bV}{\mathbf{V}}
\newcommand{\bW}{\mathbf{W}}
\newcommand{\bX}{\mathbf{X}}
\newcommand{\bY}{\mathbf{Y}}
\newcommand{\bZ}{\mathbf{Z}}

\newcommand{\bzero}{\mathbf{0}}
\newcommand{\bone}{\mathbf{1}}

 \newcommand{\blam}{\mbox{\mathversion{bold}$\lam$}}
 \newcommand{\bte}{\mbox{\mathversion{bold}$\te$}}
\newcommand{\bxi}{\mbox{\mathversion{bold}$\xi$}}
\newcommand{\boeta}{\mbox{\mathversion{bold}$\eta$}}
\newcommand{\bbe}{\mbox{\mathversion{bold}$\beta$}}
\newcommand{\bzeta}{\mbox{\mathversion{bold}$\zeta$}}
\newcommand{\bph}{\mbox{\mathversion{bold}$\ph$}}
\newcommand{\bpsi}{\mbox{\mathversion{bold}$\psi$}}
\newcommand{\beps}{\mbox{\mathversion{bold}$\eps$}}
\newcommand{\bobeta}{\mbox{\mathversion{bold}$\beta$}}
\newcommand{\bgamma}{\mbox{\mathversion{bold}$\gamma$}}
\newcommand{\bom}{\mbox{\mathversion{bold}$\om$}}

\newcommand{\bPsi}{\mbox{\mathversion{bold}$\Psi$}}
\newcommand{\bPhi}{\mbox{\mathversion{bold}$\Phi$}}
\newcommand{\bUp}{\mbox{\mathversion{bold}$\Up$}}
\newcommand{\bLam}{\mbox{\mathversion{bold}$\Lambda$}}
\newcommand{\bSig}{\mbox{\mathversion{bold}$\Sigma$}}
\newcommand{\bTe}{\mbox{\mathversion{bold}$\Theta$}}
\newcommand{\bXi}{\mbox{\mathversion{bold}$\Xi$}}
\newcommand{\bcalL}{\mbox{\mathversion{bold}$\calL$}}
\newcommand{\bDelta}{\mathbf{\Delta}}

\newcommand{\calA}{{\mathcal{A}}}
\newcommand{\calB}{{\mathcal{B}}}
\newcommand{\calC}{{\mathcal{C}}}
\newcommand{\calD}{{\mathcal{D}}}
\newcommand{\calE}{{\mathcal{E}}}
\newcommand{\calF}{{\mathcal{F}}}
\newcommand{\calG}{{\mathcal G}}
\newcommand{\calH}{{\mathcal H}}
\newcommand{\calJ}{{\mathcal{J}}}
\newcommand{\calK}{{\mathcal{K}}}
\newcommand{\calL}{{\mathcal{L}}}
\newcommand{\calM}{{\mathcal M}}
\newcommand{\calN}{{\mathcal N}}
\newcommand{\calO}{{\mathcal O}}
\newcommand{\calP}{{\mathcal{P}}}
\newcommand{\calR}{{\mathcal{R}}}
\newcommand{\calS}{{\mathcal{S}}}
\newcommand{\calT}{{\mathcal T}}
\newcommand{\calW}{{\mathcal W}}
\newcommand{\calX}{{\mathcal{X}}}
\newcommand{\calY}{{\mathcal{Y}}}
\newcommand{\calZ}{{\mathcal{Z}}}
\newcommand{\calV}{{\mathcal{V}}} 
\newcommand{\calU}{{\mathcal{U}}}
 
\newcommand{\lan}{\langle}
\newcommand{\ran}{\rangle}


\newcommand{\sumlL}{\sum_{l=1}^L}
\newcommand{\summM}{\sum_{m=1}^M}

\newcommand{\hbP}{\widehat{\bP}} 
\newcommand{\hbB}{\widehat{\bB}}
\newcommand{\hbC}{\widehat{\bC}}
\newcommand{\hbG}{\widehat{\bG}}
\newcommand{\hbH}{\widehat{\bH}}
\newcommand{\hbTe}{\widehat{\bTe}}
\newcommand{\hbD}{\widehat{\bD}}
\newcommand{\hbU}{\widehat{\bU}}
\newcommand{\hbV}{\widehat{\bV}}
\newcommand{\hbW}{\widehat{\bW}}
\newcommand{\hbQ}{\widehat{\bQ}}
\newcommand{\hbX}{\widehat{\bX}}
\newcommand{\hbZ}{\widehat{\bZ}}
\newcommand{\hbUp}{\widehat{\bUp}}
\newcommand{\hbLam}{\widehat{\bLam}}
\newcommand{\hhbZ}{\widehat{\widehat{\bZ}}}
\newcommand{\hd}{\widehat{d}}
\newcommand{\hz}{\widehat{z}}

\newcommand{\hA}{\widehat{A}}
\newcommand{\hB}{\widehat{B}}
\newcommand{\hD}{\widehat{D}}
\newcommand{\hG}{\widehat{G}}
\newcommand{\hH}{\widehat{H}}
\newcommand{\hX}{\widehat{X}}
\newcommand{\hY}{\widehat{Y}}
\newcommand{\hPsi}{\widehat{\Psi}}
\newcommand{\hU}{\widehat{U}}
\newcommand{\hV}{\widehat{V}}
\newcommand{\hW}{\widehat{W}}
\newcommand{\hSig}{\widehat{\Sig}}
\newcommand{\hs}{\hat{s}}
\newcommand{\hte}{\hat{\te}}
\newcommand{\hTe}{\widehat{\Te}}
\newcommand{\hS}{\widehat{S}}
\newcommand{\hhU}{\widehat{\widehat{U}}}
\newcommand{\hL}{\widehat{L}} 
\newcommand{\hlam}{\widehat{\lam}}
\newcommand{\hr}{\hat{r}} 
\newcommand{\hLam}{\widehat{\Lam}}
\newcommand{\hK}{\widehat{K}} 
\newcommand{\hatc}{\hat{c}}

\newcommand{\scrE}{\mathscr{E}}
\newcommand{\scrH}{\mathscr{H}} 
\newcommand{\scrX}{\mathscr{X}}

\newcommand{\barU}{\overline{U}}
\newcommand{\barX}{\overline{X}}
\newcommand{\barK}{\overline{K}}

\newcommand{\tilr}{\tilde{r}}
\newcommand{\tilh}{\tilde{h}}
\newcommand{\tilz}{\tilde{z}}
\newcommand{\tilA}{\widetilde{A}}
\newcommand{\tilC}{\widetilde{C}}
\newcommand{\tilL}{\widetilde{L}}
\newcommand{\tilR}{\widetilde{R}} 
\newcommand{\tilbG}{\widetilde{\bG}}
\newcommand{\tilbP}{\widetilde{\bP}}
\newcommand{\tilbA}{\widetilde{\bA}}
\newcommand{\tilbU}{\widetilde{\bU}}
\newcommand{\tilbV}{\widetilde{\bV}}
\newcommand{\tilbW}{\widetilde{\bW}}
\newcommand{\tilbTe}{\widetilde{\bTe}}
\newcommand{\tilbLam}{\widetilde{\bLam}}
\newcommand{\tbLam}{\widetilde{\bLam}}
\newcommand{\tOm}{\widetilde{\Om}}
\newcommand{\tH}{\widetilde{H}}
\newcommand{\tilTe}{\widetilde{\Te}}
\newcommand{\tdel}{\widetilde{\del}}

\newcommand{\tilI}{\widetilde{I}}
\newcommand{\tilX}{\widetilde{X}}
\newcommand{\tilY}{\widetilde{Y}}
\newcommand{\tilU}{\widetilde{U}}
\newcommand{\tilD}{\widetilde{D}}
\newcommand{\tilO}{\widetilde{O}}
\newcommand{\tilP}{\widetilde{P}}
\newcommand{\tilS}{\widetilde{S}}
\newcommand{\tilG}{\widetilde{G}}
\newcommand{\tilW}{\widetilde{W}}
\newcommand{\tilB}{\widetilde{B}}
\newcommand{\tilb}{\widetilde{b}}
\newcommand{\tilOm}{\widetilde{\Om}}

\newcommand{\tilscrE}{\widetilde{\scrE}}

\newcommand{\brX}{\breve{X}}
\newcommand{\brSig}{\breve{\Sig}}
\newcommand{\brB}{\breve{B}}
\newcommand{\brb}{\breve{b}}
\newcommand{\brP}{\breve{P}}

\newcommand{\bcalW}{\mbox{\mathversion{bold}$\calW$}}
\newcommand{\hbcalW}{\widehat{\bcalW}}
\newcommand{\tbcalW}{\widetilde{\bcalW}}

\newcommand{\bcalP}{\mbox{\mathversion{bold}$\calP$}}

\newcommand{\bcalV}{\mbox{\mathversion{bold}$\calV$}}
\newcommand{\hbcalV}{\widehat{\bcalV}}
\newcommand{\tbcalV}{\widetilde{\bcalV}}

\newcommand{\hcalP}{\widehat{\calP}}
\newcommand{\hcalG}{\widehat{\calG}}
\newcommand{\hcalH}{\widehat{\calH}}
\newcommand{\hcalW}{\widehat{\calW}}

\newcommand{\barbZ}{\overline{\bZ}}
\newcommand{\barbU}{\overline{\bU}}
\newcommand{\barbV}{\overline{\bV}}
\newcommand{\barbD}{\overline{\bD}}
\newcommand{\barbH}{\overline{\bH}}
\newcommand{\barbG}{\overline{\bG}}
\newcommand{\barbQ}{\overline{\bQ}}
\newcommand{\hbarbG}{\overline{\widehat{\bG}}}
\newcommand{\barbB}{\overline{\boB}}
\newcommand{\barcalH}{\overline{\calH}}
\newcommand{\barbR}{\overline{\bR}}
\newcommand{\barcalR}{\overline{\calR}}

\newcommand{\di}{\displaystyle}
 
\newcommand{\minmM}{\displaystyle \min_{m=1, ....M}\ }
\newcommand{\maxmM}{\displaystyle \max_{m=1, ....M}\ }

\newcommand{\minL}{\displaystyle \min_{l=1, ....L}\ }
\newcommand{\maxL}{\displaystyle \max_{l=1, ....L}\ }

\newcommand{\lowc}{\underline{c}}
\newcommand{\highc}{\bar{c}}
\newcommand{\lowC}{\underline{C}}
\newcommand{\highC}{\bar{C}}
\newcommand{\lowd}{\underline{d}}
\newcommand{\highd}{\bar{d}}

\newcommand{\Ctau}{C_{\tau}}
\newcommand{\tilCtau}{\widetilde{C_{\tau}}}
\newcommand{\clam}{c_{\lam}}
\newcommand{\cd}{c_{d}}
\newcommand{\Cd}{C_{d}}

\newcommand{\sinTeU}{\sin\Te(\hU,U)}
\newcommand{\sinTeV}{\sin\Te(\hV,V)}
\newcommand{\tvert}{\vert\vert\vert}
\newcommand{\tinf}{_{2,\infty}}

\newcommand{\epso}{\eps_0}
\newcommand{\epsinf}{\eps\tinf}
\newcommand{\epsE}{\eps_{\scrE}}
\newcommand{\epsEu}{\eps_{\scrE U}}
\newcommand{\epsu}{\eps_U}
\newcommand{\epsv}{\eps_V}

\newcommand{\Delo}{\Del_0}
\newcommand{\Delinf}{\Del\tinf}
\newcommand{\DelE}{\Del_{\scrE}}
\newcommand{\DelEu}{\Del_{\scrE U}}

\newcommand{\barXixi}{\overline{\Xi\,\Xi^T}}
\newcommand{\Xixi}{\Xi\,\Xi^T}
\newcommand{\barXixil}{\overline{\Xi\upl\,(\Xi\upl)^T}}
\newcommand{\Xixil}{\Xi\upl\,(\Xi\upl)^T}


\newcommand{\tDel}{\widetilde{\Del}}
\newcommand{\tDelinf}{\tDel\tinf}
\newcommand{\tDelXi}{\tDel_{\Xi}}
\newcommand{\tDelv}{\tDel_{V}}
\newcommand{\tDelu}{\tDel_{U}}
\newcommand{\tDeluv}{\tDel_{U,V}}
\newcommand{\tDeluvo}{\tDel_{U,V,0}}
\newcommand{\tDeltinf}{\tDel\tinf}
\newcommand{\tDelov}{\tDel_{0,V}}
\newcommand{\tepsuvo}{\teps_{U,V,0}}

\newcommand{\tDelo}{\widetilde{\Del}_0}
\newcommand{\tDeloo}{\widetilde{\Del}_{0,0}}
\newcommand{\tDelE}{\widetilde{\Del}_{\scrE}}
\newcommand{\tDelEo}{\widetilde{\Del}_{\scrE,0}}
\newcommand{\tDelEuo}{\widetilde{\Del}_{\scrE,U,0}}
\newcommand{\tDelEtinf}{\widetilde{\Del}_{\scrE, 2, \infty}^{(1,2)}}
\newcommand{\tDeluo}{\widetilde{\Del}_{U,0}}
\newcommand{\tDelvqinf}{\widetilde{\Del}_{V,q,\infty}}
\newcommand{\tDelqinf}{\widetilde{\Del}_{q,\infty}}
\newcommand{\tDelvtinf}{\widetilde{\Del}_{V,2,\infty}}
\newcommand{\tDelutinf}{\widetilde{\Del}_{U,2,\infty}}
\newcommand{\tDelxo}{\widetilde{\Del}_{X,0}}
\newcommand{\tDeluqinf}{\widetilde{\Del}_{U,0,q,\infty}}
\newcommand{\tDelxiuqinf}{\widetilde{\Del}_{\Xi,U,q,\infty}}
\newcommand{\tDelxuqinf}{\widetilde{\Del}_{X,U,q,\infty}}
\newcommand{\tDelxqinf}{\widetilde{\Del}_{X,q,\infty}}
\newcommand{\tDeluuo}{\tDel_{\hU,U,0}}

\newcommand{\tDelxiqinf}{\widetilde{\Del}_{\Xi,q,\infty}}
\newcommand{\tDelxiutinf}{\widetilde{\Del}_{\Xi,U,2,\infty}}
\newcommand{\tDelxutinf}{\widetilde{\Del}_{X,U,2,\infty}}

\newcommand{\tDelxtinf}{\widetilde{\Del}_{X,2,\infty}}
\newcommand{\tDelxoinf}{\widetilde{\Del}_{X,1,\infty}}
\newcommand{\tDelxio}{\widetilde{\Del}_{\Xi,0}}
\newcommand{\tDelxiuo}{\widetilde{\Del}_{\Xi,U,0}}
\newcommand{\tDelxitinf}{\widetilde{\Del}_{\Xi,2,\infty}}

\newcommand{\teps}{\tilde{\eps}}
\newcommand{\tepsy}{\teps_Y}
\newcommand{\tepso}{\teps_0}
\newcommand{\tepsov}{\teps_{0,V}}

\newcommand{\tepsoo}{\teps_{0,0}}
\newcommand{\tepsE}{\teps_{\scrE}}
\newcommand{\tepsEo}{\teps_{\scrE, 0}}
\newcommand{\tepsuoo}{\teps_{U,0,0}}
\newcommand{\tepsuo}{\teps_{U,0}}
\newcommand{\tepsvqinf}{\teps_{V,q,\infty}}
\newcommand{\tepsvtinf}{\teps_{V,2,\infty}}
\newcommand{\tepsxio}{\teps_{\Xi,0}}
\newcommand{\tepsxiuo}{\teps_{\Xi,U,0}}
\newcommand{\tepsxiqinf}{\teps_{\Xi,q,\infty}}
\newcommand{\tepsxitinf}{\teps_{\Xi,2,\infty}}

\newcommand{\tepsxo}{\teps_{X,0}}
\newcommand{\tepsuqinf}{\teps_{U,0,q,\infty}}
\newcommand{\tepsxqinf}{\teps_{X,q,\infty}}
\newcommand{\tepsutinf}{\teps_{U,0,2,\infty}}
\newcommand{\tepsxtinf}{\teps_{X,2,\infty}}
\newcommand{\tepsxoinf}{\teps_{X,1,\infty}}
\newcommand{\tepsEtinf}{\teps_{\scrE, 2, \infty}^{(1,2)}}

\newcommand{\tepstinf}{\teps\tinf}
\newcommand{\tepsxiutinf}{\teps_{\Xi,U,2,\infty}}
\newcommand{\tepsxiuqinf}{\teps_{\Xi,U,q,\infty}}
\newcommand{\tepsxutinf}{\teps_{X,U,2,\infty}}
\newcommand{\tepsEuo}{\teps_{\scrE,U,0}}
\newcommand{\tepsuuo}{\teps_{\hU,U,0}}  
 
\newcommand{\tdelo}{\tdel_0}
\newcommand{\tdelor}{\tdel_{0,r}}
 
\newcommand{\Omtau}{\Om_{\tau}}
\newcommand{\Omtauo}{\Om_{\tau,1}}
\newcommand{\Omtaut}{\Om_{\tau,2}}

\newcommand{\tilOmtau}{\tilOm_{\tau}}
\newcommand{\tilOmtauo}{\tilOm_{\tau,1}}
\newcommand{\tilOmtaut}{\tilOm_{\tau,2}}

\newcommand{\upl}{^{(l)}}
\newcommand{\minmn}{(m \wedge n)}

\newcommand{\Ruu}{R(\hU,U)}
\newcommand{\Ruul}{R(\hU,\hU\upl,U)}

\newcommand{\Ruuf}{\|R(\hU,U)\|_F}
\newcommand{\Ruulf}{\|R(\hU,\hU\upl,U)\|_F}
\newcommand{\Ruuop}{\|R(\hU,U)\|}
\newcommand{\Ruulop}{\|R(\hU,\hU\upl,U)\|}
\newcommand{\Ruutinf}{\|R(\hU,U)\|\tinf}
\newcommand{\Ruultinf}{\|R(\hU,\hU\upl,U)\|\tinf}

\newcommand{\dscrEl}{\Del \scrE\upl}

\newcommand{\nmin}{n_{\min}}
\newcommand{\nmax}{n_{\max}}


\newcommand{\upm}{^{(m)}}

\newcommand{\linL}{l \in [L]}
\newcommand{\minM}{m \in [M]}
\newcommand{\kinK}{k \in [K]}
\newcommand{\kinKm}{k \in [K_m]}
\newcommand{\linm}{l \in [m]}

\newcommand{\ZS}{Z_{\calS}} 
\newcommand{\US}{U_{\calS}}
\newcommand{\ASS}{A_{\calS, \calS}} 
\newcommand{\zS}{z_{\calS}} 
\newcommand{\hzS}{\hz_{\calS}} 
\newcommand{\Sc}{\calS^c}
\newcommand{\ASSc}{A_{\calS, \Sc}} 
\newcommand{\PSSc}{P_{\calS, \Sc}} 
\newcommand{\zSc}{z_{\Sc}} 
\newcommand{\hzSc}{\hz_{\Sc}} 
\newcommand{\USc}{U_{\Sc}}
\newcommand{\ZSc}{Z_{\Sc}} 

\newcommand{\Ysh}{Y^{\sharp}}
\newcommand{\hYsh}{\hY^{\sharp}}
\newcommand{\scrEsh}{\scrE^{\sharp}}
\newcommand{\Ush}{U^{\sharp}}
\newcommand{\Lamsh}{\Lam^{\sharp}}
\newcommand{\hUsh}{\hU^{\sharp}}
\newcommand{\hLamsh}{\hLam^{\sharp}}
\newcommand{\Wsh}{W^{\sharp}}

\newcommand{\delrs}{\del_{rs}}
\newcommand{\tdelrs}{\tilde{\del}_{rs}}


\newcommand {\colred}[1] {\textcolor{red}{#1}}
\newcommand {\colblue}[1] {\textcolor{blue}{#1}}
\newcommand {\colyellow}[1] {\textcolor{yellow}{#1}}
\newcommand {\colblack}[1] {\textcolor{black}{#1}}
\newcommand {\colgreen}[1] {\textcolor{green}{#1}}
\newcommand {\colsred}[1] {\textcolor{OrangeRed}{#1}}
 

\long\def\ignore#1{}



\title{Davis-Kahan Theorem in the two-to-infinity norm and its application to perfect clustering}

\author{Marianna Pensky,  University of Central Florida  
\thanks{Manuscript received October 15, 2025; revised  July 20, 2026. 
Corresponding author: M. Pensky (email: marianna.pensky@ucf.edu).}}


\markboth{IEEE TRANSACTIONS ON INFORMATION THEORY, VOL. ??, NO. ??,  }%
{  }

\IEEEtitleabstractindextext{ 
\begin{abstract}
Many statistical applications, such  as the Principal Component Analysis, matrix completion, tensor regression and many others,  
rely on accurate estimation of leading eigenvectors of a matrix. The Davis-Kahan theorem is known to be instrumental for bounding 
above the distances between matrices  $U$ and $\widehat{U}$ of population eigenvectors and their sample versions. 
While those distances can be measured in various metrics, the recent developments have shown advantages of evaluation 
of the deviation in the two-to-infinity norm.  The purpose of this paper  is to develop a toolbox for derivation 
of upper  bounds for the distances between $U$ and $\hU$ in the two-to-infinity norm for a variety of possible scenarios.
Although this problem has been studied  by several authors, the difference between this paper and its predecessors 
is that the upper bounds are obtained under various sets of assumptions. The upper bounds are initially derived  
with  no or mild probabilistic assumptions on the error, and  are subsequently refined, when some generic 
probabilistic assumptions on the errors hold. The paper also provides rectification  of the upper bounds 
in the cases of heavy-tailed or exponentially fast decaying errors. 
In addition, the paper suggests alternative methods for evaluation of $\hU$ and, therefore,  enables one 
to compare the resulting accuracies. As an example of an  application of the techniques in the paper, 
we derive sufficient conditions for perfect clustering in a generic setting, and then employ them in various scenarios. 
\end{abstract}

}

\maketitle

\IEEEdisplaynontitleabstractindextext

\IEEEpeerreviewmaketitle


 
\section{Introduction}
\label{sec:introduction}

\subsection{Problem formulation and  review of the results}
\label{sec:formulation}

Many statistical applications, such  as the Principal Component Analysis, matrix completion, tensor regression and many others,  
rely on accurate estimation of leading eigenvectors of a matrix. Consider matrices $U$ and $\hU$ of $r$ leading eigenvectors 
of  symmetric matrices $Y, \hY\in \RR^{n \times n}$. 
Then, the deviations between    $U$ and $\hU$ is  tackled by the Davis-Kahan theorem (\cite{Davis_Kahan_1970}),
which has been cited almost 1600 times, and this number would be much higher, if many authors did not refer to the paper's sequels, such as, e.g., 
also highly cited, \cite{10.1093/biomet/asv008}. The deviation between orthonormal bases of two subspaces is usually measured in $\sinTe$ distance. 
If $U, \hU \in \RR^{n \times r}$, $n \geq r$,  are matrices with orthonormal columns, then (see, e.g., \cite{10.1214/17-AOS1541}) 
\begin{align} \label{eq:ca-_zhang1}
& \|\sinTeU\| = \sqrt{1 - \sig_r^2(\hU^T U)}, \\  
& \|\sinTeU\|_F = \sqrt{r - \|\hU^T U\|^2_F}, \nonumber
\end{align}
where  $\|A\|$ and $\|A\|_F$ denote, respectively, the spectral and the Frobenius norm of any matrix $A$.
The Davis-Kahan theorem   developed an upper bound for the $\sinTe$-error in the  Frobenius norm,
and the follow-up papers promptly extended this result to the operational norm. 
Below, we present the version of the theorem in the common case, when matrix $Y$ has $r$ large eigenvalues,  
and the rest of eigenvalues are significantly smaller.

\begin{theorem}\label{th:Davis-Kahan}   
Let $Y, \hY \in \RR^{n \times n}$ be symmetric matrices with eigenvalues 
$\lam_1 \geq ... \geq \lam_r > \lam_{r+1} \geq ... \geq \lam_n$ and 
$\hlam_1 \geq ... \geq \hlam_r > \hlam_{r+1} \geq ... \geq \hlam_n$, respectively, and $\scrE = \hY - Y$. 
If $U, \hU \in \RR^{n \times r}$ are matrices of 
orthonormal eigenvectors corresponding to $\lam_1, ...,  \lam_r$  and $\hlam_1, ...,  \hlam_r$, respectively, 
then  
\be \label{eq:Davis-Kahan}
  \tvert \sinTeU  \tvert \leq 2\, (\lam_r - \lam_{r+1})^{-1}\,  \tvert \scrE \tvert,
\ee
where $\tvert \scrE \tvert$ is the spectral or the Frobenius norm of matrix $\scrE$. 
\end{theorem}

It turns out that the $\sinTe$ distances between the principal subspaces evaluate  the errors of the  best-case  approximation of 
matrix $U$  by $\hU$. Since those matrices are determined up to a rotation, those approximation errors are defined as 
\begin{align} \label{eq:dist}
& D_{sp} (U, \hU) = \inf_{O \in \calO_{r}} \| \hU - UO \|, \\ 
& D_{F} (U, \hU) = \inf_{O \in \calO_{r}} \| \hU - UO \|_{F},  \nonumber
\end{align}
where $\calO_r$ is the set of $r$-dimensional orthogonal matrices. It is known that (see, e.g., \cite{10.1214/17-AOS1541}) 
 \begin{align} 
 \|\sinTeU\| & \leq D_{sp} (U, \hU) \leq \sqrt{2}\, \|\sinTeU\|,  \nonumber\\
&  \label{eq:cai-zhang2}\\
\|\sinTeU\|_F & \leq D_{F} (U, \hU) \leq \sqrt{2}\, \|\sinTeU\|_F. \nonumber
\end{align} 
Although Theorem~\ref{th:Davis-Kahan} only implies the  existence of matrix $O \in \calO_r$ that provides the infimum
in \fr{eq:dist},  the   matrix $W_U \in \calO_r$, delivering the minimum of $D_{F} (U, \hU)$, is known. 
Specifically, if $U^T \hU = W_1 D_U W_2^T$ is the SVD of $U^T \hU$, then $W_U = W_1 W_2^T$  (see, e.g., \cite{procrustes2736}).
It turns out (see, e.g.,  \cite{10.1214/17-AOS1541},  \cite{Cape_L2_inf_AOS2019})  
that $W_U$ delivers an almost optimal upper bound in \fr{eq:dist} under the spectral norm also: 
\be \label{eq:D_K_Spectral}
\| \hU   - U W_U\| \leq \sqrt{2}\,   D_{sp} (U, \hU).   
\ee
In many contexts, however, one would like to derive a similar upper bound for the deviation between $U$ and $\hU$ in the 
two-to-infinity norm. For this purpose, for any matrix $A$, denote 
\be \label{eq:2inf_dist}
D_{2,\infty} (U, \hU) = \inf_{O \in \calO_{r}} \| \hU - UO \|_{2,\infty},
\ee
where $\|A\|\tinf = \displaystyle \max_i\ \|A(i,:)\|$ and $\|A(i,:)\|$  is the norm of the $i$-th row of $A$. 
Specifically, if   $\|U \|\tinf$ is small, then $D_{2,\infty} (U, \hU)$ may be significantly smaller than $D_{sp} (U, \hU)$,
which is extremely advantageous in many applications.

It is worth observing that while the upper bounds for  $D_{sp} (U, \hU)$ and  $D_{F} (U, \hU)$ 
are relatively straightforward, this is no longer true in the case of $D_{2,\infty} (U, \hU)$. 
The seminal paper of  \cite{Cape_L2_inf_AOS2019}   develops an expansion for $\hU   - U W_U$,
which allows to derive upper bounds  for $\|\hU   - U W_U\|\tinf$. 
While the paper contains a number of very useful examples,  the universal upper bound leaves a lot of room for improvement. 
Specifically, the generic upper bound in Theorem~4.2  of \cite{Cape_L2_inf_AOS2019} relies on the $l_1$-norms of the rows 
of the error matrix, which grow too fast in many practical situations.

In the last few years, many authors  (see, e.g., \cite{abbe_fan_AOS2022}, \cite{abbe_fan_aos2020},  \cite{cai_AOS2021_unbalanced_incomplete},
\cite{chen_Assymetry_helps_AOS2021}, \cite{Chen_2021},  \cite{lihua_lei_2020_generic_symmetric}, 
\cite{tsyganov2026_matrixfree_2-to-infty}, \cite{wang2024_singular_subspaces_random}, 
\cite{Xie-Bernoulli2024}, \cite{xie2025AOS},  \cite{Yan_AOS2024_MissingData}, \cite{zhou2024_illposedPCA}) 
obtained upper bounds for $\|\hU   - U W_U\|\tinf$, designed for a variety of scenarios. 
While some of those upper bounds have some correspondence to the upper bounds derived in this paper, 
the majority of those upper bounds were obtained  under  relatively strict assumptions on the error distribution and problem settings.
The  main difference between the present paper and most of the ones cited above is that   those works were written 
 with   specific applications in mind,  while the objective of this paper is to provide a universal useful tool that 
can be applied for a variety of scenarios, even in the absence of probabilistic assumptions, or in the presence of mild assumptions.
Specifically, results in this paper are derived without a common assumption that the elements of the error matrix are independent. 
Although some of the above mentioned papers contain such upper bounds,  none of them provide a comprehensive picture 
of the deviations between the true and estimated singular spaces in the two-to-infinity errors. 
We present a   detailed comparison with the existing results in Section~\ref{sec:comparison}.

The purpose of this paper is   to provide a complete toolbox for derivation of  universal upper bounds for  $\|\hU   - U W_U\|\tinf$,
in the spirit of \cite{Cape_L2_inf_AOS2019} and \cite{10.1093/biomet/asv008}.
We argue that results in \cite{Cape_L2_inf_AOS2019} can be refined and improved,  
without additional assumptions or with generic probabilistic assumptions.  That is why the paper 
should be viewed as an extension of the Davis-Kahan (and the Wedin) theorem to the case of the two-to-infinity norm
rather than a study of a specific statistical problem. In particular,   the paper   starts with the case of symmetric errors,
then handles the case of non-symmetric errors, and subsequently considers symmetrization of the problem. 
In each of these three  situations, we derive upper bounds for the errors with no probabilistic assumptions and
subsequently   provide upper bounds under generic probabilistic assumptions on the errors. 
In addition, these results are later refined if the errors are heavy-tailed or exhibit exponential decay.
Although some upper bounds are cumbersome, they are completely straightforward,
and their presence    for symmetric, non-symmetric and symmetrized versions allows one to compare precisions of those techniques.

We emphasize that our  goal is not to derive the most accurate optimal upper bound for some particular
problem of interest but rather to provide an instrument that can be applied in a variety of scenarios.
Although we examine  sufficient conditions for perfect clustering as an application of the upper
bounds constructed in the paper, this is just one example of the situation where the theories of the paper 
can be helpful. We point out  that, although this paper studies only this particular application,  
its results  can be potentially useful for many other tasks  such as, e.g., 
noisy matrix completion (see, e.g., \cite{abbe_fan_aos2020}, \cite{Chen_Fan_2019_PNAS_Matrix_Completion}),
or derivation of low-rank contextual bandits (see, e.g., \cite{Jedra2024_LowRank_Bandits}).
\\

\noindent
Specifically, this  paper delivers the following novel results:
\begin{enumerate}  

\item 
We develop upper bounds for  $\|\hU   - U W_U\|\tinf$ with   no additional assumptions,
when $U$ and $\hU$ are obtained from either a symmetric or non-symmetric matrix. 
Although those upper bounds sometimes involve a number of quantities, they are completely straightforward. 

\item  
In the case when the data and the error matrices are not symmetric, we show that symmetrizing the 
problem often leads to more accurate upper bounds for $\|\hU   - U W_U\|\tinf$.

\item
Although the main objective of the paper is to establish upper bounds  for $\|\hU   - U W_U\|\tinf$
that are valid for   any errors,   generic results are supplemented by the upper bounds, derived under mild probabilistic assumptions 
on the error matrices. Nevertheless, those assumptions are weaker than the ones, employed in
majority of papers. The upper bounds in the paper do not require independence of the elements of the error matrix,
and can be used when errors are heavy-tailed. In addition, the paper offers refinements of the results 
in the situation when the errors are sub-Gaussian or sub-exponential.

\item
One of the important novel results is formulation of the generic sufficient conditions for perfect clustering, 
with no or very few mild assumptions 
on the errors. Subsequently, these conditions are  tailored for solution of  specific problems. 
In particular, Section~\ref{sec:subsampled} derives sufficient conditions for perfect clustering 
of a  sampled sub-network, in the case when the original network is equipped by the Stochastic Block Model. 
Another success is confirming that the between-layer clustering 
algorithm in \cite{pensky2021clustering} indeed  leads to perfect clustering, 
the result that was eluding the authors for a long time.   Notably, perfect clustering is proved 
 without any additional assumptions with respect to \cite{pensky2021clustering}, and employs a
generic upper bound  on $\|\hU   - U W_U\|\tinf$, which does not rely on  assumptions on the error distribution.

\end{enumerate}

\noindent 
The rest of the paper is organized as follows. Section~\ref{sec:notations} introduces notations used in the paper. 
Section~\ref{sec:symmetric}  starts the paper with the case, where both the matrix of interest 
and the data matrix are symmetric. This is a standard setting of the Davis-Kahan theorem, 
which we extend to the case of  two-to-infinity norm errors without any additional conditions (Theorem~\ref{th:sym_up_bound}),
and with mild probabilistic assumptions on the error matrix (Theorem~\ref{th:sym_up_bound_new}). 
We show that our generic upper bounds in Theorem~\ref{th:sym_up_bound} are more accurate than the ones in \cite{Cape_L2_inf_AOS2019}.
Section~\ref{sec:nonsymmetric} studies the case, where both the matrix of interest 
and the data matrix are non-symmetric. In this section, we derive  upper bounds for $\|\hU   - U W_U\|\tinf$
with no probabilistic assumptions (Theorem~\ref{thm:nonsym_up_bound}), as well as with non-restrictive 
probabilistic assumptions on the error matrix  (Theorem~\ref{thm:nonsym_probab_up_bound_new}).
Nevertheless, in Section~\ref{sec:symmetrized}, we argue 
that symmetrizing the problem sometimes  allows to significantly improve the accuracy of $\hU$ as an estimator of $U$.
Specifically, Theorem~\ref{thm:symetrized_up_bound}  provides generic upper bounds for $\|\hU   - U W_U\|\tinf$,
while Theorem~\ref{thm:symetrized_probab_up_bound} upgrades those  bounds, when additional probabilistic assumptions on the
error matrix are imposed.

Section~\ref{sec:perfect_clust} considers application of our theories to perfect spectral clustering. 
We would like to  point out that this is just one of other numerous applications of the error bounds that have been derived 
in the previous sections.  
In particular,  Propositions~\ref{prop:perfect_clust}~and~\ref{prop:perfect_clust_sym}  in Section~\ref{sec:spec_clust} use the upper bounds 
in the previous sections to deliver sufficient conditions for   perfect spectral clustering in the cases 
of non-symmetric and symmetric data matrices, respectively. 
Section~\ref{sec:didactic} compares those conditions in the case of independent Gaussian errors. While we are keenly aware that 
this setting is very well studied in the literature, our goal in Section~\ref{sec:didactic} is not 
to derive novel results but rather to demonstrate how various approaches to derivation of  $\|\hU   - U W_U\|\tinf$,
offered in Sections~\ref{sec:nonsymmetric}~and~\ref{sec:symmetrized}, lead to different sufficient conditions for 
perfect clustering. Subsequently, Sections~\ref{sec:subsampled}~and~\ref{sec:multilayer}
employ the theories above to random networks. Section~\ref{sec:subsampled} is devoted to the 
situation where one sub-samples nodes in a very large network, equipped with communities,  
and subsequently clusters those nodes.
Section~\ref{sec:multilayer} studies a multilayer network where all layers have the same set of nodes, and 
layers can be partitioned into groups with different subspace structures.  
Section~\ref{sec:comparison}  provides a comparison of the results in the present paper with the existing ones. 
The proofs of all statements in the paper are provided in Supplementary Material. 

\renewcommand{\arraystretch}{1.2} 
\begin{table*}[t]
\centering
\setlength{\tabcolsep}{2pt}
\begin{tabular}{|p{0.2\textwidth} p{0.2\textwidth} p{0.2\textwidth} p{0.25\textwidth}|}
\multicolumn{4}{  l }{ {\bf Table 1.} \ \  Notations }\\ 
\hline   \hline
\multicolumn{4}{| l |}{ {\bf Group 1:} \  Non-random with $Y =X X^T$ }\\ 
\hline  
$\epsu = \|U\|\tinf$ &    $\epsv  =   \|V\|\tinf$ &  $\tepsy =  d_r^{-2}\, \|\diag(Y)\|_{\infty}$ &  \\
\hline 
\hline  
\multicolumn{4}{| l |}{{\bf Group 2:} \  Random with $\scrE = \hY - Y$, \  $q=1,2$ }\\ 
\hline  
$\Del_0 = |\lam_r|^{-1}\, \|\scrE\|$ &  $\Del_{q,\infty} = |\lam_r|^{-1}\, \|\scrE\|_{q,\infty}$ & 
$\DelEu =  |\lam_r|^{-1}\, \|\scrE\, U\|\tinf$ & \\
\hline 
\hline
\multicolumn{4}{| l | }{ {\bf Group 3:} \  Random with $\Xi = \hX - X$,\  $q=1,2$ }\\ 
\hline
$\tDel_0 = d_r ^{-1}\, \|\Xi\|$ & $\tDel_{q,\infty} = d_r ^{-1}\, \|\Xi\|_{q,\infty}$   & 
$\tDel_{2,\infty}^T = d_r ^{-1}\, \|\Xi^T\|_{2,\infty}$  &   $\tDeluvo =  d_r^{-1}\, \|U^T\, \Xi\, V\|$  \\
$\tDeluo   =   d_r^{-1}\, \|U^T \Xi\|$ & $\tDelov =   d_r^{-1}\, \|\Xi\, V \|$  & $\tDelvtinf = d_r ^{-1}\, \|\Xi \, V\|_{2,\infty}$  & \\
\hline  \hline
\multicolumn{4}{| l |}{ {\bf Group 4:} \  Random with $\barXixi  = \scrH(\Xi\, \Xi^T)\, \tilh + \Xi\, \Xi^T \, (1 - \tilh) $}\\ 
\hline
$\tDelxio = d_r ^{-2}\, \| \barXixi \|$ & $\tDelxiuo = d_r ^{-2}\, \| \barXixi \, U\|$ &   
$\tDelxitinf = d_r ^{-2}\, \| \barXixi \|_{2,\infty}$  & $\tDelxiutinf = d_r^{-2} \|\barXixi \, U \|_{2,\infty}$   \\
%
\hline   \hline
\multicolumn{4}{| l |}{ {\bf Group 5:} \  Random with $\tilscrE = \scrH(\hX\, \hX^T)\, \tilh + \hX\, \hX^T \, (1 - \tilh) - X\, X^T$ }\\ 
\hline
$\tDelEo =   d_r ^{-2}\, \| \tilscrE \|$ & $\tDelEuo =   d_r ^{-2}\, \| \tilscrE\, U\|$ &  &\\
\hline  \hline
 \end{tabular}
\end{table*}


\ignore{
   

\renewcommand{\arraystretch}{1.4} 
\begin{table}   [t]
\begin{center}
\begin{tabular}{|l l l  |    }
\multicolumn{3}{ l }{{\sc  Table 1.\  Notations.}}\\ 
\hline \hline
\multicolumn{3}{| l |}{ {\bf Group 1:} \  Non-random with $Y =X X^T$ }\\ 
\hline  
$\epsu = \|U\|\tinf$ &    $\epsv  =   \|V\|\tinf$ &  $\tepsy =  d_r^{-2}\, \|\diag(Y)\|_{\infty}$   \\
\hline 
\hline  
\multicolumn{3}{| l |}{{\bf Group 2:} \  Random with $\scrE = \hY - Y$, \  $q=1,2$ }\\ 
\hline  
$\Del_0 = |\lam_r|^{-1}\, \|\scrE\|$ &  $\Del_{q,\infty} = |\lam_r|^{-1}\, \|\scrE\|_{q,\infty}$ & 
$\DelEu =  |\lam_r|^{-1}\, \|\scrE\, U\|\tinf$  \\
\hline 
\hline
\multicolumn{3}{| l | }{ {\bf Group 3:} \  Random with $\Xi = \hX - X$,\  $q=1,2$ }\\ 
\hline
$\tDel_0 = d_r ^{-1}\, \|\Xi\|$ & $\tDel_{q,\infty} = d_r ^{-1}\, \|\Xi\|_{q,\infty}$   & 
$\tDel_{2,\infty}^T = d_r ^{-1}\, \|\Xi^T\|_{2,\infty}$  \\
$\tDeluvo =  d_r^{-1}\, \|U^T\, \Xi\, V\|$ &  $\tDeluo   =   d_r^{-1}\, \|U^T \Xi\|$ & 
$\tDelov =   d_r^{-1}\, \|\Xi\, V \|$ \\
%
                  & $\tDelvtinf = d_r ^{-1}\, \|\Xi \, V\|_{2,\infty}$ & \\
\hline  \hline
\multicolumn{3}{| l |}{ {\bf Group 4:} \  Random with $\barXixi  = \scrH(\Xi\, \Xi^T)\, \tilh + \Xi\, \Xi^T \, (1 - \tilh) $}\\ 
\hline
$\tDelxio = d_r ^{-2}\, \| \barXixi \|$ & $\tDelxiuo = d_r ^{-2}\, \| \barXixi \, U\|$ &  \\
$\tDelxitinf = d_r ^{-2}\, \| \barXixi \|_{2,\infty}$  & $\tDelxiutinf = d_r ^{-2}\, \|\barXixi \, U \|_{2,\infty}$ & \\
%
\hline \hline
\multicolumn{3}{| l |}{ {\bf Group 5:} \  Random with $\tilscrE = \scrH(\hX\, \hX^T)\, \tilh + \hX\, \hX^T \, (1 - \tilh) - X\, X^T$ }\\ 
\hline
$\tDelEo =   d_r ^{-2}\, \| \tilscrE \|$ & $\tDelEuo =   d_r ^{-2}\, \| \tilscrE\, U\|$ & \\
\hline
 \end{tabular}
\end{table*}
 
}


\subsection{Notations}
\label{sec:notations}

We denote $[n] = \{1, ...,n\}$, $a_n = O(b_n)$ if $a_n \leq C b_n$,  $a_n = \om(b_n)$ if $a_n \geq c b_n$,
$a_n \asymp b_n$ if $c b_n \leq a_n \leq C b_n$,  where $0<c\leq C <\infty$ are absolute constants independent of $n$.
Also, $a_n = o(b_n)$  and $a_n = \Om(b_n)$ if, respectively, $a_n/b_n \to 0$ and  $a_n/b_n \to \infty$
as $n \to \infty$. 
\ignore{
If $\xi_n$ is a sequence of random variables and $r_n$ and $\tau$ are positive deterministic quantities,
we write 
\bes
\xi_n = O_{\PP} (r_n, \tau) \quad \mbox{if}\quad  \PP(\xi_n \geq \Ctau\, r_n) \leq n^{-\tau}
\ees 
for some absolute constant $\Ctau < \infty$ and $n$ large enough.
}
We use $C$ as a generic absolute constant, and $\Ctau$ as a generic absolute constant that depends on $\tau$ only.

For any vector $v \in \RR^p$, denote  its $\ell_2$, $\ell_1$, $\ell_0$ and $\ell_\infty$ norms 
by $\|   v\|$, $\|   v\|_1$,  $\|   v\|_0$ and $\|  v\|_\infty$, respectively. 
Denote by $1_m$  the $m$-dimensional column vector with all components equal to one.

The column $j$ and the row $i$ of a matrix $A$ are denoted by $A(:, j)$ and $A(i, :)$, respectively.
For any matrix $A$,  denote its spectral, Frobenius, maximum,  $(2,\infty)$  and $(1,\infty)$ norms by, respectively, 
 $\|  A \|$, $\|  A \|_F$, $\|A\|_{\infty}$, 
$\|  A \|_{2,\infty} = \displaystyle  \max_i \|A(i,:)\|$  and $\|  A \|_{1,\infty} = \displaystyle  \max_i\|A(i,:)\|_1$. 
We are aware that the latter differs from the classical notation of the respective induced norm and 
emphasize that notation $\|  A \|_{1,\infty}$ is motivated entirely by the readers' convenience and clarity of
presentation.
Denote the $k$-th eigenvalue  and the $k$-th singular value  of $A$ by $\lam_k(A)$ and $\sig_k(A)$, respectively.
Let $\SVD_r(A)$ be $r$ left leading eigenvectors of $A$.
Let $\vect(A)$ be the vector obtained from matrix $A$ by sequentially stacking its columns. 
Denote the diagonal of a matrix $A$ by $\diag(A)$. Also, with some abuse of notations, denote the $K$-dimensional 
diagonal matrix with $a_1,\ldots,a_K$ on the diagonal by $\diag(a_1,\ldots,a_K)$, and the diagonal matrix consisting of only the diagonal of a square matrix $A$ by $\diag(A)$.
Denote 
$\calO_{n,K} = \left \{A \in \RR^{n \times K} : A^T A = I_K \right \}$,   $\calO_n=\calO_{n,n}$.
%

In what follows, we use $\Del$ and $\tDel$  with subscripts to denote various norms of the error, 
$\Del$ for  $\scrE = \hY - Y$, where matrices  $Y$ and $\hY$ are symmetric,
and $\tDel$ for norms associated with the error $\Xi = \hX - X$, where matrices $X$ and $\hX$ are  not symmetric. 
We use subscripts 0, $(1, \infty)$ and $(2, \infty)$ for, respectively, the spectral norm, the $(1, \infty)$-norm and the  $(2, \infty)$-norm. 
For   the quantities, defined using conventions above, we  denote their upper bounds (attained with high probability) by $\eps$ 
with the same subscripts as for $\Del$, and by $\teps$ with the same subscripts as for $\tDel$.
The complete list of notations is presented in Table~1.


\section{A   Davis--Kahan theorem in the two-to-infinity norm: symmetric case. }
\label{sec:symmetric}

Consider symmetric matrices $Y, \hY \in \RR^{n \times n}$ and denote $\scrE = \hY - Y$. 
Then, for any $r <n$, one has the following eigenvalue expansions
\be \label{eq:main_rep}
Y = U \Lam  U^T + U_{\perp} \Lam_{\perp} U_{\perp}^T, \quad 
\hY = \hU \hLam  \hU^T + \hU_{\perp} \hLam_{\perp} \hU_{\perp}^T, 
\ee 
where $U, \hU \in \calO_{n,r}$,  $U_{\perp}, \hU_{\perp} \in \calO_{n,n-r}$,
$\Lam = \diag(\lam_1, ...,  \lam_r)$, $\hLam = \diag(\hlam_1, ...,  \hlam_r)$,  
$\Lam_{\perp}= \diag(\lam_{r+1}, ...,  \lam_n)$ and $\hLam_{\perp} = \diag(\hlam_{r+1}, ...,  \hlam_n)$.
As before, consider
\be \label{eq:W_u}
W_U = W_1 W_2^T \quad \mbox{where}\quad U^T \hU = W_1 D_U W_2^T.
\ee
One of the main results of \cite{Cape_L2_inf_AOS2019} is the expansion of the error as
\be \label{eq:cape-expan}
\begin{array} {ll}
\hU & - U W_U n  = (I - U U^T) \scrE U W_U \hat{\Lambda}^{-1}  \\
& + (I - U U^T) \scrE (\hU - U W_U) \hat{\Lambda}^{-1} \\
& + (I - U U^T) Y (\hU - U U^T \hU) \hat{\Lambda}^{-1}  \\
&  + U (U^T \hU - W_U),
\end{array}
\vspace{-2mm}
\ee 
which allows one to obtain a straightforward upper bound for $\|\hU - U W_U \|\tinf$.
Assume that, for some absolute constant  $c_{\lam}$, one has
\be \label{eq:lamr_r}
\lam_r - \lam_{r+1} \geq c_{\lam}  |\lam_r|, \quad c_{\lam}>0.
\ee
For $q=1,2$, denote 
\begin{align} \label{eq:errors}
& \Del_0 = |\lam_r|^{-1}\, \|\scrE\|, \quad \Del_{q,\infty} = |\lam_r|^{-1}\, \|\scrE\|_{q,\infty}, \\
& \DelEu =  |\lam_r|^{-1}\, \|\scrE\, U\|\tinf, \quad \epsu =  \|U\|\tinf, \nonumber
\end{align}
where, for any matrix $B$, one has $\|B\|_{q,\infty} = \displaystyle \max_i \|B(i,:)\|_q$.
In \fr{eq:errors}, $\Del_0$, $\Del_{q,\infty}$ and $\DelEu$ are random variables, while 
$\epsu$ is a fixed quantity that depends on $n$. We assume those quantities to be bounded with high probability.
\\

\noindent
{\bf Assumption A1 \ (Group 1 in Table 1).}  For any $\tau>0$,
there exists a constant $\Ctau$ and deterministic quantities $\epso$, $\eps_{q,\infty}$, $\epsEu$, 
that depend on $n$, $r$, and possibly $\tau$,   such that simultaneously, with probability at least $1 - n^{-\tau}$, one has
\be \label{eq:probab_errors}
\Del_0 \leq \Ctau  \epso, \  \Del_{q,\infty} \leq \Ctau \eps_{q,\infty},\  \DelEu \leq \Ctau \epsEu,
%
\ee
for $q=1,2$ and $n$ large enough. Here, we use $\Ctau$ as a generic absolute constant that depends on $\tau$ only and can take different values 
at different places. 

\noindent
Note that Assumption A1 and a similar Assumption A3 later do not require the elements of error matrix to 
follow any thin-tailed distributions since the quantities in \fr{eq:probab_errors} can depend on the constant $\tau$.
In those assumptions we are merely trying to avoid fixing the acceptable probability as, e.g., $1 - n^{-1}$,
or $1 - n^{-2}$,  or $1 - n^{-10}$,  as it is done in some other papers. 
Specifically, Assumption A1 holds for    heavy-tailed errors.
%
%
It is easy to see that  $\epsu \leq 1$ and  $\Delinf \leq \Delo$. Also,  by  Proposition~6.5 of  \cite{Cape_L2_inf_AOS2019}),
 $\DelEu \leq \min(\Delinf, \epsu\, \Del_{1,\infty})$, hence,  $\epsEu \leq \min(\epsinf, \epsu\, \eps_{1,\infty})$.
Expansion \fr{eq:cape-expan} implies the following upper bounds.

\begin{theorem}\label{th:sym_up_bound}  
Let  $Y, \hY \in \RR^{n \times n}$ have the eigenvalue expansions \fr{eq:main_rep} and $\scrE = \hY - Y$. 
Let \fr{eq:lamr_r} hold. If $\Delo \leq 1/4$, then 
\begin{align} 
\|\hU  & - U W_U\|\tinf \leq    \lkr   \frac{4}{3}   + \frac{2\ }{3\, \clam}  + \frac{1}{\clam^2} \rkr \Delo\, \epsu \nonumber \\
& + 
\frac{8\, \Delo}{3\, \clam}    \lkr \Delinf + \frac{|\lam_{r+1}|}{|\lam_r|} \rkr + \frac{4}{3} \DelEu.  \label{eq:sym_up_bound}
\end{align}
If, in addition, \fr{eq:probab_errors} is valid with $q=2$ and   $\epso \leq 1/4$, then,  with probability at least $1 - n^{-\tau}$, one has
\begin{align}
\|\hU - U W_U\|\tinf & \leq  \Ctau \lkr   \epso\, \epsu + \epso\, \epsinf \right. \nonumber \\
& + \left.  |\lam_r|^{-1}\, |\lam_{r+1}| \, \epso  +   \epsEu \rkr   
 \label{eq:probab_sym_up_bound}
\end{align} 
Here, $\DelEu \leq \min(\Delinf, \epsu\, \Del_{1,\infty})$, and  hence,  $\epsEu \leq \min(\epsinf, \epsu\, \eps_{1,\infty})$.
%
%
\end{theorem}

\medskip

Note that, since we made absolutely no assumptions on the values of $\epso$, $\eps_{q,\infty}$  and  $\epsEu$
in \fr{eq:probab_errors}, Theorem~\ref{th:sym_up_bound} applies to any errors that are bounded with high probability.
Also observe that, if $\rank(Y)=r$, so that  $\lam_{r+1}=0$  and $\clam =1$, then due to 
$\max(\|\scrE\|, \|\scrE\|\tinf) \leq \|\scrE\|_{1,\infty}$,  one has $\max(\Delo, \Delinf, \DelEu) \leq \Del_{1,\infty}$, and 
\be \label{eq:cape_like}
\|\hU - U W_U\|\tinf \leq  7\, \epsu\, \Del_{1,\infty}.
\ee 
Observe that this upper bound  is more accurate than  the one in Theorem~4.2 of \cite{Cape_L2_inf_AOS2019},
which states the infimum of the  approximation error
\bes 
\inf_{O \in \calO_r} \|\hU - U O \|\tinf  \leq  14\, \epsu\, \Del_{1,\infty}
\ees
under a  stronger (due to $\Del_{1,\infty} \geq \Delo$) condition $\Del_{1,\infty} \leq 1/4$. 
Unfortunately,  in many situations the upper bound \fr{eq:cape_like} is not  useful. 
Observe that, not only  $\eps_{1,\infty} \geq \epso$, but, in addition,
$\eps_{1,\infty}$ can be significantly higher than $\epso$ or $\epsEu$. 
For example, if $\scrE$ has independent standard Gaussian entries, 
then $\epso \asymp |\lam_r|^{-1}\,\sqrt{n}$, $\epsEu \asymp |\lam_r|^{-1}\,\sqrt{r}\, \log n$ 
and $\eps_{1,\infty}\asymp |\lam_r|^{-1}\, n$, so that 
$\epsEu \asymp \epso \ll \eps_{1,\infty}$, if $r \ll n$.
For this reason, in a general situation, one should use the upper bound \fr{eq:sym_up_bound} 
rather than \fr{eq:cape_like}.

As we have mentioned, the upper bound \eqref{eq:probab_sym_up_bound} holds under a variety of  assumptions.
Below, we provide a corollary of Theorem~\ref{th:sym_up_bound} in the case when the above the 
diagonal entries of matrix $\scrE$ are  independent heavy-tailed random variables.

\begin{cor}\label{cor:sym_heavy_tailed}  
Let  $Y, \hY \in \RR^{n \times n}$ have the eigenvalue expansions \fr{eq:main_rep} and $\scrE = \hY - Y$. 
Let $\scrE(i,j)$ be independent zero mean variables for $1 \leq i \leq j \leq n$ with $\EE \lkv \scrE(i,j)\rkv^2 \leq \sig^2$
and  $\EE \lkv \scrE(i,j)\rkv^{2s} \leq \nu_{2s}$, $s \geq 2$. If $n$ is large enough, so that $\Delo \leq 1/4$, then, with probability at least $1 - n^{-\tau}$, one has
\be \label{eq:probab_sym_heavytail}
\|\hU - U W_U\|\tinf \leq  \Ctau   \delrs  \lkr \epsu  n^{\frac{1}{2s}} + 
  \frac{|\lam_{r+1}|}{|\lam_r|} 
  +   \delrs \rkr  
  \vspace{-2mm}
\ee 
Here, $\displaystyle \delrs = |\lam_r|^{-1}\, n^{\frac{\tau}{2s}}\, \lkr \sig \sqrt{n} +  (n\,\nu_{2s})^{\frac{1}{2s}} \rkr$.
\end{cor}

If elements of matrix $\scrE$ have faster decline, the error bounds can be improved.
To this end, let us compare the magnitudes of the terms in \fr{eq:sym_up_bound}. For simplicity, we 
consider the case when $|\lam_r|^{-1}\, |\lam_{r+1}|$ is very small or zero. 
Then, we need to analyze three terms: $\Delo\, \epsu$, $\Delo\, \Delinf$ and $\DelEu$.
There is nothing one can do to remove the last term, $\DelEu$. Indeed, as it follows from 
the proof of Theorem~\ref{th:sym_up_bound}, this term comes from
$\| \scrE U W_U \hat{\Lambda}^{-1}  \|\tinf$, and, if $|\lam_1|/|\lam_r|$  is bounded above by a constant,
then  $\|\scrE U W_U \hat{\Lambda}^{-1}\|\tinf  \geq C\, \DelEu$. 
The relationship between $\Delo\, \epsu$ and  $\Delo\, \Delinf$ can vary depending on the nature of matrices $Y$ and $\scrE$.
It is always true that $\sqrt{r/n} \leq \epsu \leq 1$ and $\Delinf \leq \Delo$ but those inequalities allow for large 
variations of quantities. However, while the term $\Delo\, \epsu$ appears multiple times in the derivation of the 
upper bound \fr{eq:sym_up_bound} and is hard to eliminate, 
the   term $\Delo\, \Delinf$ can be reduced under additional conditions on the error. 
\\

\noindent
{\bf Assumption A2.\ }   For any fixed $\tau >0$, there exists an absolute constant $\Ctau$ that depends on $\tau$ only,
such that, for any matrix $G \in \RR^{n \times r}$ and  for some deterministic quantities $\eps_1$ and $\eps_2$, that  depend on $n$ and $r$, 
but not on matrix $G$ and $\tau$,
one has 
\begin{align} \label{eq:assump_A2}
\PP \lfi  \|\scrE\, G\|\tinf \right.   \leq \Ctau\, |\lam_r|\, \lkv \eps_1\, \|G\|_F \right. & + \left. \eps_2\, \|G\|\tinf \rkv \, \left. \rfi\\
& \geq 1 - n^{-\tau}. \nonumber
\end{align}
In addition,  $\epso$, $\epsEu$ and $\eps_{q,\infty}$, $q=1,2$,  in \fr{eq:probab_errors} depend on $n$ and $r$, but not on $\tau$.

 \medskip

Note that some version of  Assumption A2 is always  valid,
as long as  $\epso$, $\epsEu$ and $\eps_{2,\infty}$ are independent of $\tau$. 
 Indeed, since $\|\scrE\, G\|\tinf \leq \|\scrE\|\tinf \, \|G\|$,
\fr{eq:assump_A2} holds with $\eps_1 = \epsinf$ and $\eps_2 = 0$, in which case Theorem~\ref{th:sym_up_bound_new} 
reduces to Theorem~\ref{th:sym_up_bound} provided $r = O(1)$. Alternatively, 
it also holds with $\eps_1 = 0$ and $\eps_2 = \eps_{1,\infty}$.  
However   Assumption A2 is designed for the situation where elements of matrix $\scrE$ are Bernstein-type, 
sub-Gaussian or sub-exponential, in which case one can provide specific bounds for those quantities. 
In particular, the following  statement is true.

\begin{lem}\label{lem:lemma_Assump2}  
Let rows of $\scrE$ be such that $\EE\lkv (\scrE(i,:))^T  \scrE(i,:) \rkv = \Sig$. \\
a)\ If rows of $\scrE$ are sub-Gaussian with $\|\scrE(i,:)\, u\|_{\psi_2} \leq K\, \sqrt{u^T \Sig u}$
for any fixed vector $u$, then Assumption A2 holds with $|\lam_r|\, \eps_1 = K\, \sqrt{\log n\, \|\Sig\|}$
and $\eps_2 =0$.\\
b)\ If rows of $\scrE$ are sub-exponential  with $\|\scrE(i,:)\, u\|_{\psi_1} \leq K\, \sqrt{u^T \Sig u}$
for any fixed vector $u$, then Assumption A2 holds with $|\lam_r|\, \eps_1 = K\, \log n\,  \sqrt{\|\Sig\|}$
and $\eps_2 =0$.\\
c)\ If the elements of the top half of matrix $\scrE$ are independent $(v, H)$-Bernstein variables, i.e., 
$\EE\lkv |\scrE(i,j)|^k \rkv \leq 0.5\, v\, k! \, H^{k-2}$ for all integers $k \geq 2$ and $i \leq j$, then
Assumption A2 holds with $|\lam_r|\, \eps_1 = \sqrt{v\, \log n}$, 
 $|\lam_r|\, \eps_2 = H, \log n$.\\
\end{lem}

\noindent 
In order to use condition \fr{eq:assump_A2} we apply the ``leave-one-out'' analysis. 
For any  $l \in [n]$, define 
\be \label{eq:scrEupl}
\scrE\upl(i,j) = \lfi
\begin{array}{ll}
\scrE(i,j), & \mbox{if}\quad i \neq l, j \neq l\\
0, & \mbox{if}\quad i = l\ \mbox{or}\   j = l.
\end{array}\right.
\ee 
The following statement provides an improved upper bound under Assumption A2. 

\begin{theorem}\label{th:sym_up_bound_new} 
Let conditions of Theorem \ref{th:sym_up_bound} and Assumption A2 hold. 
Let matrix $\hY$ be such that, for any  $l \in [n]$,  row $\scrE(l,:)$ of $\scrE$ and $\scrE\upl$
are independent from each other. 
If  
\be \label{eq:eps_conditions}
\epso= o(1),\quad \eps_1 = o(1), \quad \eps_2 = o(1) \quad \mbox{as} \quad  n \to \infty,
\ee
 then, for $n$ large enough, with probability at least $1 - 2 \, n^{-\tau}$, one has 
\begin{align} \label{eq:sym_up_bound_new}
 \|\hU - U W_U\|\tinf & \leq  \Ctau\, \lkr \epso\, \epsu + \epso\, \eps_1\, \sqrt{r} \right.\\
& + \left. |\lam_r|^{-1}\, |\lam_{r+1}| \, \epso + \epsEu \rkr . \nonumber
\end{align}
%
\end{theorem}


\section{A   Davis--Kahan theorem in the two-to-infinity norm: non-symmetric case }
\label{sec:nonsymmetric}

Now consider the case when one has an arbitrary    matrix $X \in \RR^{n \times m}$, its estimator $\hX \in \RR^{n \times m}$
and $\Xi = \hX - X$. Denote $\minmn = \min (m,n)$.
Then, for any $r < \minmn$, one has the following SVD expansions
\be \label{eq:main_nonsym}
X = U D  V^T + U_{\perp} D_{\perp} V_{\perp}^T, \quad 
\hX = \hU \hD  \hV^T + \hU_{\perp} \hD_{\perp} \hV_{\perp}^T,
\ee 
where $U, \hU \in \calO_{n,r}$,   $V, \hV \in \calO_{m,r}$,
$U_{\perp}, \hU_{\perp} \in \calO_{n,\minmn-r}$,   $V_{\perp}, \hV_{\perp} \in \calO_{m,\minmn-r}$, 
%
$D = \diag(d_1, ...,  d_r)$, $\hD = \diag(\hd_1, ...,  \hd_r)$,  
$D_{\perp}= \diag(d_{r+1}, ...,  d_{\minmn})$ and 
$\hD_{\perp} =   \diag(\hd_{r+1}, ...,  \hd_{\minmn})$.
Here, 
\be \label{eq:sing-Val}
d_k = \sigma_k(X), \quad \hd_k = \sigma_k(\hX),
\ee 
with $d_1 \geq \ldots \geq d_{\minmn}$  and $\hd_1 \geq \ldots \geq \hd_{\minmn}$.
Similarly to the symmetric case, define  $W_V = W_3 W_4^T$,  where  $V^T \hV = W_3 D_V W_4^T$ is the SVD of $V^T \hV$.
Then, \cite{Cape_L2_inf_AOS2019} provides the following expansion of the difference between 
the true and estimated left eigenbases $\hU$ and $U$:
\begin{align}
\hU & - U W_U = (I - U U^{T}) \Xi V W_V \hat{D}^{-1} \nonumber \\
 & + (I - U U^{T}) \Xi (\hat{V} - V W_V) \hat{D}^{-1} \label{eq:nonsymmetr}\\
& + (I - U U^{T}) X (\hat{V} - V V^{T} \hat{V}) \hat{D}^{-1} 
  + U (U^{T} \hU - W_U).  \nonumber
\end{align}
Consider quantities in {\bf Group 3} of Table 1, for $q=1,2$, 
\begin{align}   
& \tDel_0 = d_r ^{-1}\, \|\Xi\|, \  \tDeluvo =  d_r^{-1}\, \|U^T\, \Xi\, V\|, \nonumber  \\ 
& \tDelvtinf =  d_r ^{-1}\, \|\Xi \, V \|_{2,\infty},  \  \tDelqinf =  d_r ^{-1}\, \|\Xi \|_{q,\infty}, \label{eq:non-sym-err}
%
\end{align}

\noindent
{\bf Assumption A3 \ (Part of Group  3). }  
For any $\tau>0$, there exist a constant $\Ctau$ and deterministic quantities $\teps_{**}$
that depend on $n$, $m$, $r$ and possibly $\tau$,   such that simultaneously, with probability at least $ 1 - n^{-\tau}$, 
for $n$ and $m$ large enough,  all random quantities $\tDel_{**}$ in \fr{eq:non-sym-err}  are bounded   above by  
$\teps_{**}$ with the same respective sub-scripts,
i.e.  
\begin{align} \label{eq:probab_nonsym_err}
& \tDelo \leq \Ctau\, \tepso, \ \tDeluvo \leq \Ctau\,\tepsuvo,\\  
& \tDelvtinf \leq \Ctau\, \tepsvtinf,   \  \tDeltinf \leq \Ctau\,  \tepstinf . \nonumber
\end{align}
Then, in the spirit of Theorem~\ref{th:sym_up_bound}, one can derive an upper bound
for $\|\hU - U W_U\|\tinf$.

\begin{theorem}\label{thm:nonsym_up_bound}  
Let  $X, \hX \in \RR^{n \times m}$ have the  SVD expansions \fr{eq:main_nonsym} and $\Xi = \hX - X$. 
Let  
\be \label{eq:d_r}
d_r - d_{r+1} \geq \cd \, d_r, \quad \cd>0.
\ee
If $\tDelo \leq 1/4$, then 
\begin{align} \label{eq:nonsym_up_bound}
\|\hU - U W_U\|\tinf & \leq  C\, \lkv \epsu\, (\tDeluvo +   \tDelo^2) \right. \\
& + \left.
\tDelvtinf + \tDelo\, (\tDeltinf + d_{r+1}\, d_r^{-1})  \rkv.  \nonumber
\end{align}
Here,   $\tDelvtinf \leq \min(\tDeltinf, \tDel_{1,\infty}\, \epsv)$. 
If, in addition, Assumption A3 holds and $\tepso < 1/4$, then 
\begin{align}
\label{eq:probab_nonsym_up_bound}
\PP \lfi \|\hU - U W_U\|\tinf \right. & \leq  \Ctau \lkv  \epsu\, (\tepsuvo +   \tepso^2) +
\tepsvtinf \right. \\
& +  \left. \left. \tepso\, (\tepstinf + d_{r+1}\, d_r^{-1}) \rkv    \rfi 
\geq 1 - n^{-\tau}. \nonumber
\end{align}
\end{theorem}

\medskip

\noindent
Similarly to the case of symmetric errors, we provide a corollary of Theorem~\ref{thm:nonsym_up_bound}
for the case of heavy-tailed errors.

\begin{cor}\label{cor:nonsym_heavy_tailed}  
Let  $X, \hX \in \RR^{n \times n}$ have the eigenvalue expansions \fr{eq:main_nonsym} and $\Xi = \hX - X$. 
Let $\Xi(i,j)$ be independent zero mean variables for $i \in [n]$, $j \in [m]$
with $\EE \lkv \Xi(i,j)\rkv^2 \leq \sig^2$ and  $\EE \lkv \Xi(i,j)\rkv^{2s} \leq \nu_{2s}$, $s \geq 2$. 
For $k = 1,2,\ldots$, denote
\bes
\tdelrs(k) = d_r^{-1}\, n^{\frac{\tau}{2s}} \lkr \sig \sqrt{k} +  k^{\frac{1}{2s}}\,\nu_{2s}^{\frac{1}{2s}}\rkr.
\ees
If $n$ and $m$ are large enough, so that $\tDelo \leq 1/4$, then, with probability at least $1 - n^{-\tau}$, one has
\begin{align} 
\|\hU - U W_U\|\tinf & \leq  \Ctau \, \lkv   \tdelrs (n+m) \, \lkr \epsu +  
n^{\frac{1}{2s}}\, \tdelrs (m) \right.  \right. \nonumber\\
&   \left.  \left. + d_r^{-1} d_{r+1} \rkr   +  n^{\frac{1}{2s}}\,   \tdelrs (r)  \rkv 
\label{eq:probab_nonsym_heavytail}
\end{align}
\end{cor}

\medskip

\noindent
The upper bound in Theorem~\ref{thm:nonsym_up_bound} can be improved if the rows of matrix $\Xi$  
satisfy an assumption similar to Assumption A2. In this case, we can replace the term $\tepso\,  \tepstinf$
in \fr{eq:probab_nonsym_up_bound} by a tighter upper bound.
\\

\noindent
{\bf Assumption A4.\ } 
Assume that, for any fixed $\tau >0$, there exists an absolute constant $\Ctau$ that depends on $\tau$ only,
such that, for any matrix $G$ and  some deterministic quantities $\teps_1$ and  $\teps_2$,  that  depend on $n$, $m$, $r$, 
but not on $\tau$, and matrix $G \in \RR^{m \times r}$, 
 one has  with probability at least $1 - n^{-\tau}$
\be \label{eq:assump_A4}
\|\Xi\, G \|\tinf \leq \Ctau\, d_r  \,  \lkv  \teps_1\, \|G \|_F + \teps_2\, \|G \|\tinf \rkv 
\ee
In addition,  all quantities in the right  sides of inequalities in \fr{eq:probab_nonsym_err}
depend on $n$, $m$ and $r$, but not on $\tau$.
\\

\noindent 
Note that, similarly to the case of Assumption A2, some version of  Assumption A4 is always  valid,
as long as   all quantities in the right  sides of inequalities in \fr{eq:probab_nonsym_err}
depend on $n$, $m$ and $r$, but not on $\tau$.
 Indeed, since $\|\Xi\, G\|\tinf \leq \|\Xi \|\tinf \, \|G\|$,
\fr{eq:assump_A2} holds with $d_r\, \teps_1 = \tepstinf$ and $\teps_2 = 0$.  
Nevertheless,   Assumption A4 is designed for the case where elements of matrix $\Xi$ are Bernstein-type, 
sub-Gaussian or sub-exponential, in which case one can provide specific bounds for those quantities.

\smallskip

\begin{lem}\label{lem:lemma_Assump4}  
Let rows of $\Xi$ satisfy $\EE\lkv (\Xi(i,:))^T\,  \Xi(i,:) \rkv = \Sig$. \\
a)\ If rows of $\Xi$ are sub-Gaussian with $\|\Xi(i,:)\, u\|_{\psi_2} \leq K\, \sqrt{u^T \Sig u}$
for any fixed vector $u$, then Assumption A4 holds with $d_r\, \teps_1 = K\, \sqrt{\log n\, \|\Sig\|}$
and $\teps_2 =0$.\\
b)\ If rows of $\Xi$ are sub-exponential  with $\|\Xi(i,:)\, u\|_{\psi_1} \leq K\, \sqrt{u^T \Sig u}$
for any fixed vector $u$, then Assumption A4 holds with $d_r\, \teps_1 = K\, \log n\,  \sqrt{\|\Sig\|}$
and $\teps_2 =0$.\\
c)\ If elements of   matrix $\Xi$ are independent $(v, H)$-Bernstein variables, i.e., 
$\EE\lkv |\Xi(i,j)|^k \rkv \leq 0.5\, v\, k! \, H^{k-2}$ for all integers $k \geq 2$ and $i \neq j$, then
Assumption A2 holds with $d_r\, \teps_1  = \sqrt{v\, \log n}$, 
 $d_r\, \teps_2  = H\, \log n$.
\end{lem}

\smallskip 

In what follows, we assume that both $m$ and $n$ are large and that, in addition, for some absolute constant $\tau_0$
\be \label{eq:m_n_rel}
m \leq n^{\tau_0}.
\ee
Then, the following statement holds.

 
\begin{theorem}\label{thm:nonsym_probab_up_bound_new}  
Let conditions of Theorem \ref{thm:nonsym_up_bound} hold, and  Assumptions A3, A4 and \fr{eq:m_n_rel} be valid. 
Let rows of matrix $\Xi = \hX - X$ be independent and $\tepso = o(1)$ as $n, m \to \infty$.
Then, for $n$ and $m$ large enough, with probability at least $1 - 2\, n^{-\tau}$, one has 
 \begin{align}  
\|\hU - U W_U\|\tinf   & \leq   \Ctau \, \lkv \tepsvtinf + \sqrt{r}\, \tepso (\teps_1 + \teps_2 + d_r^{-1}\, d_{r+1}) \right. \nonumber \\ 
& + \left.  \epsu (\tepsuvo + \tepso^2)   \rkv.  \label{eq:nonsym_up_bound_new}  
\end{align}
\end{theorem}
 

\begin{cor}\label{cor:nonsym_subgaus}  
Let  $X, \hX \in \RR^{n \times n}$ have the eigenvalue expansions \fr{eq:main_nonsym} and $\Xi = \hX - X$. 
Let rows of $\Xi$ be independent sub-Gaussian with $\EE\lkv (\Xi(i,:))^T  \Xi(i,:) \rkv = \Sig$ where $\|\Sig\|\leq \sig$. 
If rows of $\Xi$ satisfy  $\|\Xi(i,:)\, u\|_{\psi_2} \leq K\, \sqrt{u^T \Sig u}$
for any fixed vector $u$ and $\tepso = o(1)$ as $n,m \to \infty$, 
then, for $n$ and $m$  large enough, such that $\tDelo \leq 1/4$,  with probability at least $1 - 2\,n^{-\tau}$, one has 
\begin{align} 
\|\hU & - U W_U\|\tinf   \leq  \Ctau \, \lkv   \frac{\sig}{d_r}\, (\sqrt{r} + \sqrt{\log n})  \right. \nonumber\\
& + \frac{d_{r+1}}{d_r}\, \frac{\sig}{d_r}\, (\sqrt{n} + \sqrt{m}) \label{eq:probab_nonsym_subgaus} \\ 
& + \left. \frac{\sig^2}{d_r^2}\, (\sqrt{n} + \sqrt{m})\,   \lkr \sqrt{r\, \log n} + \epsu (\sqrt{n} +   \sqrt{m})\rkr
\rkv. \nonumber
\end{align}
\end{cor}



While the upper bounds   \fr{eq:probab_nonsym_up_bound} and \fr{eq:nonsym_up_bound_new} may be very useful in some cases,  
they both require $\tDel$ to be small when $n$ and $m$ grow. One of the  ways  to  obtain  more accurate 
upper bounds for $\|\hU - U W_U\|\tinf$ in the absence of this condition is to symmetrize the problem. 
Specifically, one can construct an estimator of $Y = X X^T$ and use its leading eigenvectors as 
$\hU$. This may not work very well if the magnitudes of the first $r$ singular values of $X$ 
vary significantly. However, if  for some  absolute constant $\Cd < \infty$ one has
\be \label{eq:singval_cond}
d_1 \leq \Cd\, d_r,
\ee
in some cases, one can reap significant benefits from symmetrizing the problem, as it was shown in, e.g., 
\cite{abbe_fan_AOS2022} and \cite{zhou2024_illposedPCA}.


\section{A   Davis--Kahan theorem in the two-to-infinity norm: symmetrized solution }
\label{sec:symmetrized}

Note that the error   $\|\hU - U W_U\|\tinf$ in the non-symmetric case relies heavily  on the error $\tDelo$.
In some cases, this error may not tend to zero fast enough, or may not tend to zero altogether. In these situations, one can 
use a symmetrized solution proposed below.

Consider, as before,  matrices $X \in \RR^{n \times m}$,  $\hX \in \RR^{n \times m}$, 
 $\Xi = \hX - X$, and let \fr{eq:main_nonsym} be valid. Consider the eigenvalue decomposition 
\be \label{eq:Y_def} 
\hspace*{-2mm} Y \! = \! X X^T = U D^2  U^T \! + \! U_{\perp} D^2_{\perp} U_{\perp}^T, \  
\Lam \! = \! D^2,  \,  \Lam_{\perp} \! = \! D_{\perp}^2, 
\ee  
with $U  \in \calO_{n,r}$  and $U_{\perp} \in \calO_{n,n-r}$,
so \fr{eq:main_rep} holds with $\Lam = D^2$, $\Lam_{\perp} = D_{\perp}^2$.
One of possible  estimators for  $Y$ is $ \hX\, \hX^T$. Then, 
\be \label{eq:new_E}
 \hX\, \hX^T - Y  = \Xi\, \Xi^T + \Xi\, X^T +  X\, \Xi^T.
\ee
Note, however,  that although we do not impose any assumptions on the matrix $\Xi$, in many applications,
its elements are independent  zero mean random variables. In this case, one has $\EE(\Xi\, X^T) = \EE(X\, \Xi^T) = 0$
but $\EE(\Xi\, \Xi^T) = D_{\Xi} \neq 0$, where $D_{\Xi}$ is the diagonal matrix with elements
$D_{\Xi} (i,i) = \EE \|\Xi(i,:)\|^2$. Let $D_Y = \diag(Y)$ be the diagonal of the matrix $Y$.
Then,  $D_{\Xi}$ constitutes the ``price'' of estimating $D_Y$. If $D_{\Xi}$ is larger than $D_Y$,
which happens, e.g.,  in the case of sparse random networks \cite{lei2021biasadjusted},
the errors are reduced, if matrix $\hX\, \hX^T$ is {\it hollowed}, i.e., its diagonal is set to zero.
It is known that removing the diagonal is often advantageous for estimation of eigenvectors
(see, e.g., \cite{abbe_fan_AOS2022},  \cite{ndaoud_aos2022}).

For any square matrix $A \in \RR^{n \times n}$ we denote   its hollowed version by $\scrH(A) = A - \diag(A)$.
It is easy to see that  operator $\scrH$ is linear and that for $q = 1, 2$,
\be \label{eq:hollow_propert}
\|\scrH(A)\| \leq 2\, \|A\|, \quad \|\scrH(A)\|_{q,\infty} \leq \|A\|_{q,\infty}.
\ee
Consider an estimator $\scrH(\hX\, \hX^T)$ of $X\, X^T$, and observe that 
$[\hX\, \hX^T -Y] - [\scrH(\hX\, \hX^T)-Y] = \diag (\hX\, \hX^T)$, a nonnegative definite matrix, 
which means that replacing  $\hX\, \hX^T$ by $\scrH(\hX\, \hX^T)$ may be potentially beneficial. 
Indeed, let matrix $\Xi$ have independent rows with $\EE(\Xi(i,:))=0$ and $\EE\|\Xi(i,:)\|^2=\sig_i^2$, $i \in [n]$. 
Denote $\Sig = \diag(\sig_1^2, \ldots, \sig_n^2)$ and observe that
\be \label{eq:tilscrE}
\EE(\hX \hX^T) \!=\! X X^T \!+ \! \Sig, \  \EE(\hX  \hX^T) \!=\! X X^T \!-\! \diag(X X^T).
\ee
Therefore, both  $\hX\, \hX^T$ and  $\scrH(\hX\, \hX^T)$ are biased estimators of $Y = X X^T$, and 
the decision, whether to apply the hollowing operator or not, depends on which of the biases in \fr{eq:tilscrE}
dominates, and also on their nature. 
For example if $\sig_i = \sig$ for all $i \in [n]$, matrix $X X^T + \Sig  = X X^T + \sig^2 I$ has the same collection of 
eigenvectors as $X X^T$ but strongly heterogeneous noise may be extremely detrimental to estimation of $U$.


In order  to treat both $\hX\, \hX^T$ and $\scrH(\hX\, \hX^T)$ simultaneously, 
we consider the indicator $\tilh$ of hollowing, such that $\tilh=1$ if  
$\scrH(\hX\, \hX^T)$ is used, and $\tilh=0$ otherwise. Denote
\be \label{eq:hY_neq}
\hY = \scrH(\hX\, \hX^T)\, \tilh + \hX\, \hX^T \, (1 - \tilh),   
\ee
and write the eigenvalue decomposition of $\hY$ as in \fr{eq:main_rep}:
\be \label{eq:hY_eigen}
\hY = \hU \hLam  \hU^T + \hU_{\perp} \hLam_{\perp} \hU_{\perp}^T, \quad 
\hU \in \calO_{n,r}, \   \hU_{\perp} \in \calO_{n,n-r}.
\ee
Then $\scrE = \hY - Y$ can be partitioned as 
\be \label{eq:new_sym_E}
\tilscrE =  \hY - Y = 
\tilscrE_1 + \tilscrE_2 + \tilscrE_3 + \tilscrE_d,
\ee 
where  $\tilscrE_1$, $\tilscrE_2$, $\tilscrE_3$ and $\tilscrE_d$ are   components of the error,
the last one being a diagonal matrix:
\begin{align} \label{eq:scrE_components}
& \tilscrE_1 = \barXixi, \ 
\tilscrE_2 = \Xi\, X^T,   \ 
\tilscrE_3    =    X\, \Xi^T,\\
& \tilscrE_d = -   \tilh\, \lkv \diag(Y) +  2 \, \diag(\Xi\, X^T) \rkv. \nonumber
\end{align}
Here,
\be \label{eq:barXixi}
\barXixi= \scrH(\Xi\, \Xi^T)\, \tilh + \Xi\, \Xi^T \, (1 - \tilh).
\ee
Now, as before, one can plug   $\tilscrE$ into the expansion \fr{eq:cape-expan}
and examine the components. For  this purpose,   we denote
\begin{align} 
& \tDelxio = d_r ^{-2} \| \barXixi \|, \ 
\tDeluo   =   d_r^{-1} \|U^T \Xi\|, \   \tDelov =   d_r^{-1} \|\Xi\, V \|, \nonumber \\
& \tDel_{2,\infty}^T = d_r ^{-1}\, \|\Xi^T\|_{2,\infty},\  
\tDelEo =   d_r ^{-2}\, \| \tilscrE \|, 
  \label{eq:non-sym-err_more} \\
& \tDelxiutinf   =    d_r ^{-2}\, \|\barXixi \, U \|_{2,\infty}, \   
\tDelEuo =   d_r ^{-2}\, \| \tilscrE\, U\|. \nonumber 
\end{align}
Also,    similarly to the symmetric case, we assume that quantities in \fr{eq:non-sym-err_more} 
are bounded above by some non-random quantities  with high probability.

\smallskip

\noindent
{\bf Assumption A3* \ (Groups 3,4 and 5). }  
For any $\tau>0$, there exist a constant $\Ctau$ and deterministic quantities $\teps_{**}$
that depend on $n$, $m$, $r$ and possibly $\tau$,   such that simultaneously, with probability at least $ 1 - n^{-\tau}$, 
for $n$ and $m$ large enough,  all random quantities $\tDel_{**}$ in Groups 3,4 and 5 in Table~1 are 
bounded by above by  $\teps_{**}$ with the same respective sub-scripts,
i.e.  
\begin{align} 
& \tDel_0 \leq \Ctau  \tepso, \  \tDeluo \leq \Ctau \tepsuo, \  \tDelov \leq \Ctau \tepsov, \nonumber \\
&   \tDeluvo  \leq \Ctau \tepsuvo, \  \tDel_{q,\infty} \leq \Ctau \teps_{q,\infty},  \ \tDel_{2,\infty}^T  \leq \Ctau \teps_{2,\infty}^T,  \nonumber\\
&    
\tDelxitinf \leq \Ctau \tepsxitinf,  \  \tDelvtinf \leq \Ctau \tepsvtinf, 
   \label{eq:new_probab_errors} \\  
& \tDelxiutinf  \leq \Ctau \tepsxiutinf, \   \tDelxiuo  \leq \Ctau \tepsxiuo, 
\nonumber\\
& \tDelxio  \leq \Ctau \tepsxio, \ 
  \tDelEo \leq \Ctau \tepsEo, \  \tDelEuo \leq  \Ctau \tepsEuo. \nonumber
\end{align}


\noindent
Note that Assumption A3* presents an expanded version of Assumption A3.
Here, we use $\Ctau$ as a generic absolute constant that depends on $\tau$ only and can take different values 
at different places. Then, the following statement holds.


\begin{theorem}\label{thm:symetrized_up_bound}  
Let  $X  \in \RR^{n \times m}$ have the  SVD expansion  \fr{eq:main_nonsym} and $\Xi = \hX - X$. 
Denote 
\bes 
\tepsy = d_r^{-2}\, \max_{i \in [n]}\,  Y(i,i) = d_r^{-2}\, \|\diag(Y)\|_{\infty}.
\ees
Consider the estimator  $\hY$ defined in  \fr{eq:hY_neq} and assume that its eigenvalue expansion is given by 
\fr{eq:hY_eigen}.
If 
\be \label{eq:new_cond_sym}
\tilh\  \tepsy \leq 1/4, \quad \tDelEo \leq 1/2
\ee
and conditions  \fr{eq:d_r} and \fr{eq:singval_cond} hold, then, 
%
%
\begin{align} 
  \|\hU & - U W_U\|\tinf   \leq   C\, \lfi  \tDelxiutinf +  \tDelvtinf  \right. \nonumber\\
  & +  
 d_{r+1}\, d_r^{-1}\, \lkr \tDeluo + \tDeltinf    \rkr + \tilh\, \tepsy  \epsu 
 \label{eq:symetrized_up_bound} \\
& +      \min(\tDelEo, \sqrt{r}\, \tDelEuo) \lkv \tDelxitinf + \epsu \right. \nonumber \\
& + \left. \left. (d_{r+1}\, d_r^{-1})^2  
+ d_{r+1}\, d_r^{-1}\, \tDelo + \tilh\, \tepsy \rkv \rfi. \nonumber
\end{align}
Here,
\be \label{eq:extra_bounds}
\begin{split} 
& \tDelEo \leq C \lkr \tDelxio + \tDelov +  \frac{d_{r+1}}{d_r}  \tDelo + \tilh  \tepsy \rkr,   \\
&\tDelEuo \leq C \lkr \tDelxiuo + \tDelov +  \frac{d_{r+1}}{d_r}  \tDeluo + \tilh   \tepsy\rkr.
\end{split} 
\ee
Moreover, if \fr{eq:new_probab_errors} is valid and $\tepsEo \leq 1/2$,  
then, with probability at least $1 - n^{-\tau}$, one has
\be   \label{eq:probab_symetrized_up_bound}
\begin{split}  
& \|\hU - U W_U\|\tinf   \leq   \Ctau \, \lfi  \tepsxiutinf +  \tepsvtinf  \right. \\
&  + \left.  d_{r+1}\, d_r^{-1}\, \lkr \tepsuo + \tepstinf    \rkr + \tilh\, \tepsy  \epsu \right. \\
& +    \min(\tepsEo, \sqrt{r}\, \tepsEuo) \lkv \tepsxitinf + \epsu \right.\\
& +    \left.   \left. (d_{r+1}\, d_r^{-1})^2  + d_{r+1}\, d_r^{-1}\, \tepso + \tilh\, \tepsy \rkv \rfi.  
\end{split}
\ee
\end{theorem}


We point out  that  one of the advantages of symmetrization is that one does not need $\tDelo$ to be small any more,
which is the requirement of Theorems~\ref{thm:nonsym_up_bound} and \ref{thm:nonsym_probab_up_bound_new}. 
Indeed in the upper bound  \fr{eq:symetrized_up_bound}, $\tDelo$ appears only in the product  with $d_{r+1}\, d_r^{-1}$, 
which may be sufficiently small to offset  $\tDelo$  when it is large. 
Note also that \fr{eq:new_cond_sym} requires $\tepsy \leq 1/4$ 
in the hollowed case. This is very reasonable since one would not use $\tilh =1$ unless $\tepsy$ is small.

The upper bounds in Theorem~\ref{thm:symetrized_up_bound} do not exploit finer features of the error matrix $\Xi$ 
and are similar to the upper bounds in Theorems~\ref{th:sym_up_bound} and \ref{thm:nonsym_up_bound}.
These upper bounds, however, can be improved under additional assumptions on the matrix   $\Xi$.
The following condition is a somewhat stronger version of Assumption~A4 in the previous section (since it requires more quantities to 
be independent of $\tau$).
 
\smallskip
 
\noindent
{\bf Assumption A4*.\ } 
Assume that, for any fixed $\tau >0$, there exists an absolute constant $\Ctau$ that depends on $\tau$ only,
such that, for any matrix $G$ and  some deterministic quantities $\teps_1$ and  $\teps_2$,  that  depend on $n$, $m$, $r$, 
but not on $\tau$, and matrix $G \in \RR^{m \times r}$, 
 one has  
\be \label{eq:assump_A4*}
\hspace*{-2.5mm} \PP \Big\{\|\Xi  G \|\tinf \leq \Ctau  d_r    \lkv  \teps_1  \|G \|_F + \teps_2 \|G \|\tinf \rkv 
\Big\} \geq 1 \!- \! n^{-\tau}.
\ee
In addition,  all quantities in the right  sides of inequalities in \fr{eq:new_probab_errors}
depend on $n$, $m$ and $r$, but not on $\tau$.

\smallskip

\begin{theorem}\label{thm:symetrized_probab_up_bound}  
Let conditions of Theorem \ref{thm:symetrized_up_bound} hold, and  Assumptions A3*, A4* and \fr{eq:m_n_rel} be valid. 
Let rows of matrix $\Xi = \hX - X$ be independent, and let, for simplicity, $d_{r+1} = 0$.
If, as $n, m \to \infty$, one has
\be \label{eq:teps_conditions}
\begin{split}
& \tepsEo= o(1),\ \sqrt{r}\, \teps_1 (\tepso +1) = o(1), \\
& \teps_2 (\tepstinf^T + \epsv) = o(1),  \  (1 - \tilh)   \tepstinf = o(1),
\end{split}
\ee
then, for $n$ and $m$ large enough, with probability at least $1 - n^{-\tau}$, one has 
 \be  \label{eq:symmetrized_up_bound_new}  
\|\hU - U W_U\|\tinf   \leq   \Ctau \, \lkr \tdel_1 + \epsu\, \tdel_{1,U} \rkr,
\ee
where
\begin{align}  
 & \tdel_1   = \tepsxiutinf +  \tepsvtinf  + \tepsuuo   \lkv \sqrt{r}\, \teps_1 (\tepso +1) + \right.  \nonumber \\
  &\ \ \ \  +   \left. \teps_2 (\tepstinf^T + \epsv)\rkv +
 \tilh\, \tepsy + (1 - \tilh) \, \tepstinf^2, \hspace{4mm} \ \  \label{eq:tdel1}   \\
&  \tdel_{1,U}  =  \tepsuuo + \tepsEo \, \lkv \tepsEo +  \teps_1 (\tepso +1) \right. \nonumber \\
& \ \ \ \ + \left. \teps_2 (\tepstinf^T + \epsv)\rkv.  \label{eq:tdel1u}
\end{align}
\end{theorem}

\medskip

\begin{rem} {\bf Symmetrization by Hermitian dilation.\ }
{\rm Note that one can symmetrize matrix $X$ and its estimator $\hX$ by introducing symmetric matrices 
\bes
\Ysh = \lkr \! 
\begin{array}{c c}
\!  0 \!  & \! X \!  \\
\!  X^T \!  & \!   0 \! 
\end{array} \! \rkr, \  \ 
\hYsh = \lkr \! 
\begin{array}{c c}
\! 0 \!  & \!  \hX \! \\
\! \hX^T \!  & \!  0 \! 
\end{array} \!  \rkr, \ \ 
\scrEsh = \lkr \! 
\begin{array}{c c}
\! 0 \!  & \!  \Xi \! \\
\! \Xi^T \! & \!  0 \! 
\end{array}\!  \rkr.
\ees
In this case, the SVDs of $\Ysh$ and $\hYsh$ are of the form 
$\Ysh = \Ush \Lamsh (\Ush)^T + \Ush_{\perp} \Lamsh_{\perp} (\Ush_{\perp})^T$ and 
$\hYsh = \hUsh \hLamsh (\hUsh)^T + \hUsh_{\perp} \hLamsh_{\perp} (\hUsh_{\perp})^T$ with 
\be \label{eq:Ush_hUsh} 
\Ush = \frac{1}{\sqrt{2}}\, \lkr 
\begin{array}{c c}
U  & U \\
V &  -V
\end{array} \rkr, \ 
\hUsh = \frac{1}{\sqrt{2}}\, \lkr 
\begin{array}{c c}
\hU  & \hU \\
\hV &  -\hV
\end{array} \rkr.
\ee 
Now, apply Theorem~\ref{th:sym_up_bound}  with $\scrE$ and $U$ replaced with
$\scrEsh$ and $\Ush$, respectively, and observe that \fr{eq:Ush_hUsh} yields
$\|\hUsh - \Ush \Wsh_{\Ush} \|\tinf  = 2\, \max \lkr  \|\hU - U\, W_U\|\tinf,  \|\hV - V\, W_V\|\tinf \rkr$.
Due to \cite{tropp2015introduction}, one has
$\|\scrEsh\| = \|\Xi\|$,
$|\lam_r| = d_r$, $|\lam_{r+1}| = d_{r+1}$, $\eps_{\Ush} = \max(\epsu, \epsv)$,
$\|\scrEsh\|\tinf = \max(\tDeltinf, \tDeltinf^T)$ and 
$\|\scrEsh\, \Ush\|\tinf = \max(\|\Xi \, V\|\tinf, \|\Xi^T\, U\|\tinf)$.
Since also  $\max(a, b) \asymp a+ b$ for $a,b>0$,    obtain  that
\bes   
\begin{split}
& \max \lkr  \|\hU - U\, W_U\|\tinf, \! \|\hV - V\, W_V\|\tinf \rkr \!   \leq \! C  \lkv
 (\tDeltinf + \right. \\
 &   \tDeltinf^T  \! +  \! d_r^{-1}\, d_{r+1})  \tDelo  
%
\! +  \! \left.(\epsu + \epsv)  \tDelo \! +  \! \tDelvtinf \! +  \! \tDelutinf^T \rkv,  
\end{split} 
\ees
where $\tDelutinf^T = d_r^{-1}\, \|\Xi^T\, U\|\tinf$. 
It is easy to see that the Hermitian dilation essentially replaces all quantities in Theorem~\ref{th:sym_up_bound} 
by the maximums with respect to $X$ and $X^T$, so that, the Hermitian dilation upper bound 
is always higher 
(and may be infinitely larger) than the upper bound in Theorem~\ref{th:sym_up_bound}.
Therefore, unless one is interested in simultaneous estimation of $\hU$ and $\hV$, 
the Hermitian dilation does not lead to accuracy improvement.

}
\end{rem}


\section{Perfect spectral clustering using the two-to-infinity norm bounds and its applications to random networks. }
\label{sec:perfect_clust}

\subsection{Sufficient conditions for perfect spectral clustering}
\label{sec:spec_clust}

In the last decade, evaluation of accuracy of clustering techniques came to the frontier of the statistical science.
Recently a number of papers studied precision of the k-means clustering algorithm (or its versions, like k-medoids). 
Since data are usually  contaminated by noise, it needs to be pre-processed prior to using the k-means 
algorithm (\cite{giraud2018RelaxedKmeans}, \cite{H_Zhou_AOS2021_Spec_Clust}).  
Therefore,   various techniques for pre-processing   data were developed,
 such as  Semidefinite Programming (SDP) (\cite{giraud2018RelaxedKmeans},  \cite{NIPS2017_6776}), 
or spectral analysis (\cite{abbe_fan_AOS2022},   \cite{H_Zhou_AOS2021_Spec_Clust}, \cite{ndaoud_aos2022}).
In particular,  it turns out that spectral methods in combination with k-means/medoid  clustering 
algorithms produce  very accurate clustering assignments in a variety of problems, from Gaussian mixture models to 
random networks (\cite{abbe_fan_AOS2022}, \cite{giraud_2024_comp_gap}, \cite{giraud2018RelaxedKmeans},     
\cite{lei2021biasadjusted}, \cite{lei2015}).

Theoretical assessments of clustering precision rely on various error metrics. For example, 
\cite{giraud2018RelaxedKmeans} and \cite{NIPS2017_6776}
use the $l_1$-norm of the difference between the membership matrix and its SDP-based estimator for derivation of the clustering precision. 
The accuracy of approaches that use variants of the SVD are usually based on the operational norm of the induced errors
(\cite{lei2021biasadjusted},  \cite{H_Zhou_AOS2021_Spec_Clust}). While this is totally justifiable in the case when the original errors 
are Gaussian or sub-Gaussian, as it is assumed  in the above cited papers, in the situations where the distributions of errors 
are arbitrary,  it is sometimes very difficult to construct tight upper bounds for the operational norm.

Consider a version of the k-means setting,  where rows of matrix $X \in \RR^{n \times m}$ take $r$
different values $\Te(k,:)$, $k \in [r]$. Hence, there exists  a clustering function $z: [n] \to [r]$ such that 
$X(i,:) = \Te(z(i), :)$, $i \in [n]$. In this case, $X$ can be presented $X = Z \Te$,
 where $\Te \in \RR^{r \times m}$ and  $Z \in \{0,1\}^{n \times r}$ is a clustering matrix, 
such that $Z(i,k) = 1$ if $z(i) = k$, and $Z(i,k) = 0$  otherwise. 
In this scenario, data come in the form of $\hX \in \RR^{n \times m}$,
and the goal is to estimate the clustering function $z$.    
In what follows, we denote the size of the $k$-th cluster by $n_k$, 
$\nmax = \displaystyle \max_k n_k$ and $\nmin = \displaystyle \min_k n_k$.

Since  clustering is unique only up to a permutation of cluster labels, 
denote the set of $r$-dimensional permutation functions of $[r]$ by $\aleph(r)$.
For simplicity, let $r$ be known, and let $\hz:[n] \to [r]$ be an estimated clustering assignment. 
The number of  errors of a clustering assignment $\hz$ with respect to the true clustering function $z$,
and the associated error rate are then defined, respectively, as
\be \label{eq:clust_er}
\begin{split}
& \calN_n (\hz, z) = \min_{\phi \in \aleph(r)}\ \sum_{l=1}^n I \lkr \phi(\hz(i) \rkr \neq z(i)), \\
& \calR_n (\hz, z) = n^{-1}\, \calN_n (\hz, z).
\end{split}
\ee
The estimated clustering $\hz$ is {\it  consistent} if $\calR_n (\hz, z) \to 0$ as $n \to \infty$.
If  $\calN_n (\hz, z) \to 0$ as $n \to \infty$, then clustering is called {\it strongly consistent}. 
In the case of a  strongly consistent clustering algorithm, for $n$ large enough, one obtains
$\calN_n <1$, which is equivalent to $\calN_n =0$. In this case, $\hz = \phi(z)$ for some $\phi \in \aleph(r)$,
and one achieves {\it perfect clustering}. It turns out that application of two-to-infinity norm allows to 
establish conditions for strongly consistent clustering under  rather generic assumptions.

Assume that one measures $\hX = X + \Xi$, where $X$ is the unknown true matrix. 
We intentionally do not impose any additional restrictions on $\Xi$, as it is done in majority of papers, where $\Xi$
is often assumed to have  independent Gaussian or sub-Gaussian rows.
For simplicity, consider the situation where $\rank(\Te) = r$, the smallest and the largest singular values of $\Te$ 
are of the same magnitude and that clusters are balanced, so that, for some absolute constants 
$C_\sig$ and $c_0$, one has 
\be \label{eq:clust_assump}
\sig_r(\Te) \geq C_\sig \sig_1(\Te), \quad \nmax   \leq c_0^2 \nmin. 
\ee
Note that one can remove some of the assumptions and generalize our theory to a less restrictive setting,
but this will make presentation more cumbersome.

Denote $D_z = Z^T Z = \diag(n_1, ..., n_r)$,   where $n_k$ is the number of elements in the $k$-th cluster,
and observe that $U_z = Z \, D_z^{-1/2} \in \calO_{n,r}$. Then $X = U_z \sqrt{D_z}\, \Te$.
If $\sqrt{D_z}\, \Te = U_{\Te} D V^T$   is the SVD of $\sqrt{D_z}\, \Te$, where  $U_{\Te} \in \calO_r$, $V \in \calO_{m,r}$, then 
the SVD of $X$ can be written as 
\be \label{eq:X_SVD}
X = U D V^T, \quad U = U_z U_{\Te} \in \calO_{n,r}, \ \ \ V \in \calO_{m,r}.
\ee
In this case, one has $U(i,:) = U(j,:)$ if $z(i)=z(j)$ and 
\be \label{eq:U_rel}
\|U(i,:) - U(j,:)\| \geq \sqrt{2}\, (\nmax)^{-1/2}\ \ \mbox{if}\ \ z(i)  \neq z(j),
\ee
where $z:[n]  \to [r]$ is the true clustering function.
In addition, consider $Y = X X^T$ and its eigenvalue decomposition 
\be  \label{eq:new_Y}
Y = X X^T = U \Lam U^T, \quad \Lam = D^2,
\ee 
which coincides with \fr{eq:Y_def}, where $\Lam_{\perp} = 0$, $\lam_{r+1} = 0$.

Estimate $X$ by $\hX$,  or   $Y$ by $\hY$  defined in \fr{eq:hY_neq}, and recall that $\hX$ and $\hY$ have the 
SVDs,  given in \fr{eq:main_nonsym} and  \fr{eq:hY_eigen}, respectively. 
After that,  use  the $(1+a)$-approximate k-means clustering to rows of $\hU$ 
to obtain the final clustering assignments. There exist efficient algorithms 
for solving the $(1+a)-$approximate k-means problem  
(see, e.g., \cite{1366265}). The process is summarized as Algorithm~\ref{alg:spec_clust}.

%
\begin{algorithm} [t] 
\caption{\ Spectral clustering algorithm}
\label{alg:spec_clust}
\begin{flushleft} 
{\bf Input:} Matrix $\hX \in \RR^{n \times m}$; number of clusters $r$;
 parameter $a >0$ \\
{\bf Output:} Estimated clustering function  $\hz: [n] \to r$ \\
{\bf Steps:}\\
{\bf 1:} Find $\hU = \SVD_r(\hX)$, the $r$  left leading eigenvectors of $\hX$; \\
\hspace*{5mm}         {\bf or}  construct $\hY$ using formula \fr{eq:hY_neq} and find  $\hU = \SVD_r(\hY)$.\\
{\bf 2:} Cluster  $n$ rows of  $\hU$ into $r$ clusters using   $(1+a)$-approximate k-means clustering. Obtain \\
\hspace*{5mm}         estimated clustering function $\hz$.   
\end{flushleft} 
\end{algorithm}
%


It turns out that the accuracy of Algorithm~\ref{alg:spec_clust}  relies on the closeness of $U$ 
and $\hU$ in the two-to-infinity norm. Specifically the following statement holds.


\begin{lem}  \label{lem:perf_clust}  
Let conditions \fr{eq:clust_er}-\fr{eq:new_Y} be valid. 
If, as $n \to \infty$, 
\be \label{eq:lem_perf_clus}
\sqrt{r}\, D_{F} (U, \hU) = o(1), \quad
D_{2,\infty} (U, \hU) = o(\epsu),
\ee
where $D_{F} (U, \hU)$ and $D_{2,\infty} (U, \hU)$ are defined in \fr{eq:dist} and \fr{eq:2inf_dist}, respectively,
then,   when $n$ is large enough, 
clustering is  perfect with probability at least $1 - C\, n^{-\tau}$. 
\end{lem}
 

%

\noindent
Combining Lemma~\ref{lem:perf_clust} with the results in Theorems~\ref{thm:nonsym_up_bound},
\ref{thm:nonsym_probab_up_bound_new},  \ref{thm:symetrized_up_bound}  and \ref{thm:symetrized_probab_up_bound},
we obtain the following statement.


\begin{prop}\label{prop:perfect_clust}   
Let $X = Z \Te$,  where $\Te \in \RR^{r \times m}$ and  $Z \in \{0,1\}^{n \times r}$ is a clustering matrix, 
such that $Z(i,k) = 1$ if row $i$ of $X$ is in the $k$-th cluster, and $Z(i,k) = 0$  otherwise,  $i \in [n]$, $k \in [r]$. 
Let $Y = X\, X^T$, so that  $X$ and $Y$ have the SVDs  \fr{eq:X_SVD} and \fr{eq:new_Y}, respectively.
Let $\hX$ be an estimator of $X$, $\hU$ be obtained using Algorithm~\ref{alg:spec_clust} and, in addition 
assumptions \fr{eq:m_n_rel} and  \fr{eq:clust_assump} hold  for some absolute constants $\tau_0$, $C_\sig$ and $c_0$.

If $\hU = \SVD_r(\hX)$ and conditions  of Theorem~\ref{thm:nonsym_up_bound} hold, 
then, when $n$ is large enough, 
clustering is perfect with probability at least $1 - C\, n^{-\tau}$, 
provided that, as $n \to \infty$
\be \label{eq:perf_cl1}
\sqrt{r}\, \tepso = o(1), \quad \epsu^{-1} \lkr \tepsvtinf + \tepso\,  \tepstinf \rkr = o(1).
\ee

If $\hU = \SVD_r(\hX)$ and conditions  of Theorem~\ref{thm:nonsym_probab_up_bound_new}  hold, 
then, when $n$ is large enough, 
clustering is perfect with probability at least $1 - C\, n^{-\tau}$, 
provided that, as $n \to \infty$
\be \label{eq:perf_cl1_1}
\sqrt{r}\, \tepso = o(1), \  \epsu^{-1} \lkr \tepsvtinf + 
\sqrt{r} \tepso  (\teps_1 + \teps_2) \rkr = o(1).
\ee

If $\hU = \SVD_r(\hY)$ and  Assumptions of Theorem~\ref{thm:symetrized_up_bound}    hold,
then, when $n$ is large enough, 
clustering is  perfect with probability at least $1 - C\, n^{-\tau}$,  
provided that, as $n \to \infty$, one has $\sqrt{r}\, \tepsEo = o(1)$,   \ $\tilh\, \tepsy = o(1)$, and 
\be \label{eq:perf_cl2}
\begin{split}
& \epsu^{-1} \lkv \tepsxiutinf +  \tepsvtinf \right. \\
& +\left.  \min(\tepsEo, \sqrt{r}\, \tepsEuo) \, ( \tepsxitinf   +\tilh\, \tepsy)  \rkv 
= o(1).
\end{split}
\ee

If $\hU = \SVD_r(\hY)$, Assumptions of Theorem~\ref{thm:symetrized_probab_up_bound}  hold  
and, in addition, rows of matrix $\Xi = \hX - X$ are independent, 
then, when $n$ is large enough, clustering is  perfect with probability at least $1 - C\, n^{-\tau}$,  provided that, as $n \to \infty$,
\be \label{eq:perf_cl3}
\sqrt{r}   \tepsEo = o(1), \quad \tilh  \tepsy = o(1), \quad  
\epsu^{-1}\, \tdel_1 = o(1),
\ee
where $\tdel_1$ is defined in \fr{eq:tdel1}.
\end{prop}


Note that, in a less common case, when one needs to cluster a symmetric matrix, 
one can use a similar approach. Indeed, consider the situation where,
for some clustering function $z:  [n] \to [r]$, the elements of a symmetric matrix $Y$ in 
\fr{eq:main_rep} are of the form $Y(i,j) = Q(z(i), z(j))$  for some matrix $Q \in \RR^{r \times r}$,
so that $Y = Z\, Q\, Z^T$, where $Z$ is the clustering matrix, which
corresponds to the clustering function $z$.
Introducing matrices  $D_z$ and $U_z$, similarly to the non-symmetric case considered above, 
and writing the  eigenvalue decomposition   $\sqrt{D_z}\, Q \sqrt{D_z}= U_Q \, \Lam \, U_Q^T$, where $U_Q \in \calO_r$,
derive an eigenvalue decomposition of  $Y$,   similarly to \fr{eq:new_Y}: 
\be \label{eq:Y_SVD_cl}
Y = U \Lam U^T, \quad U = U_z U_{Q} \in \calO_{n,r}.
\ee
Then, combination of Lemma~\ref{lem:perf_clust} and Theorems~\ref{th:sym_up_bound} and \ref{th:sym_up_bound_new} 
yields the following statement.


\begin{prop}\label{prop:perfect_clust_sym}  
Let $Y = Z Q Z^T$,  where $Q \in \RR^{r \times r}$. Let  $Z \in \{0,1\}^{n \times r}$ be a clustering matrix, 
such that $Z(i,k) = 1$ if row $i$ of $Y$ is in the $k$-th cluster, and $Z(i,k) = 0$  otherwise, 
$i \in [n]$, $k \in [r]$.  Let the SVD of $Y$ be given by \fr{eq:Y_SVD_cl} and, in addition,
the second inequality in \fr{eq:clust_assump} holds.
%
%
Let $\hY$ be an estimator of $Y$, and  $\hU= \SVD_r (\hY)$.  

If Assumption A1 is valid, then,   when $n$ is large enough, 
clustering is perfect with probability at least $1 - C\, n^{-\tau}$, 
provided that, as $n \to \infty$
\be \label{eq:perf_cl1_sym}
\sqrt{r}\,  \epso = o(1), \quad \epsu^{-1} \lkr \epsinf \epso + \epsEu   \rkr = o(1).
\ee 
If, in addition, Assumption A2 holds and  matrix $\tilscrE = \hY- Y$ is such that, 
for any  $l \in [n]$,   rows $\tilscrE(l,:)$ of $\tilscrE$ and matrix 
$\tilscrE\upl$, defined in \fr{eq:scrEupl}, are independent from each other, then 
when $n$ is large enough, clustering is perfect with probability at least $1 - C\, n^{-\tau}$, 
provided that, as $n \to \infty$
\be \label{eq:perf_cl2_sym}
\begin{split}
& \sqrt{r}\,  \epso= o(1),\quad \eps_1 = o(1), \quad \eps_2 = o(1), \\ 
& \epsu^{-1} \lkr \epso\, \eps_1\, \sqrt{r} + \epsEu   \rkr = o(1).
\end{split}
\ee 
\end{prop}


\begin{rem}
{\rm 
Note that assumptions that quantities in \fr{eq:perf_cl1}-\fr{eq:perf_cl3} and 
\fr{eq:perf_cl1_sym}, \fr{eq:perf_cl2_sym} tend to zero as $n \to \infty$ are sufficient conditions.
Indeed, according to the Lemma~\ref{lem:abbe_fan} and our subsequent reasoning, it is sufficient that
those quantities are bound above by some small  (but unknown in practice) constants. Since the latter 
is hard to ensure, we impose  slightly stronger conditions in \fr{eq:perf_cl1}-\fr{eq:perf_cl3} and 
\fr{eq:perf_cl1_sym}, \fr{eq:perf_cl2_sym}.

Also observe that, in this paper, we study the case where one can obtain clustering assignments 
by partitioning rows of $\hU$.  This is not generally true in the $k$-means setting where 
the number of distinct rows of matrix $X$ may be higher than its rank. 
In the latter case, one needs to multiply $\hU$ by the estimated diagonal matrix of the singular values, which leads to 
different bounds on the errors.
}
\end{rem}


\subsection{A didactic example: the case of independent Gaussian errors }
\label{sec:didactic}

In order to examine the usefulness of various parts of Proposition~\ref{prop:perfect_clust},
below we study perfect clustering when the error matrix $\Xi = \hX - X \in \RR^{n \times m}$  has independent $\calN(0, \sig^2)$
Gaussian entries. We are keenly aware that this scenario has been studied extensively in a multitude of papers
(see, e.g., \cite{abbe_fan_AOS2022},   \cite{Chen_2021}, \cite{H_Zhou_AOS2021_Spec_Clust},  \cite{ndaoud_aos2022} 
and \cite{zhou2024_illposedPCA}), where more nuanced results were derived  under, sometimes, the weaker condition that
elements of $\Xi$ are independent sub-Gaussian. However, each of the papers listed above studied only one of many possible scenarios 
in this problem. The objective of this section is not to  derive new results but  to demonstrate, 
how the usefulness of various techniques, proposed in Sections~\ref{sec:nonsymmetric} and  \ref{sec:symmetrized},
depends on the settings of the model. Specifically, we are interested in exploring, what conditions for 
the perfect clustering are, if we use or do not use symmetrization  or/and  Assumptions~{\bf A4} and {\bf A4*}. 
While upper bounds in Sections~\ref{sec:nonsymmetric} and  \ref{sec:symmetrized} are obtained under 
mild conditions, the assumption that errors are independent Gaussian in this subsection is  motivated 
exclusively by the simplicity of evaluation of all quantities, that appear in the respective Theorems and Propositions,
and is not utilized in any other way.

As before, we assume that    $X \in \RR^{n \times m}$ can be presented as $X = Z \Te$,
where $\Te \in \RR^{r \times m}$ and  $Z \in \{0,1\}^{n \times r}$ is a clustering matrix,
which we would like to recover. We furthermore assume that one observes $\hX = X + \Xi$, that 
inequalities in \fr{eq:clust_assump} are valid, and  that
\be \label{eq:m_n_te}
\log m \asymp \log n, \quad r^2/n  \to 0, \quad r^2/m \to 0. 
\ee 
Since 
$ \sig_r(\Te) \leq \min \|\Te(i,:)\| \leq \max \|\Te(i,:)\| \leq \sig_1 (\Te),$
conditions \fr{eq:clust_assump} and \fr{eq:m_n_te} imply that, for $\te = m^{-1/2} \max \|\Te(i,:)\|$, one has 
\be \label{eq:d_r_epsu_gauss}
\begin{split}
& \|\Te \|\tinf  \asymp \sqrt{m}\,  \te, \quad  \epsu \asymp \sqrt{r}/\sqrt{n}, \\
& d_1 \asymp d_r = \sig_r(X) \asymp \te\, \sqrt{m\, n}/\sqrt{r}.
\end{split}
\ee
Now, depending on the relationship between parameters $m$, $n$, $\sig$ and $\te$, one can use 
Algorithm~\ref{alg:spec_clust} for clustering with $\hU = \SVD_r(\hX)$ or $\hU = \SVD_r(\hY)$.
In order to discuss the pros and the cons of each of the choices, we evaluate the quantities that appear 
in the conditions \fr{eq:perf_cl1}, \fr{eq:perf_cl2} and \fr{eq:perf_cl3} of Proposition~\ref{prop:perfect_clust}.


\begin{lem} \label{lem:err_bounds_gauss}  
Let $X,  \hX \in \RR^{n \times m}$ and $\Xi = \hX - X$  have independent $\calN(0, \sig^2)$ Gaussian entries. 
Let \fr{eq:clust_assump},  \fr{eq:m_n_te} and \fr{eq:d_r_epsu_gauss}  hold.
If $\hU = \SVD_r(\hX)$, then,  with probability at least $1 - n^{-\tau}$, one has
\be  \label{eq:tepso_tepstinf}
\begin{split}
& \tepso \asymp \frac{\sig\, \sqrt{r}}{\te}\, \lkr \frac{1}{\sqrt{m}} + \frac{1}{\sqrt{n}} \rkr,
\quad \tepstinf \asymp \frac{\sig\, \sqrt{r}}{\te}\, \frac{\sqrt{\log n}}{\sqrt{n}},\\
&   \hspace{25mm} \tepsvtinf \asymp \frac{\sig\, \sqrt{r}}{\te}\, \frac{\sqrt{r\, \log n}}{\sqrt{m\, n}}.
\end{split}
\ee
If $\hU = \SVD_r(\hY)$,   where $\hY$ is defined in \fr{eq:hY_neq} with $\tilh=1$, i.e.,
$\hY = \scrH(\hX\, \hX^T)$, then  $\tepsy \asymp r/n$ and,  with probability at least $1 - n^{-\tau}$
one has
\be  \label{eq:gauss_symmetrized_bounds} 
\begin{split}
 & \tepsxiutinf \leq \Ctau \, \frac{\sig^2\, r}{\te^2}\,  \frac{\log n\, \sqrt{r}}{n\, \sqrt{m}}, \ 
 \tepsxitinf \leq \Ctau \,   \frac{\sig^2\, r \, \log n}{\te^2\, \sqrt{m\, n}}\\ 
 & \hspace{20mm} \tepsEo \leq \Ctau \, \lkv \frac{\sig^2\, r}{\te^2}\, \frac{\log n}{m} + \frac{r}{n} \rkv.
\end{split}
\ee 
Finally,  \fr{eq:assump_A4} and \fr{eq:assump_A4*} in Assumptions~A4 and A4* are satisfied with  
\be \label{eq:teps12}
\teps_1 \leq \Ctau \,  \, \frac{\sig\, \sqrt{r\, \log n}}{\te\, \sqrt{m\, n}}, \quad \quad \teps_2=0.
\ee
\end{lem}

  \medskip



\noindent
Using Lemma~\ref{lem:err_bounds_gauss}  and Proposition~\ref{prop:perfect_clust}, 
one can derive sufficient conditions for perfect clustering, summarized in the following statement. 
%


\begin{prop} \label{prop:didactic}  
Let conditions \fr{eq:m_n_te} hold and the upper bounds for the quantities in Table 1 be given by Lemma~\ref{lem:err_bounds_gauss}.
If one uses Algorithm~\ref{alg:spec_clust} with $\hU = \SVD_r(\hX)$, then condition (N1) in \fr{eq:nes_suf_no_sym} is 
necessary   for consistent clustering while condition (S1) is sufficient for perfect clustering: as $m,n \to \infty$
\be \label{eq:nes_suf_no_sym}
\begin{split}
& ({\rm N1}): \, \frac{\sig  \sqrt{r}}{\te \sqrt{\min(m,n)}}=o(1); \ \\
&  ({\rm S1}): \, \frac{\sig, \sqrt{r \log n}}{\te \sqrt{\min(m,n)}} \lkr 1 + \frac{\sig}{\te} \rkr =o(1). 
\end{split}
\ee
If one uses Algorithm~\ref{alg:spec_clust} with $\hU = \SVD_r(\hY)$ with  $\hY = \scrH(\hX\, \hX^T)$, then the necessary condition for 
consistent clustering  is
\be \label{eq:nes_sym}
\frac{\sig^2\, r\,\log n}{\te^2\, m} = o(1), \quad m,n \to \infty.
\ee
The sufficient conditions for the perfect clustering in this case are 
\be \label{eq:suf_sym}
\begin{split}
& \frac{\sig^2\, r\, \log n}{\te^2\, \sqrt{m\, n}}  \lkr 1 + \frac{(r + \log n)\, \sqrt{n} }{\sqrt{m}}\rkr = o(1); \\
& \frac{\sig^2\,  \log n}{\te^2} \, \frac{r^{3/4}}{m ^{3/4}} = o(1),  
\quad m,n \to \infty,
\end{split}
\ee
where only the first condition in \fr{eq:suf_sym} is required,  if Assumption A4*  is  satisfied.
%
\end{prop}


\smallskip

Note  that, if $\sig  = O(\te)$, then sufficient condition (S1) in \fr{eq:nes_suf_no_sym}  is always true and 
 there is no need for symmetrization. However, if $\sig \gg \te$, symmetrization may be useful. 
Observe that condition \fr{eq:nes_sym} is weaker than condition (N1) in \fr{eq:nes_suf_no_sym} when $n \ll m$,
so that, one expects that symmetrization leads to accuracy improvement in this case.
Specifically, if Assumption A4*  holds, then sufficient conditions for perfect clustering become
\bes
\begin{split}
& \frac{\sig^2\, r\, \log n}{\te^2\, \sqrt{m\, n}} = o(1) \quad \mbox{if} \quad \sqrt{n} (r  + \log n) = O(\sqrt{m}),\\
& \frac{\sig^2\, r\, \log n (r + \log n)}{\te^2\, m} = o(1)\quad \mbox{otherwise}.
 \end{split}
\ees
If Assumption A4* does not hold, then one needs to add the second condition in \fr{eq:suf_sym}.


In order to obtain a deeper insight into  whether to use Algorithm~\ref{alg:spec_clust}  with or without 
symmetrization, and which upper bounds in  Proposition~\ref{prop:perfect_clust} is better to utilize, 
consider a simple case when 
\bes
 r= O(1), \quad n = m^{\ga}, \quad \sig/\te \asymp m^{\nu}. 
\ees
Then, in the absence of symmetrization, condition (S1) in \fr{eq:nes_suf_no_sym}   is equivalent to
$\nu  < \min(1/4, \ga/4)$. If one applies symmetrization, then it follows from \fr{eq:suf_sym} 
that perfect clustering is guaranteed by  $\nu < \min(3/8,  (1   + \ga)/4)$, which is weaker  than the 
condition in the non-symmetric case.
Finally, if Assumption A4* is taken into account, then sufficient condition for perfect clustering 
becomes $\nu < \min(1/2,  (1   + \ga)/4)$, which is the weakest condition than all previous ones.

In conclusion, this example demonstrates, how comparisons of methods for estimating $\hU$ and  
of various error bounds constructed in this paper, allow  one to choose the most advantageous ones. 
Specifically, in the case of Gaussian errors, symmetrization with hollowing is beneficial 
for any combination of $n$ and $m$ but the full advantage can be exploited only if one employs Assumption A4*.


\subsection{Perfect clustering in a sub-sampled  network }
\label{sec:subsampled}

Consider a binary undirected stochastic network on $n$ nodes, that can be partitioned into $r$ communities. 
Let $z:[n] \to [r]$ be a clustering function,  such that $z(i) = k$ if node $i$ belongs to community $k$.
Additionally,  assume that the network is equipped with the Stochastic Block Model ({\bf SBM}) (see, e.g., \cite{JMLR:v18:16-480}), so that 
there exists a matrix $Q \in [0,1]^{r \times r}$ of block connection probabilities, such that 
the probability of connection between nodes $i$ and $j$ is fully determined by the communities to which they belong:
$P(i,j) =  Q(z(i), z(j))$. In this setting, one observes an   adjacency matrix $A \in \{0,1\}^{n \times n}$
where, for $1 \leq i < j \leq n$, elements $A(i,j)$ of $A$ are independent Bernoulli variables  with 
$\PP \lfi A(i,j) = 1 \rfi = P(i,j)$. Here, $P^T = P$ and $A^T = A$. Since usually networks are sparse, 
i.e.,  probabilities of connections become smaller as the network size $n$ grows, 
the network is equipped with a  sparsity factor $\rhon = o(1)$ as $n \to \infty$, 
where $\rhon$ is defined  by
\be \label{eq:P0Q0}
\begin{split}
&  P = \rhon\,  P_0, \quad \|P_0\|_{\infty} = 1, \\ 
& Q = \rhon\, Q_0, \quad P_0(i,j) =  Q_0(z(i), z(j)).
 \end{split}
\ee 
The main question of interest in this setting is recovery of the community assignment $z$. 
The problem of community detection in the SBM was addressed in an abundance of publications, 
under  a variety of assumptions (see, e.g.,  \cite{JMLR:v18:16-480},  \cite{abbe_bandeira_2016},  
\cite{DBLP:journals/corr/AminiL14}, \cite{rohe2011spectral}, \cite{Ander_Zhang_IEEE2024_SBM} 
among others). At present, perhaps the 
most popular method of community detection is spectral clustering that was studied in, e.g., 
\cite{lei2015} and \cite{rohe2011spectral}.
However, this procedure becomes prohibitively  computationally expensive when the number of nodes is huge. 
For this reason, recently several authors suggested a variety of 
approaches for reduction of computational costs. Majority of those proposals start with  sub-sampling 
a group of nodes,  and then partitioning those nodes into communities. This process may be repeated several times in order 
to obtain community assignment of all nodes, as in, e.g., \cite{chakrabarty2023sonnet}, \cite{mukherjee2021two}
and \cite{bhadra2025scalablecomm_detect}. 
In this section, we address the first part of this process: sub-sampling of nodes with the subsequent community assignment.

We would like to remind the reader  that our goal here is to formulate sufficient conditions for strongly consistent 
clustering in a sub-sampled network. As such, we are not interested in assessment of a sharp threshold 
for possibility of community detection, as it is done in, e.g., \cite{JMLR:v18:16-480},  \cite{abbe_bandeira_2016}
or \cite{Ander_Zhang_IEEE2024_SBM}, under the assumption that the connection probabilities take only two distinct values.
Instead, we would like to provide a practitioner with a tool for evaluation, how the  sample size should be chosen under 
generic regularity conditions.

In what follows, we assume that a set $\calS$ of  $m$ nodes is sampled uniformly at random. Denote by $\Sc$
the set of remaining nodes. 
The goal here is to estimate community assignments of the $m$ nodes in $\calS$. 
It appears that   many papers  estimate community assignment on the basis of solely 
the $(m \times m)$ portion $\ASS \in \{0,1\}^{m \times m}$ of matrix $A$,   as it is done in, e.g., 
\cite{chakrabarty2023sonnet} or \cite{mukherjee2021two}.
However, in a very sparse network, this may either require  to sample a large number of nodes, or to risk obtaining inaccurate results.
Indeed, consider the situation when one uses only the sub-matrix    $\ASS  \in \{0,1\}^{m \times m}$ for clustering.
Then,   it is well known that, if $m\, \rhon$ is bounded above by a constant, then community assignment is inconsistent, 
while   $m\, \rhon > C \log n$, for a sufficiently large constant $C$, leads to perfect clustering of $m$ nodes into communities. 
As it is easy to see, these restrictions lead to a lower bound on $m$.

For this reason, we are going to utilize  the $m \times (n- m)$ sub-matrix $\ASSc$  of matrix $A$ for clustering.
We denote $\hX = \ASSc \in \{0,1\}^{m \times (n-m)}$ and $X = \EE \hX = \PSSc$ and show 
that using    matrix   $\hX$ instead of $\ASS$ allows   to reduce this lower bound on $m$. 

Let $\zS: [m] \to [r]$ and $\zSc: [n-m] \to [r]$ be the reductions of the clustering function $z:[n] \to [r]$  
to the $m$ sub-sampled nodes
and $(n-m)$ nodes in $\Sc$. Denote the clustering matrices corresponding to $\zS$ and $\zSc$ by, respectively,  $\ZS \in \{0,1\}^{m \times r}$ and 
$\ZSc \in \{0,1\}^{(n-m) \times r}$. Then,   $X = \ZS Q \ZSc^T$. 
Denote the community sizes for the whole network, and the sub-networks based on $\calS$ and on $\Sc$ by,
respectively, $n_k$, $m_k$ and $N_k$, $k \in [r]$.

Let the SVDs of $X$ and $\hX$ be given in \fr{eq:main_nonsym}.
It is easy to see that, for a sparse network, the number of sub-sampled nodes 
$m$ should grow with $n$, when one is estimating $U$ by $\hU$. 
The rate of growth, however, depends on the methodology which one uses. 
Recall that, if one samples just a square symmetric 
sub-matrix $\ASS$ with rows and columns in $S$, then one needs $m$ to be large enough, so that $m \rhon \to \infty$ as $n \to \infty$. 
Moreover,  even if one utilizes the $m \times (n-m)$-dimensional matrix $\ASSc$ but employs techniques 
in Section~\ref{sec:nonsymmetric}, the condition $m \rhon \to \infty$ still cannot be avoided. Indeed, if $m = o(n)$, one has
$\| \Xi\| \asymp \|\Xi\|\tinf \asymp \sqrt{n\, \rhon}$, and therefore, $\tepso \asymp (m \rhon)^{-1}$, which leads 
to the requirement $m\, \rhon \to \infty$ as  $n \to \infty$. 
Nevertheless, this condition is not needed anymore, if one applies symmetrization described in Section~\ref{sec:symmetrized}.

To this end, consider $Y = X X^T$ with the eigenvalue decomposition \fr{eq:new_Y}, 
and construct its estimator $\hY$ of the form \fr{eq:hY_neq} with $\tilh=1$. 
Subsequently, apply Algorithm~\ref{alg:spec_clust} and obtain estimated clustering assignment 
$\hzS: [m] \to [r]$. In this setting, it is necessary to  impose conditions that guarantee correctness of  the 
Algorithm~\ref{alg:spec_clust}. In particular, similarly to \fr{eq:clust_assump}, assume that 
for matrix $Q_0$ in  \fr{eq:P0Q0} and some absolute constants $C_\sig$ and $c_0$, one has 
\be \label{eq:SBM_clust_assump}
\begin{split}
& \sig_r(Q_0) \geq C_\sig \sig_1(Q_0), \\ 
& \nmax = \max_k n_k   \leq c_0^2 \nmin = \min_k n_k. 
\end{split}
\ee
Then, the following statement holds.
 

\begin{prop} \label{prop:SBM_clust}  
Let condition \fr{eq:SBM_clust_assump} hold.  Let 
$m  \to \infty$,    $m = o(n)$ and $r/m = o(1)$,    as  $n \to \infty$. 
Let, in addition,  as  $n \to \infty$,  $r^6 \rhon / \log n = o(1)$ and also
\be \label{eq:new_SBM_cond}
\begin{split}
& \frac{r^3\, (\log  n)^4}{n^3 \rhon^4}  = o(1), \quad
\frac{r\, \sqrt{r}\, \log  n}{\rhon\, \sqrt{m\, n}} = o(1), \\
&  \hspace{25mm} \frac{(\log n)^5 \, r^3}{\rhon^5 \, m\, n^3} = o(1).
\end{split}
\ee
Then, if $n$ is large enough, with probability at least $1 - n^{-\tau}$,
estimated community assignment $\hzS$, obtained by Algorithm~\ref{alg:spec_clust} 
with $\hY$ of the form \fr{eq:hY_neq} with $\tilh=1$, coincides with the true  
community assignment $\zS$ up to a permutation of community labels.
\end{prop}


Using Proposition~\ref{prop:SBM_clust}, we can  confirm  that   using matrix $\ASSc$ instead of   matrix $\ASS$
allows   one to reduce the value of $m$.  Indeed, consider the situation, where $r$ is fixed and  
$\rhon \asymp n^{-\alpha}$. It is known  that the strongly consistent community assignment,  based on 
the complete data,  requires $\al < 1$.
%
However, according to  the first condition  in \fr{eq:new_SBM_cond},
one needs  $\al < 3/4$ for perfect clustering. Now, if    $m  \asymp n^{\beta}$, then 
the second and the third conditions in \fr{eq:new_SBM_cond} lead  to 
%
$\beta >  \max(2\, \al -1, 5\, \al -3).$ 
%
In comparison, if   matrix $\ASS$ were utilized, one would need $\beta > \al$,
which is a stronger condition, since $\al > \max(2\, \al -1, 5\, \al -3)$ for $\al < 3/4$.
For instance, if $\al = 1/2$, then using 
 $\ASS$  leads to the requirement that $\beta > 1/2$ while conditions of 
Proposition~\ref{prop:SBM_clust}   are satisfied for any positive value of $\beta$.


\begin{rem} {\bf Computational complexity.\ }
{\rm 
For a sparse matrix $B$, the computational complexity $CC(B,r)$ of evaluating its $r$ left singular vectors 
is $CC(B,r) = O(r\, \nnz(B))$, where $\nnz(B)$ is the number of nonzero elements of matrix $B$. 
Let $\rhon \asymp n^{-\alpha}$.  Denote by $m_0$   the number of sub-sampled nodes when $\ASS$ is used,  and by $m$
  the number of sub-sampled nodes in the case of $\ASSc$. 
Consider  $1/2<\alpha <1$, since $m_0^2 = O(n)$ for $\alpha \leq 1/2$.

Then, using  $\ASS$ requires $m_0  =  n^{\alpha}\, \polylog(n)$, where we denote  any power of $\log n$ by $\polylog(n)$.
Since $\nnz(\ASS) = O(\rhon\, m_0^2)$, derive that
$CC(\ASS,r) = O(r\, n^{\alpha}\, \polylog(n)).$
On the other hand, if one uses $\ASSc \in \{0,1\}^{m \times (n-m)}$ and $\hY = \scrH(\ASSc \ASSc^T)$, 
then the average number of nonzero elements in $\hY$ is
$\nnz(\hY) = O \lkr m^2\, [1 - (1 - \rhon^2)^n] \rkr = O \lkr  m^2\, n\, \rhon^2  \rkr.$
If $m = n^\beta\, \polylog(n)$ with $\beta =  \max(2\, \al -1, 5\, \al -3)$, then 
$\nnz(\hY) = O (n^\gamma\,  \polylog(n))$ where $\gamma = \max(2 \alpha -1, 8 \alpha -5)$.
Therefore,
\bes
\begin{split}
CC(\hY,r) & = O\lkr r\, n^{\max(2 \alpha -1, 8 \alpha -5)}\, \polylog(n)\rkr \\
& >  n^{\alpha}\, \polylog(n) \quad 
\mbox{for} \quad  \alpha > 1/2.
\end{split}
\ees
The latter means that Section~\ref{sec:subsampled} provides an instructive didactic example but is not recommended for applications. 
For a comprehensive treatment of sub-sampling based clustering on the basis of $\ASS$, see \cite{bhadra2025scalablecomm_detect}.
}
\end{rem}


\subsection{Perfect clustering of layers in a diverse multilayer  network}
\label{sec:multilayer}

Consider an $L$-layer undirected  network on the same set of $n$ vertices, 
with symmetric matrices of connection probabilities
in each layer $\linL$. 
We assume that   the layers of the network follow the  so called   Generalized Random Dot 
Product Graph ({\bf GRDPG})  model introduced by \cite{GDPG}. 
GRDPG assumes that the matrix of connection probabilities $P$ can be 
presented as $P =  H  I_{p,q}  H^T$, where $H \in \RR^{n \times K}$ is the 
latent position matrix and $I_{p.q}$ is the diagonal matrix with $p$ ones 
and $q$ negative ones on the diagonal, where $p + q = K$. Matrix $H$ is assumed to be such that  $P \in [0,1]^{n \times n}$.
If $H = U D_H V_H^T$ is the   SVD  of $H$, then $P$ can be alternatively 
presented as $P = U Q U^T$, where $Q = D_H V_H^T I_{p,q}  V_H D_H$. 
Then, $U$ is the basis of the ambient subspace of the GRDPG network, and $Q$ is the loading matrix.
It is known that the GRDPG generalizes a multitude of random network models, including the SBM,  
 studied in the previous section.

In this paper, we examine the case, where matrices  of probabilities of connections $P^{(l)} \in [0,1]^{n \times n}$,
 $\linL$,  can be partitioned into   $M$  groups with the common subspace structure, or community assignment. 
The latter means that  there exists a label function $z: [L] \to [M]$, which identifies to which of $M$ groups a layer belongs.
Specifically, we assume that    each group of layers is embedded in its own ambient subspace, 
but   all  loading matrices can be different. Then,   $P\upl$, $l \in [L]$, are given by
\be \label{eq:DIMPLE_GDPG}
P\upl = U\upm Q\upl (U\upm)^T, \quad m = z(l), \   m \in [M],
\ee
where $Q\upl = (Q\upl)^T$, and $U\upm \in \calO_{n, K_m}$ is a  basis matrix of the ambient subspace of the   $m$-th group of layers.
Here, $U\upm$ and  $Q\upl$ are such that  all entries of $P\upl$ are in $[0,1]$. 
This setting was extensively studies in \cite{pensky2021clustering}.
In this context, one observes   adjacency matrices   $A\upl$ such that $A\upl (i,j)$ are 
independent Bernoulli variables with
\bes
\begin{split}
& A{\upl} (i,j) = A\upl (j,i),  \quad \mbox{for} \quad 1 \leq i < j \leq n, \ \linL, \\
& \PP (A\upl(i,j)=1)=P \upl (i,j).  
\end{split}
\ees
The key objective in this setting is to recover the layer clustering function $z: [L] \to [M]$, since  estimation of $U\upm$, $\minM$,
can be subsequently carried out by some sort of averaging.

For simplicity, we assume that the rank $K\upl$ of each matrix $P\upl$ is known and that matrices $Q\upl$
in \fr{eq:DIMPLE_GDPG} are of full rank. Here, of course,
$K\upl = K_m$ when $z(l) = m$, but we are not going to use this information for clustering.
In order to estimate  the clustering function $z$,  observe that, by using the SVD $Q\upl =  O_Q\upl S_Q\upl (O_Q\upl)^T$ of $Q\upl$,
 matrices $P\upl$ in \eqref{eq:DIMPLE_GDPG} can be presented as 
\be  \label{eq:expans1}
P\upl =  \tilU\upl     S_Q\upl   (\tilU\upl)^T, \quad \tilU\upl = U\upm   O_Q\upl, 
\ee 
where $m = z(l)$, $\linL$, $\tilU\upl \in \calO_{n,K_m}$, and $S_Q\upl \in \calO_{K_m}$ and $S_Q\upl$ are diagonal matrices. 
In order to extract common information from matrices  $P\upl$, we furthermore consider the immediate SVD of  $P\upl$ with $m = z(l)$, $\linL$,
\be \label{eq:svd1} 
P\upl = U_{P,l} \Lam_{P,l} (U_{P,l})^T, \quad U_{P,l} \in \calO_{n,K_m},
\ee 
and relate it to the expansion \eqref{eq:expans1}. 
Due to  $\tilU\upm     \in  \calO_{n,K_m}$,   expansion \eqref{eq:expans1} is 
just another way of writing the SVD of $P\upl$. Hence, up to the $K_m$-dimensional rotation matrix $O_Q\upl$,
matrices $U\upm$ and $U_{P,l}$ are equal to each other, when $z(l)=m$.

Since finding an appropriate rotation matrix for each $\linL$ is cumbersome and computationally expensive, 
we build the between-layer clustering on the basis of matrices
\be \label{eq:main_rel}
U_{P,l} (U_{P,l})^T \! = \! U\upm   O_Q\upl (U\upm   O_Q\upl)^T  \! = \! U\upm (U\upm)^T, 
\ee
that depend  on $l$ only via $m = z(l)$, and are uniquely defined for $\linL$.
For this purpose, we consider the matrix $X \in \RR^{L \times n^2}$
with rows  $\Te(m,:)$, $m = z(l)$, $\linL$:
\be \label{eq:X_Te}
X(l, :) = \Te(m,:) = \vect(U\upm (U\upm)^T).
\ee   
It is easy to see that $X = Z \Te$ where $Z \in \{0,1\}^{L \times M}$ is a clustering matrix, 
such that $Z(l,m) = 1$ if $X(l, :) = \Te(m,:)$ and $Z(l,m)  = 0$  otherwise.

Since in reality, neither $U\upm$ nor  $U_{P,l}$ in \fr{eq:main_rel} are known,
we construct their data-driven proxies. 
Toward that end, we consider the SVDs of the adjacency matrices $A\upl$, $\linL$,  of the layers.
Let $\hU\upl$ be the matrices of $K\upl$ leading singular vectors of 
$A\upl$. Now consider  matrix $\hX \in \RR^{L \times n^2}$ with rows 
\be \label{eq:hU_upl_PW}
\hX(l, :) = \vect(\hU\upl (\hU\upl)^T), \quad \hU\upl = \SVD_{K\upl} (A\upl). 
\ee 
We use $\hX$ for estimating the clustering assignment $z: [L] \to [M]$.
Specifically, similarly to \cite{pensky2021clustering},  we apply Algorithm~\ref{alg:spec_clust} 
with $r=M$, $n=L$, and $\hU = \SVD_M(\hX)$.


In order to evaluate the clustering errors, we impose assumptions, that are similar to the ones in \cite{pensky2021clustering}.
Let $L_m$ be the number of layers of type $\minM$. Following \fr{eq:clust_assump} and \cite{pensky2021clustering}, we assume that 
clusters are balanced, that subspace dimensions  $K_m$ are of similar magnitude  and that matrix $\Te \in \RR^{M \times n^2}$ is well conditioned.
Therefore, we suppose that, for $K = \max K_m$ and some absolute positive constants  $C_\sig$, $C_K$, $\lowc$ and $\highc$,   one has
\be \label{eq:PW_clust_assump}
\begin{split}
& \sig_M(\Te) \geq C_\sig \sig_1(\Te), \quad C_K K \leq K_m \leq K, \\ 
& \lowc \, L/M \leq L_m \leq \highc \, L/M, \quad \minM.
\end{split}
\ee
In addition, as it is customary for  network data, we assume that the network is sparse, with the common  
sparsity factor $\rhon$, such that, for $\linL$,
\be \label{eq:sparsity}   
\begin{split}
& P\upl = \rhon\,  P_0\upl, \ \ \| P_0\upl\|_{\infty} \leq \highC, \\ 
& \rhon  \geq C_\rho\, n^{-1}\, \log n, \quad \| P_0\upl\|^2_F \geq  C_{0,P}^2\, K^{-1}\,  n^2,   
\end{split}
\ee
for some constants $\highC$, $C_\rho$ and $C_{0,P}$.  
In particular, the last inequality in \fr{eq:sparsity}  implies that, while elements of the  matrices $P_0\upl$ are bounded above 
by a constant, a fixed proportion of them are above a multiple of $K^{-1/2}$. 
We should comment that one can assume that sparsity factors are layer-dependent  
but this will make exposition here  less transparent. 
Also, as in \cite{pensky2021clustering},  we assume that matrices $Q\upl$ are also well conditioned, so that 
for some absolute constant $C_\lam \in (0,1)$, one has for $m=z(l)$
\be \label{eq:minbQl}   
\min_{\linL} \lkv \sig_{K_m}\lkr  Q\upl \rkr \Big/\sig_1 \lkr  Q\upl\rkr \rkv \geq C_\lam. 
\ee
Finally, similarly to \cite{pensky2021clustering}, in this paper, we study the case,
 where $L$ is large but is bounded above by some fixed power of $n$, i.e., 
\be \label{eq:L_n_tauo}
L \leq n^{\tau_0}, \quad \tau_0 < \infty.
\ee
We emphasize that  conditions \fr{eq:PW_clust_assump}--\fr{eq:L_n_tauo} are just a re-formulation 
of assumptions in \cite{pensky2021clustering} in the  notations of this paper.
The theoretical results however are very different. 


Recall that the between-layer clustering algorithm in  \cite{pensky2021clustering}
is just a version of Algorithm~\ref{alg:spec_clust} above with $r=M$, $n=L$, and $\hU = \SVD_M(\hX)$, where
$\hX$ defined in \fr{eq:hU_upl_PW}. Theoretical results in  \cite{pensky2021clustering}
rely on the upper bound for the spectral norm of the error matrix $\Xi = \hX - X$,
similarly to how it is done in, e.g.,    \cite{lei2021biasadjusted}, \cite{lei2015} and \cite{H_Zhou_AOS2021_Spec_Clust}. 
Observe that, although rows of matrix $\Xi$ are independent, its elements are not, 
and they are not necessarily  sub-Gaussian or sub-exponential. 
Consequently, one does not have a good control of the spectral norm $\|\Xi\|$ of matrix $\Xi$,
which leads to  exaggeration  of clustering errors.
In particular, under assumptions above, \cite{pensky2021clustering}  obtained the following results.

\begin{prop} \label{prop:error_between}  
{\bf (Theorem 1 of \cite{pensky2021clustering}). }
If assumptions \fr{eq:PW_clust_assump}--\fr{eq:L_n_tauo} hold, then,
 for any positive $\tau$ and some absolute constant $\Ctau>0$ one has, when $n$ 
is large enough
\be \label{eq:error_between_PW}
\PP \lfi \calR_n  (\hz, z)  \leq 
\frac{\Ctau K^2}{n \rho_n}  \rfi \geq 1 -  L\,  n^{- \tau}  \geq 1 -   n^{-(\tau-\tau_0)}.
\ee
Here, $\calR_n (\hz, z)$ is defined in \fr{eq:clust_er}. 
\end{prop}


In contrast to \cite{pensky2021clustering}, we use Proposition~\ref{prop:perfect_clust}
to assess clustering errors. 
Then, perfect clustering is guaranteed by conditions in \fr{eq:perf_cl1}. 
It turns out that, under mild assumptions, these conditions are satisfied,
and one obtains the following statement.

\begin{prop} \label{prop:error_between_new}  
Let conditions of Proposition~\ref{prop:error_between} hold 
and, in addition, 
\be \label{eq:new_between_cond}
\lim_{n \to \infty}  \,  (n \, \rhon)^{-1} \, (K\,M^2 \, \log^2 n + K^2)  =0.
\ee
Then, if $n$ is large enough, the between-layer clustering is  perfect 
with probability at least $1 - n^{-\tau}$.
\end{prop}


While Proposition~\ref{prop:error_between} only states that clustering is consistent,
Proposition~\ref{prop:error_between_new} ensures that, as $n$ grows, one achieves 
perfect clustering with high probability.  This is the precision guarantee  that was missing in 
\cite{pensky2021clustering}. Note that similar results hold when one considers a signed 
version of the same setting, featured in  \cite{pensky2024signed}.
However, \cite{pensky2024signed} applied centering to matrices $A\upl$ removing the means 
to achieve perfect clustering, Nevertheless, as  Proposition~\ref{prop:error_between_new} shows, 
perfect clustering can be obtained using singular vectors of matrices $A\upl$, $l \in [L]$.


 
\section{Comparison with the existing results}
\label{sec:comparison}

It is difficult to provide a comparison of the existing body of work  with the results in the present paper,
due to the fact that, as we mentioned before, majority of authors studied the bounds 
under much more stringent conditions, and with a specific application in mind.
To the  best of our knowledge, \cite{Cape_L2_inf_AOS2019}  is the only paper which had construction 
of generic upper bounds as a goal.

In the last few years, many authors  (see, e.g., \cite{abbe_fan_AOS2022}, \cite{AgterbergLubbertsPriebe2022},  
\cite{cai_AOS2021_unbalanced_incomplete}, \cite{chen_Assymetry_helps_AOS2021}, \cite{Chen_2021}, 
\cite{ChengWeiChen2021},  \cite{lihua_lei_2020_generic_symmetric}, 
\cite{wang2024_singular_subspaces_random}, 
\cite{Xie-Bernoulli2024}, \cite{xie2025AOS}, \cite{Yan_AOS2024_MissingData}, \cite{zhou2024_illposedPCA}) 
obtained upper bounds for $\|\hU   - U W_U\|\tinf$, designed for a variety of situations. 
However, those upper bounds were usually obtained for special scenarios, and, very often, 
under  relatively strict assumptions on the error distribution and problem settings.

For example, \cite{abbe_fan_AOS2022},  \cite{Chen_2021}  and \cite{Xie-Bernoulli2024} require the errors to be sub-Gaussian,
and \cite{Xie-Bernoulli2024}, in addition, examines the case of weak signals.   
\cite{xie2025AOS} construct uniform upper bounds on the entrywise differences under the assumptions that errors 
are independent and either sub-Gaussian or sparse Bernoulli variables.
\cite{wang2024_singular_subspaces_random} studies only the case of Gaussian errors. The authors of 
\cite{cai_AOS2021_unbalanced_incomplete} consider the case of a non-symmetric matrix 
where one dimension is much larger than another, noise components are independent and may be missing at random.
 \cite{chen_Assymetry_helps_AOS2021} examine the case where errors are  independent and bounded,    
the true matrix is symmetric while the error matrix is not. The main purpose of \cite{lihua_lei_2020_generic_symmetric}
is to design precise two-to-infinity norm perturbation bounds for symmetric sparse matrices. The focus of the author is 
on sharpening existing results and obtaining new ones for various random graph settings. 
 \cite{Yan_AOS2024_MissingData} studies PCA in the presence of missing data when the noise components are 
independent and  heteroskedastic. 
The objective of \cite{zhou2024_illposedPCA} is to design a new algorithm that improves the precision of the 
common SVD, when the dimensions of the observed matrix are unbalanced, so that the column space of the matrix is 
estimable in two-to-infinity norm but not in spectral norm. The authors study the case where the entries of the 
error matrix are independent and are bounded above by a fixed quantity with high probability.

In comparison, the goal of the present paper is to provide a ``toolbox'' for derivation of upper bounds
on $\|\hU-U W_U)\|\tinf$ under various sets of assumptions.  The fact that an extensive collection of the 
upper bounds is gathered in one place is one of the significant outcomes of our work. 
Majority of the results are not necessarily stronger than the ones of our predecessors, 
since many of the upper bounds obtained previously were proven to be statistically optimal, 
their improvement is impossible.  However, the upper bounds in the paper 
are obtained under weaker assumptions. We emphasize that our generic statements do
not impose the condition that  the entries of the error matrix are independent.

Below we provide a comparative summary of our results.

Theorems~\ref{th:sym_up_bound} is an incremental improvement on the result of \cite{Cape_L2_inf_AOS2019}.
Theorem~\ref{thm:nonsym_up_bound} appears in the literature as an 
intermediate results (it can be obtained by manipulations of the expansions in \cite{Cape_L2_inf_AOS2019}), 
or they are proved under some additional assumptions or conditions.
For instance, \cite{lihua_lei_2020_generic_symmetric}, whose goal is to improve the bounds in the case 
of sparse random networks that are equipped with the SBM structure,  proves a version of Theorem~\ref{th:sym_up_bound}
under some total variation conditions. Subsequently, this bound is improved by a correction of the diagonal of the data matrix,
and is applied to various versions of the  random networks.  
On the other hand, our goal is establishment of the Davis-Kahan theorem for statisticians in two-to-infinity norm.
As such, the matrix $\hU$ is found by a straightforward SVD rather than its fancy modification.
The upper bounds in Theorem~\ref{th:sym_up_bound_new}  are somewhat similar to the ones derived in \cite{abbe_fan_aos2020}.  
However, the latter bounds are derived under less flexible conditions and require a choice of a problem-dependent function $\phi$  
that may not be straightforward.   

Theorem~\ref{thm:nonsym_probab_up_bound_new}  can be viewed as a refinement of the results of 
\cite{abbe_fan_aos2020} and  \cite{Chen_2021}  since it assumes independence of the rows 
of the error matrix instead of independence of all entries (which is somewhat similar to 
\cite{AgterbergLubbertsPriebe2022}) and avoids an explicit assumption on the fraction $d_r/d_1$
of singular values.  
Theorems~\ref{thm:symetrized_up_bound}~and~\ref{thm:symetrized_probab_up_bound}
can be compared to  the papers that   advocate diagonal deletion or diagonal recovery such as
\cite{AgterbergLubbertsPriebe2022}, \cite{cai_AOS2021_unbalanced_incomplete}, \cite{Yan_AOS2024_MissingData}, 
\cite{ZhangCaiWu2022}, \cite{zhou2024_illposedPCA}. Among them, all authors except 
\cite{AgterbergLubbertsPriebe2022} assume that the error matrices have independent entries, 
while  \cite{AgterbergLubbertsPriebe2022} studies the case where the error matrix has   sub-Gaussian rows, 
components of which become independent after being de-correlated. 
On the contrary, Theorems~\ref{thm:symetrized_up_bound}~and~\ref{thm:symetrized_probab_up_bound}  are designed 
for a much general situation with the setback that the upper bounds require evaluation of many 
quantities. In addition, our paper studies only the  cases where the diagonal is either deleted or kept,
and does not examine what happens when one   recovers the diagonal before applying the SVD.
We admit that estimation of the diagonal using, e.g. HeteroPCA of \cite{cai_AOS2021_unbalanced_incomplete}
may lead to the tighter upper bounds in a variety of scenarios.

While the upper bounds in the paper are generic, they are rather tight. 
For example, consider comparison of Theorem~\ref{th:sym_up_bound_new} (which does not make an assumption 
that the entries of the error matrix are independent) to the new result of \cite{xie2025AOS}.
Just for simplicity, we assume that matrix $\scrE$ has independent Gaussian entries $\scrE(i,j)\sim N(0, \sig^2)$ for 
$1 \leq i \leq j \leq n$. In this case, it is easy to check that Assumption A2 holds with 
$\eps_1 = \sig \, \sqrt{\log n}\, |\lam_r|^{-1}$ and $\eps_2=0$.
Following assumptions of \cite{xie2025AOS}, we set $\lam_{r+1}=0$ and note that, with probability at least $1 - c\, n^{-\tau}$,
one has
\bes
\epso \asymp \sig \, \sqrt{n\, \log n}\, |\lam_r|^{-1}, \quad 
\epsEu \asymp \sig \, \sqrt{r\, \log n}\, |\lam_r|^{-1}.
\ees 
Then, plugging those upper bounds into \fr{eq:sym_up_bound_new} and observing that $\epsu \geq \sqrt{r}/\sqrt{n}$,
under the condition that $\epso = o(1)$ (which is also present in the paper of \cite{xie2025AOS}), we derive that, 
 with probability at least $1 - c\, n^{-\tau}$,
\be \label{eq:like-xie}
 \|\hU - U W_U\|\tinf  \leq \Ctau \, \epsu \sig \, \sqrt{n\, \log n}\, |\lam_r|^{-1}.
\ee  
Observe that inequality in \fr{eq:like-xie} coincides with the result of \cite{xie2025AOS},
where \fr{eq:like-xie} has slightly smaller power of $\log n$. We emphasize that, although 
the errors are independent Gaussian, Theorem~\ref{th:sym_up_bound_new} is not aware of this 
fact: we used the normality and independence assumption 
only to bound individual quantities in Theorem~\ref{th:sym_up_bound_new}.

One more example of the tightness of the bounds is provided by the derivation of the sufficient conditions 
for perfect clustering in the case of the i.i.d. Gaussian errors, which we presented in Section~\ref{sec:didactic} 
as a  didactic example. 
Specifically, below we compare our conditions for perfect clustering with the lower bounds derived in 
\cite{giraud2018RelaxedKmeans} in the Gaussian case. Let, for simplicity, $r = O(1)$, since the bounds in 
\cite{giraud2018RelaxedKmeans} are not tight in $r$ (\cite{giraud_2024_comp_gap} 
later refined their bounds to include $r$ in the case when $m \geq n$).  Then,  
under the assumptions in Section~\ref{sec:didactic}, in the notations of this paper,
\cite{giraud2018RelaxedKmeans} derived the following lower bound for the  probability of 
misclassifying an element $i \in [n]$: 
\be \label{eq:giraud_lower}  
\PP \lkr \hz(i) \neq z(i) \rkr \geq C  \exp \lfi - c  \min\lkr    \frac{\te^4 n  m}{\sig^4}, \frac{\te^2 m}{\sig^2} \rkr \rfi.
\ee  
Therefore, the necessary conditions that perfect clustering occur with high probability are  
\be \label{eq:giraud_necces}
 \frac{\sig^4 \, \log n}{\te^4\, m\, n} = O(1), \quad \frac{\sig^2 \, \log n}{\te^2\, m} = O(1).
\ee 
Now, compare   conditions in \fr{eq:giraud_necces}  with the sufficient conditions  in \fr{eq:suf_sym}  
of  Proposition~\ref{prop:didactic}.
Recalling that Assumption A4* holds and that we use $o(1)$ in Proposition~\ref{prop:didactic} to indicate that the quantity
is bounded by a small enough constant, the sufficient conditions in \fr{eq:suf_sym} become
\be \label{eq:our_suffic}
\frac{\sig^4 \, \log^2 n}{\te^4\, m\, n} = O(1), \quad \frac{\sig^2 \, \log^2 n}{\te^2\, m} = O(1).
\ee 
Hence   sufficient conditions \fr{eq:our_suffic}  coincide with the necessary conditions 
in \fr{eq:giraud_necces}   up to a $\log n$ factor,
which means that conditions \fr{eq:our_suffic}  are within at most $\log n$ factor of  optimality.

Another advantage of our paper is that ``the complete toolbox'' approach allows one to compare 
different techniques and to choose the best one. For example, \cite{wang2024_singular_subspaces_random} constructs very accurate upper bounds 
on $\|\hU - U W_U\|\tinf$, since the proof explicitly uses the fact that the errors are i.i.d. standard Gaussian. 
However, the author requires that the operational norm is smaller by a constant factor than the lowest singular value,
which, in our notations, is equivalent to  $\Delo = O(1)$. 
The latter, due to $\sig =1$,  demands that $r \te^{-2} (m^{-1} + n^{-1}) = O(1)$ which may not be true if $\te$ is small.
Section~\ref{sec:didactic}, with its comparisons of various techniques, offers an immediate remedy to this difficulty.
Indeed, if $n \ll m$, one can use symmetrization with the subsequent hollowing. 
Let $n \ll m$, and, as it is set in Section~\ref{sec:didactic},  $r = O(1)$, $n = m^\ga$  and $\te \asymp m^{-\nu}$, 
where $\ga < 1$ and $\nu>0$. Then the upper bounds in \cite{wang2024_singular_subspaces_random} can be employed only if 
$\nu \leq \ga/2$, while the upper bounds in our paper are valid if $\nu < (\ga + 1)/4$, which is always 
larger than $\ga/2$ for $\ga <1$.

In summary, although our paper uses established techniques such as leave-one-out decoupling, 
row-wise concentration, diagonal deletion and Gram symmetrization, 
and exact clustering through row-wise eigenspace control, it applies them systematically 
for a comprehensive set of cases and scenarios. Not only many of the upper bounds are derived  
 under milder conditions but they are also accompanied by refinements designed for 
various types of errors (Gaussian, sub-Gaussian, heavy-tailed).
However, if one has a setting that completely coincides with a specific model  studied 
in depth in another paper, which is dedicated to this exclusive scenario, 
it may be easier to use ready-made upper bounds rather than obtaining them 
for each individual quantity in our statements.

We would also like to mention that while our results are quite comprehensive, 
they do not include the case when  the incoherence condition or a large 
eigengap assumption \fr{eq:lamr_r} or \fr{eq:d_r} is violated. 
Consideration of such scenarios requires the finer random-matrix control of 
\cite{chen_Assymetry_helps_AOS2021}  and \cite{ChengWeiChen2021}.

\section*{Acknowledgments} 

The   author of the paper gratefully acknowledges partial support by National Science Foundation 
(NSF) grants DMS-2014928 and   DMS-2310881
 



\section{Proofs of the main statements of the paper}
\label{sec:proofs}

\renewcommand{\theequation}{P.\arabic{equation}}
 \setcounter{equation}{0}


\subsection{Proofs of statements in Section~\ref{sec:symmetric}.  } 
\label{sec:proofs_symmetric}

\noindent
{\bf Proof of Theorem~\ref{th:sym_up_bound}.\ }  
Note that, by Weyl's theorem,   one has
\bes
\hat{\lambda}_r = \lambda_r(\hat{Y}) \geq   \lambda_r - \| \scrE \|,
\ees
so that  
$\lnor \hLam^{-1} \rnor =  |\hlam_r|^{-1} \leq \lkr |\lam_r| - \| \scrE \|\rkr^{-1} = 
|\lam_r|^{-1} \, \lkv  |\lam_r| /( |\lam_r| - \| \scrE \|)\rkv.$
Thus,  
\be \label{eq:hLam_inverse}
\lnor  \hLam^{-1} \rnor \leq |\lam_r|^{-1} \,  ( 1 - \Delo)^{-1} \leq 4/3 \, |\lam_r|^{-1}.
\ee 
Observe that 
\be \label{eq:R1234}
\|\hU - U W_U\|_{2, \infty} \leq R_1 + R_2 + R_3 + R_4. 
\ee
Here, 
\begin{align*}
R_1 & =  \|( I - UU^T) \scrE U W_U \hat{\Lambda}^{-1}  \|\tinf \\
  &  \leq
 \|  U\,U^T \scrE U W_U \hat{\Lam}^{-1}   \|\tinf + 
 \|  \scrE U W_U \hat{\Lambda}^{-1}    \|\tinf \\
& \leq    \| U \|\tinf\,  \| \scrE  \|  \| U\, W_U \|  \| \hat{\Lambda}^{-1}  \|  + 
 \| \scrE\, U \|\tinf    \| \hat{\Lambda}^{-1}  \|.
\end{align*} 
Therefore, 
\be \label{eq:R1}
R_1 \leq 4/3 \,  (\Delo\, \epsu + \DelEu).
\ee 
Now, we derive an upper bound for $R_2$:
\begin{align*}
R_2 & =  \left\| \left( I - U\, U^T \right) \scrE \left( \hU - UW_U \right) \hat{\Lambda}^{-1} \right\|_{2, \infty} \\
  &  \leq 
\| U \|_{2, \infty} \| \scrE \| \, \| \hU - U\, W_U \| \lnor \hat{\Lambda}^{-1}  \rnor\\
&  + 
\| \scrE \|\tinf \, \| \hU - U\, W_U \| \lnor \hat{\Lambda}^{-1}  \rnor,
\end{align*} 
so that, due to \fr{eq:ineq3}, one has
\be \label{eq:R2}
R_2 \leq 8/3\, \clam^{-1}\, \Delo (\Delo\, \epsu +   \Delinf).
\ee 
Now consider 
\begin{align*}
R_3 & = \left\| \left( I - UU^T \right) Y \left( \hU - UU^T \hU \right) \hat{\Lambda}^{-1} \right\|_{2, \infty} \\
& =
\lnor U_\perp U_\perp^T  U_\perp \Lambda_\perp U_\perp^T (\hU - UU^T \hU) \hat{\Lambda}^{-1} \rnor\tinf \\
& \leq \|\Lambda_\perp\| \, \lnor \hat{\Lambda}^{-1} \rnor \, \|\hU - U\, U^T \, \hU\|,
\end{align*} 
so, by \fr{eq:ineq2}, obtain
\be \label{eq:R3}
R_3 \leq 8/3\,   \clam^{-1} \, |\lam_{r+1}|\,  |\lam_r|^{-1} \, \Delo.
\ee 
Finally, $R_4   =  \left\| U \left( U^T \hU - W_U \right) \right\|_{2, \infty}$ and, by \fr{eq:ineq1}, derive that 
\be \label{eq:R4}
R_4 \leq  4\,   \clam^{-2}\, \epsu\, \Delo^2.
\ee 
Finally, combining \fr{eq:R1234}--\fr{eq:R4} and taking into account that $\Delo \leq  1/4$, obtain   \fr{eq:sym_up_bound}.
Inequality  \fr{eq:probab_sym_up_bound} is the direct consequence of \fr{eq:probab_errors} and  \fr{eq:sym_up_bound}. 

\medskip


\noindent
{\bf Proof of Corollary~\ref{cor:sym_heavy_tailed}. \  }.   
It follows from 
\cite{bandeira2016,Latala2005,Seginer2000}
that, for any $t>0$ 
\be \label{eq:heavy_tail_op_norm} 
\PP \lfi \|\scrE\| \leq  C_s \, t\, \lkr \sig \sqrt{n} +  (n\,\nu_{2s})^{\frac{1}{2s}} \rkr \rfi \geq 1 - t^{-2s}.
\ee  
Also, for any matrix $G \in \RR^{n \times m}$, any  $i \in [n]$ and any $t_1>0$, one has 
\bes
\PP \lfi \| \scrE(i,:)\, G \|  \leq C_{2s}  t_1 \!   \lkr \sig \|G\|_F + \nu_{2s}^{\frac{1}{2s}} \|U^T\|_{2,2s} \rkr \! \rfi 
\geq 1 \!- \! t_1^{-2s} 
\ees
Here, for any matrix $G$, the mixed norm $\|G\|_{2,2s}$  is defined as 
\bes
\|G\|_{2,2s} = \lkr \sum \|G(:,j)\|^{2s} \rkr^{1/(2s)}.
\ees
Noting that $\|U^T\|_{2,2s} \leq   n^{1/(2s)}\, \epsu$ and applying the union bound over   $i \in [n]$, derive 
\be \label{eq:heavy_tail_tinf_norm} 
\hspace*{-2.5mm} \PP \lfi \|\scrE   U\| \leq C_{2s}   t_1\, \lkr \sig \sqrt{r} +  \epsu (n \nu_{2s})^{\frac{1}{2s}} \rkr \rfi 
\geq 1 \! - \! n  t_1^{-2s}.
\ee  
Set $t = C\,n^{\frac{\tau}{2s}}$ and $t_1 = C\,n^{\frac{\tau + 1}{2s}}$,
where the constant $C$ is such that $3\, t^{-2s} + n \, t_1^{-2s} = n^{-\tau}$,
and plug \eqref{eq:heavy_tail_op_norm} and \eqref{eq:heavy_tail_tinf_norm} into \eqref{eq:probab_sym_up_bound}.
Obtain, with probability at least $1 - n^{-\tau}$, that
\bes 
\begin{split}
\|\hU - U W_U\|\tinf & \leq  \Ctau \,   \delrs\, \lkr \epsu  + 
  |\lam_r|^{-1}\, |\lam_{r+1}|    +   \delrs \rkr \\
  &  + |\lam_r|^{-1}\, n^{\frac{\tau + 1}{2s}}\,
 \lkr \sig \sqrt{r} + \epsu\, (n\,\nu_{2s})^{\frac{1}{2s}} \rkr. 
 \end{split}
\ees
Since $\epsu^{-1} \leq \sqrt{n}/\sqrt{r}$, obtain that 
\bes
|\lam_r|^{-1}\, n^{\frac{\tau + 1}{2s}}\,
 \lkr \sig \sqrt{r} + \epsu\, (n\,\nu_{2s})^{\frac{1}{2s}} \rkr \leq \Ctau \,   \delrs\,   \epsu\, n^{1/(2s)},
\ees 
which yields \eqref{eq:probab_sym_heavytail}. 

\medskip


\noindent
{\bf Proof of Theorem~\ref{th:sym_up_bound_new}.\  }   
Denote the sets, on which \fr{eq:probab_errors} and \fr{eq:assump_A2} are true, by, respectively, $\Omtauo$ and $\Omtaut$. 
Denote $\Omtau = \Omtauo \cap \Omtaut$ and observe that $\PP(\Omtau) \geq 1 - 2\, n^{-\tau}$.

Note that, due to \fr{eq:eps_conditions} and \fr{eq:hLam_inverse}, one has   $\lnor \hLam^{-1} \rnor \leq 4/3 \, |\lam_r|^{-1}$
 for $\om \in \Omtauo$.
Also, since $\epso = o(1)$, for $\om \in \Om_{\tau,1}$, one has $\|\sinTeU\|\leq 1/\sqrt{2}$ for $n$ large enough. 
Then, by  \fr{eq:ineq1}, obtain that 
$\|U^T \hU - W_U \|\leq 1/2$, and since $W_U \in \calO_r$, by Weyl's theorem, one has 
$\sig_r(U^T\, \hU) \geq   1/2$. Therefore, by Weyl's theorem, 
\be \label{eq:inv_norm}
 \hspace*{-2.5mm}\|(U^T \hU)^{-1}\|\leq 2, \  \|(\hU^T  U)^{-1}\|\leq 2, \  \|\hLam^{-1}\| \leq 2 |\lam_r|^{-1}.  
\ee
Consider the expansion \fr{eq:cape-expan} and observe that
\bes
\begin{split}
  \hU & - U\, W_U = (\hU\, \hU^T \, U - U)(\hU^T \,  U)^{-1} \\
& + 
U[I - (U^T \, \hU)(\hU^T \,  U)](\hU^T \,  U)^{-1} + U\, (U^T\, \hU - W_U).
\end{split}
\ees
Plugging the latter into the second term of \fr{eq:cape-expan}, derive
\be \label{eq:expansion}
\begin{split}
& \hU - U W_U   = (I - U U^T) \scrE U \lkv U^T \, \hU  \right. \\
& + \left.  \lkr I -  U^T \, \hU\, \hU^T \,  U) \rkr \,(\hU^T \,  U)^{-1} \rkv \,  \hat{\Lambda}^{-1}  \\ 
& + (I - U U^T) \scrE\, (\hU\, \hU^T \, U - U)(\hU^T \,  U)^{-1}   \,  \hat{\Lambda}^{-1} \\ 
& + (I - U U^T) Y (\hU - U U^T \hU) \hat{\Lambda}^{-1}  \\
& + U (U^T \hU - W_U).  
\end{split}
\ee
Then, one has
\begin{align*}
\|\hU & - U W_U\|\tinf   \leq   2 \, |\lam_r|^{-1} \lfi  \|\scrE\| \, \epsu  + 
\|\scrE\, U\|\tinf \right.  \\
& + 2 \, \epsu\, \|\scrE\|\, \|I - (U^T \, \hU)(\hU^T \,  U)\| \\
& +   2\, \|\scrE\, U\|\tinf\, \|I - (U^T \, \hU)(\hU^T \,  U)\| \\
+ & 2\, \epsu\, \|\scrE\| \, \|\hU\, \hU^T \, U - U \| \\
& +  \left.  2\, \|\scrE\, (\hU\, \hU^T \, U - U)\|\tinf  
+ 2\, |\lam_{r+1}|\, \|\hU - U U^T \hU\| \rfi \\
& + \epsu\, \|U^T\hU - W_U\|.
\end{align*}
Hence, due to $\epso = o(1)$ and \fr{eq:ineq4}, for $\om \in \Omtauo$, one has 
\be \label{eq:for20}
\begin{split}
  \|\hU & - U W_U\|\tinf \leq  4\, |\lam_r|^{-1}\, \|\scrE\, (\hU\, \hU^T \, U - U)\|\tinf\\
& + \Ctau \, \lkr \epso\, \epsu +    \epsEu
+ |\lam_r|^{-1}\, |\lam_{r+1}| \, \epso  \rkr.  
\end{split}
\ee
Now,   use the following lemma.

\begin{lem}\label{lem:lemma1}  
Let conditions of Theorem \ref{th:sym_up_bound_new} hold. Then,  for $\om \in \Omtauo$, one has
\be 
 \|\hU\, \hU^T\, U - U \|\tinf   \leq 4\,  \|\hU - U\, W_U\|\tinf + \Ctau\, \epso^2 \, \epsu.
\label{eq:lemma11}
\ee
Also, for  $\om \in \Omtauo \cap \Omtaut$, the following inequality holds
\be \label{eq:lemma12} 
\begin{split}
&  |\lam_r|^{-1}\, \|\scrE\, (\hU\, \hU^T \, U - U)\|\tinf  \leq \\
& \hspace{4mm}  \Ctau \lfi  (\epsEu + \epso\, \epsu) (\epso + \eps_2) 
+ \sqrt{r}\, \epso\, \eps_1 \right. + \\
&  \hspace{4mm} \left. (\epso^2 + \epso\, \eps_1 + \eps_2)\,  \|\hU\, \hU^T\, U - U \|\tinf \rfi.
\end{split}
\ee
\end{lem}

Combining \fr{eq:lemma11} and \fr{eq:lemma12}, plugging them into \fr{eq:for20} and removing the smaller order terms, obtain
\begin{align*}
\|\hU & - U W_U\|\tinf  \leq  \Ctau \, \lfi    \epso\, \epsu +   \epsEu
+ |\lam_r|^{-1}\, |\lam_{r+1}| \, \epso \right.   \\
& \left. + \sqrt{r}\, \epso \eps_1  \rfi  +  4\, (\epso^2 + \epso\, \eps_1 + \eps_2)\, \|\hU - U W_U\|\tinf.
\end{align*}
Adjusting the coefficient for $\|\hU - U W_U\|\tinf$ in a view of  \fr{eq:eps_conditions}, arrive at \fr{eq:sym_up_bound_new}.


\subsection{Proofs of statements in Section~\ref{sec:nonsymmetric}} 
\label{sec:proofs_nonsymmetric}

\noindent
{\bf Proof of Theorem~\ref{thm:nonsym_up_bound}.\ }   
Using Weyl's theorem for singular values obtain, similarly 
to the proof of Theorem~\ref{th:sym_up_bound}, that
$\hd_r \geq d_r - \|\Xi\|$, 
so that 
$\|\hD^{-1} \| =  \hd_r^{-1} \leq  
d_r^{-1} \, \lkv d_r/( d_r -  \|\Xi\|)\rkv.$
Thus,  
\be \label{eq:hD_inverse}
\lnor \hD^{-1} \rnor \leq d_r^{-1} \,  ( 1 - \tDelo)^{-1} \leq C \, d_r^{-1}.
\ee 
Also, relationships \fr{eq:cai-zhang2} and \fr{eq:D_K_Spectral} are valid for both $U,\hU$ and $V, \hV$.

Note that again, $\|\hU - U W_U\|_{2, \infty} \leq R_1 + R_2 + R_3 + R_4$,
where 
\beqns
R_1 & = & \| (I - UU^T) \Xi VW_V \hD^{-1}  \|\tinf, \\
R_2 & = & \| (I - UU^T) \Xi (\hV - VW_V) \hD^{-1} \|\tinf, \\
R_3 & = & \| (I - UU^T) X (\hV - VV^T \hV) \hD^{-1}  \|\tinf, \\
R_4 & = & \| U (U^T \hU - W_U) \|\tinf.
\eeqns
Then, it is easy to see that 
\beqns
R_1 &\leq & \lkv \|U\|\tinf\,   \| U^T \Xi V  \|  + \| \Xi V \|\tinf \rkv \, \| \hD^{-1}  \|\\
&\leq & C\, (\epsu\, \tDeluvo + \tDelvtinf), \\
R_2 &\leq & \lkv \|U\|\tinf\,   \| U^T \Xi  \|  + \| \Xi \|\tinf  \rkv \, \|\hV - V W_V\|  \| \hD^{-1}  \| \\
&\leq & C \, (\epsu\, \tDelo + \tDeltinf)\, \|\sinTeV\|,   \\               
R_3 &\leq & \| D_\perp  \|\,   \| \hD^{-1}  \| \,   \| \hV - VV^T \hV  \| \\
&\leq & C\, d_{r+1}\, d_r^{-1} \, \|\sinTeV\|, \\
R_4 &\leq & C\, \epsu\, \|\sinTeU\|^2
\eeqns
In the expressions above,  the $\sinTe$ distances $\|\sinTeU\|$ and $\|\sinTeV\|$
can be bounded above using   the Wedin theorem 
which in our case appears as 
\be \label{eq:Wedin}
\begin{split}
\max \lkr \|\sinTeU\|,  \right. & \left. \| \sinTeV\| \rkr   \leq \\
    &  C\, d_r^{-1}\, \|\Xi\| \leq C \, \tDelo.
\end{split}
\ee 
Combining the upper bounds for $R_1$, $R_2$, $R_3$ and $R_4$ with \fr{eq:Wedin}, derive that
\begin{align*}
\|\hU & - U W_U\|\tinf \leq   C\, \lkv \epsu\, \tDeluvo + \tDelvtinf \right. \\
& +  \left. (\epsu\, \tDelo + \tDeltinf)\, \tDelo + 
d_{r+1}\, d_r^{-1} \, \tDelo + \epsu \, \tDelo^2 \rkv,
\end{align*}
which is equivalent to \fr{eq:nonsym_up_bound}.  Validity of \fr{eq:probab_nonsym_up_bound} follows directly from 
 \fr{eq:nonsym_up_bound} and \fr{eq:probab_nonsym_err}.

\medskip


\noindent
{\bf Proof of Corollary~\ref{cor:nonsym_heavy_tailed}.\ }  
It follows from \cite{bandeira2016,Latala2005,Seginer2000}  and symmetrization argument   
that, for any $t>0$ 
\be \label{eq:heavy_tail_nonsym_op} 
\begin{split}
& \PP \lfi \|\Xi \| \leq  C_s \, t\, \tdelrs(n+m)  \rfi \geq 1 - t^{-2s},  \\ 
& \PP \lfi \|U^T \Xi V\| \leq  C_s \, t\, \tdelrs(r)  \rfi \geq 1 - t^{-2s}.
\end{split}
\ee 
Also, similarly to the Proof of Corollary~\ref{cor:sym_heavy_tailed}, for  any $t_1>0$, derive
\be \label{eq:heavy_tail_nonsym_tinf} 
\begin{split}
& \PP \lfi \|\Xi \|\tinf \leq C_{2s} \, t_1\, \tdelrs(m)   \rfi \geq 1 - n\, t_1^{-2s}, \\
&\PP \lfi \|\Xi\, V \|\tinf \leq C_{2s} \, t_1\, \tdelrs(m)   \rfi  \geq 1 - n\, t_1^{-2s}.
\end{split}
\ee  
Setting  $t = C\,n^{\frac{\tau}{2s}}$ and $t_1 = C\,n^{\frac{\tau + 1}{2s}}$,
where the constant $C$ is such that $5\, t^{-2s} + n \, t_1^{-2s} = n^{-\tau}$,
and plugging \eqref{eq:heavy_tail_nonsym_op}  and \eqref{eq:heavy_tail_nonsym_tinf}  
into \eqref{eq:probab_nonsym_up_bound}, obtain, with probability at least $1 - n^{-\tau}$, that
\begin{align*} 
& \|\hU   - U W_U\|\tinf  \leq  \Ctau    \lkv \epsu\, \tdelrs(r) + \epsu\, (\tdelrs(n+m))^2
\right.\\
& \left. + n^{\frac{1}{2s}}\, \tdelrs(r)  + n^{\frac{1}{2s}}   \tdelrs(n+m) \tdelrs(m) +
\tdelrs(n+m)   \frac{d_{r+1}}{d_r}  \rkv.
\end{align*}
Since $\epsu = o(n^{1/(2s)})$, the first term is of the smaller order.
Combining the terms, obtain \fr{eq:probab_nonsym_heavytail}.

\medskip


\noindent
{\bf Proof of Theorem~\ref{thm:nonsym_probab_up_bound_new}.\  }   
Denote the sets, on which \fr{eq:probab_nonsym_err} and \fr{eq:assump_A4} are true, by, respectively, $\tilOmtauo$ and $\tilOmtauo$. 
Denote $\tilOmtau = \tilOmtauo \cap \tilOmtauo$ and observe that $\PP(\tilOmtau) \geq 1 - 2\, n^{-\tau}$.
It follows from the proof of Theorem~\ref{thm:nonsym_up_bound} and \fr{eq:hD_inverse}    that 
\begin{align*}
\|\hU & - U W_U\|_{2, \infty}   \leq  \tilR + d_r^{-1}\, \| \Xi V \|\tinf   
+ \epsu\, d_r^{-1}\,  \| U^T \Xi V  \| \\
& + \epsu\, d_r^{-1}\,  \| U^T \Xi\|\,  \|\hV - V W_V\|\\
& + d_r^{-1}\,   d_{r+1} \, \|\hV - VV^T \hV\| + \| U (U^T \hU - W_U) \|\tinf,
\end{align*}
where
$\tilR = \|\Xi  (\hV - V W_V) \hD^{-1} \|\tinf \leq C d_r^{-1}  \|\Xi  (\hV - V W_V)  \|$.

Applying the upper bounds, as in the proof of Theorem~\ref{thm:nonsym_up_bound} and Wedin theorem \fr{eq:Wedin},
and removing the smaller order terms, derive that 
\be \label{eq:base_ineq_f5}
\begin{split}
\|\hU & - U W_U\|_{2, \infty}   \leq  \tilR  + C\, \lkv \tDelvtinf + d_r^{-1}\,   d_{r+1} \, \tDelo  \right. \\
& + \left. \epsu\, (\tDeluvo + \tDelo^2)\rkv.
\end{split}
\ee
In order to derive an upper bound for $\tilR$, we use the ``leave one out'' method.
Specifically,  fix   $l \in [n]$, and decompose   $\Xi$ as 
\be \label{eq:Xiupl-th5}
\Xi = \Xi\upl + e_l \Xi(l,:),  
\ \ \  \Xi\upl(i,:)\! = \! \lfi 
\begin{array}{l}  
\Xi(i,:),\  \mbox{if}\  i\neq l, \\
0, \ \ \ \ \  \ \ \mbox{if}\   i= l,
\end{array} \right.
\ee 
and $e_l$ is the $l$-th canonical  vector in $\RR^n$.
Denote $\hX\upl = X + \Xi\upl$  and consider the SVD of $\hX\upl$:
\bes
\hX\upl = \hU\upl \hD\upl  (\hV\upl)^T + \hU_{\perp}\upl \hD_{\perp}\upl (\hV_{\perp}\upl)^T, 
\ees
where $\hU\upl \in \calO_{n,r}$  and $\hV\upl \in \calO_{m,r}$.
Since $\|\Xi\upl\| \leq \|\Xi\|$, one has 
\be \label{eq:Upl_rel_nonsym}
\begin{split}
 & \|\hD\upl - D\| \leq \| \hD -  D\|, \\
& \| \sinTe(\hU\upl, U)\| \leq \|\sinTeU\|, \\
& \| \sinTe(\hV\upl, V)\| \leq \|\sinTeV\|.
\end{split}
\ee
Due to $\hV - V W_V =(\hV \hV^T  V - V)]\, (\hV^T  V)^{-1} + V \lkv I_r - (V^T \hV)(\hV^T  V)\rkv (\hV^T  V)^{-1}
+  V(V^T \hV - W_V)$ and the fact that $\|(\hV^T  V)^{-1}\|\leq 2$ for $m$ and $n$ large enough, derive
 \be \label{eq:tilR0_tilR1-th5}
 \begin{split}
 \tilR & = \|\Xi\, (\hV - V W_V) \hD^{-1} \|\tinf \leq C\, (\tilR_0  + \tilR_1) \\
 & + C \, d_r^{-1}\, \|\Xi V\|\tinf\, \|\sinTeV\|^2,
 \end{split}
\ee
 where 
\begin{align} \label{eq:tilR0_tilR1}
& \tilR_0 = \max_{l \in [n]}  d_r^{-1}\, \left\| \Xi(l,:) \lkv \hV   \hV^T V - V\rkv \right\|,\\
\label{eq:tilR1}
& \tilR_1 = d_r^{-1}  \left\|\Xi V \, \lkr I_r - V^T \hV \hV^T V \rkr \right\|\tinf \leq \tDelvtinf.
\end{align}
Hence, for $m$ and $n$ large enough
\be \label{eq:tilR_upper}
 \tilR \leq C\, (\tilR_0 + \tDelvtinf).
\ee
Now observe that 
\be \label{eq:tilR01_02}
\tilR_0 \leq  \tilR_{01} + \tilR_{02}
\ee
with 
\be \label{eq:R01_R02-for}
\begin{split}
& \tilR_{01} = \max_{l \in [n]} \lnor \Xi(l,:) \lkv \hV\upl (\hV\upl)^T V - V\rkv \rnor,\\
& \tilR_{02} = \max_{l \in [n]} \|\Xi\| \, \lnor \lkv \hV \hV^T - \hV\upl (\hV\upl)^T \rkv \, V \rnor_F
\end{split}
\ee 
Start with the second term. Note that, by Wedin theorem (\cite{Wedin1972PerturbationBI}),
\be  \label{eq:lem2_for11-th5}
\hspace*{-2mm} \| [\hV \hV^T \! - \! \hV\upl (\hV\upl)^T]  V \|_F \leq \frac{C}{|d_r|} \lnor (\hX \!- \! \hX\upl) \hV\upl\rnor_F
\ee
Here, $(\hX - \hX\upl) \hV\upl = e_l \Xi(l,:)\hV\upl$. 
Since \\ $\rank(e_l \Xi(l,:)\hV\upl)=1$, derive that
\bes
\|(\hX - \hX\upl) \hV\upl\|_F = \|\Xi(l,:)\hV\upl||.
\ees
Denote $H = \hV^T V$, $H\upl = (\hV\upl)^T\, V$. 
Then,  for $n$ and $m$ large enough, $\|H^{-1}\| \leq 2$ and   $\|(H\upl)^{-1}\| \leq 2$, and 
\be \label{eq:th5_for21}
\begin{split}
\lnor  \Xi(l,:) \hV\upl \rnor & \leq 2\, \| \Xi(l,:) [ \hV\upl (\hV\upl)^T \, V -   V] \| \\
& +  2\, \|\Xi(l,:)\,  V \|.
\end{split}
\ee
Due to independence between $\Xi(l,:)$ and $\hV\upl$, for $\om \in \tilOmtau$, one has
\begin{align*}
& \|\Xi(l,:) [\hV\upl (\hV\upl)^T\, V - V] \|   \leq \\
& \ \ \Ctau |d_r| \lkr \teps_1 \|[\hV\upl (\hV\upl)^T - \hV \hV^T]\, V\|_F + \right. \\
& \ \ \teps_1 \,  \|\hV \, \hV^T \, V - V\|_F  +  \teps_2 \,  \|\hV \, \hV^T  - V\|\tinf + \\
& \ \  \left. \teps_2 \,  \| [\hV\upl (\hV\upl)^T  - \hV \, \hV^T]\, V - V\|\tinf 
  \rkr. 
\end{align*}
Plugging the last inequality into \fr{eq:th5_for21}   and noting that, for $\om \in \tilOmtau$, one has 
$\| \Xi(l,:) V\| \leq \Ctau |d_r|\,\tepsvtinf$, derive 
\begin{align*} 
& \| [\hV\upl (\hV\upl)^T  - \hV \, \hV^T]\, V \|_F  \leq     \Ctau\, \lkv  \tepsvtinf \right. \\  
&  \ \ + (\teps_1 + \teps_2)\,  \|[\hV\upl (\hV\upl)^T - \hV \hV^T]\, V\|_F        \\
& \ \ +   \left. \teps_1\,  \|\hV \, \hV^T\, V  - V\|_F + 
 \teps_2  \, \|\hV \, \hV^T\, V  - V\|\tinf  \rkv.  
\end{align*}
Combining the terms under the condition that $\Ctau\, (\teps_1 + \teps_2) <1/2$, derive that   
for $\om \in \tilOmtau$ and $n$ and $m$ large enough
\be \label{eq:th5_for25}
\begin{split}
& \| [\hV\upl (\hV\upl)^T  - \hV \, \hV^T]\, V \|_F   \leq     \Ctau\, \lkv  \tepsvtinf \right. \\
& \ \  +   \left.   \teps_1\,  \|\hV \, \hV^T\, V  - V\|_F + 
 \teps_2  \, \|\hV \, \hV^T\, V  - V\|\tinf  \rkv.  
 \end{split}
\ee 
Therefore, due to independence of $\Xi(l,:)$ and $\hV\upl$, 
the upper bound for $\tilR_{01}$ in  \fr{eq:R01_R02-for}  is of the form
\begin{align*}  
\tilR_{01}  & \leq   \Ctau\, \lkv \teps_1 \, \|[\hV\upl (\hV\upl)^T - \hV \hV^T]\, V\|_F \right. \\
& + \teps_1 \, \|\hV \, \hV^T\, V  - V\|_F 
+    \teps_2\, \|\hV \, \hV^T\, V  - V\|\tinf \\
& +  \left.
 \teps_2\, \|[\hV\upl (\hV\upl)^T - \hV \hV^T]\, V\|\tinf  \rkv.
\end{align*} 
Plugging the last inequality into \fr{eq:th5_for25}, using 
\bes
\|\hV \, \hV^T\, V  - V\|\tinf \leq \|\hV \, \hV^T\, V  - V\|_F \leq \Ctau \, \sqrt{r} \, \tepso,
\ees
 and combining the terms, obtain 
\be \label{eq:tilR01_upper}
\tilR_{01}  \leq \Ctau \, (\teps_1 + \teps_2) (\tepsvtinf + \sqrt{r} \, \tepso). 
\ee 
Using \fr{eq:th5_for25}, construct an upper bound for  $\tilR_{02}$ in  \fr{eq:R01_R02-for} 
\be \label{eq:tilR02_upper}
\tilR_{02}  \leq \Ctau \,  \tepso \lkv \tepsvtinf + \sqrt{r} \, \tepso \,(\teps_1 + \teps_2) \rkv. 
\ee 
Removing the smaller order terms, for $m$ and $n$  large enough and $\om \in \tilOmtau$, arrive at
\be  \label{eq:tilR0_upper}
\tilR   \leq \Ctau\, \lkv \tepsvtinf + \sqrt{r} \, \tepso \,(\teps_1 + \teps_2)\rkv.
\ee
Combination of \fr{eq:base_ineq_f5}, \fr{eq:tilR_upper} and \fr{eq:tilR0_upper} yields \fr{eq:nonsym_up_bound_new}.

\medskip


\noindent
{\bf Proof of Corollary~\ref{cor:nonsym_subgaus}.\ }  
It follows from \cite{vershynin2018_book} that 
\begin{align*}
& \tepso \leq   \Ctau \, d_r^{-1}\, \sig\, (\sqrt{n} + \sqrt{m}),  \\ 
& \tepsuvo \leq   \Ctau \, d_r^{-1}\, \sig\, (\sqrt{r} + \sqrt{\log n}),\\
&  \tepsvtinf \leq   \Ctau \, d_r^{-1}\, \sig\, (\sqrt{r} + \sqrt{\log n}),\\
& \teps_1 \leq   \Ctau \, d_r^{-1}\, \sig\, \sqrt{r\, \log n}, \quad \teps_2 =0.
\end{align*}
 Plugging those quantities into \fr{eq:nonsym_up_bound_new}   and removing the smaller order terms, obtain
\fr{eq:probab_nonsym_subgaus}.

 



\subsection{Proofs of statements in Section~\ref{sec:symmetrized}}
\label{sec:symmetrized_proofs}

\noindent
{\bf Proof of Theorem~\ref{thm:symetrized_up_bound}.\  }  
Note that,  under conditions \fr{eq:new_cond_sym}, one has $\tDelEo \leq 1/2$, so that,  by Weyl's theorem, 
$\hlam_r   \geq 0.5\, d_r^2$ and 
\be \label{eq:hLam_inv_new}
\|\hLam^{-1} \|  \leq 2\, d_r^{-2}.
\ee
Denote 
\be  \label{eq:tDeluuo_tDelxtinf}
\begin{split}
& \tDeluuo = \min(\tDelEo, \sqrt{r}\, \tDelEuo), \\
& \tDelxtinf = d_r ^{-2}\, \|\Xi \, X^T\|_{2,\infty}.
\end{split}
\ee 
Here, due to \eqref{eq:main_nonsym}, one has
\be \label{eq:tDelxtinf_upper}
\tDelxtinf  \leq \tDelvtinf + d_{r+1}\, d_r^{-1}\, \tDeltinf.
\ee
By Davis-Kahan theorem,  obtain $\|\sinTeU\| \leq c_d^{-1} \, \tDelEo$ and also
\bes
\begin{split}
\|\sinTeU\| & \leq \|\sinTeU\|_F \leq \sqrt{r}\, c_d^{-1} d_r^{-2} \|\scrE\, U\|\\
& \leq \sqrt{r}\, c_d^{-1} \tDeluo.
\end{split}
\ees
Therefore, 
\be \label{eq:new_sinTe}
\|\sinTeU\| \leq \frac{\min(\tDelEo, \sqrt{r}  \tDelEuo)}{c_d} =   \frac{\tDeluuo}{c_d}.
\ee
Plugging \fr{eq:new_sym_E} into expansion \fr{eq:cape-expan},   derive that   \fr{eq:R1234} holds with 
$R_1, R_2, R_3$ and $R_4$ defined as before, but $\scrE$ replaced with $\tilscrE$. First, we derive 
new upper bounds for $R_1$ and $R_2$.

Note that 
\be \label{eq:R1new}
R_1 \!  = \!   \|( I - UU^T) \tilscrE U W_U \hat{\Lambda}^{-1} \|\tinf \leq R_{11} \!  + \! R_{12} \!  + \!  R_{13}.
\ee
Here,   
\bes
R_{11} =  \| U U^T \, (\tilscrE_1 + \tilscrE_2 + \tilscrE_d)\, U W_U \hat{\Lambda}^{-1}\|\tinf  \leq C\, \tDeluuo\, \epsu,
\ees
\begin{align*}
& R_{12} =   \|(\tilscrE_1 + \tilscrE_2 + \tilscrE_d)\, U W_U \hat{\Lambda}^{-1} \|\tinf \\
& \leq 
C  d_r^{-2} \lkv \|\barXixi  U\|\tinf \!  + \!  \|\Xi X^T U\|\tinf  \!  + \!  \|\tilscrE_d\|\tinf \|U\|\tinf \rkv \\
& \leq  C \lkv \tDelxiutinf  + \tDelvtinf + d_{r+1}\, d_r^{-1}\, \tDeltinf  \right. \\
& + \left. \tilh\   \epsu\,  (d_r^{-2}\, \|\diag(\Xi\, X^T)\|\tinf + \tepsy)\rkv,
\end{align*}
due to $\|\Xi X^T U\|\tinf \leq \tDelxtinf$ and \eqref{eq:tDelxtinf_upper}. Furthermore,
\bes
\begin{split}
R_{13} & = \| (I - U U^T)\,  X\, \Xi^T\, U W_U \hat{\Lambda}^{-1}\|\tinf \\
& \leq  C\, d_r^{-2}  \|U_\perp D_\perp V_\perp^T \Xi^T U \|\tinf \leq C   d_{r+1} d_r^{-1} \tDeluo,
\end{split}
\ees
where $\tDeluo$ is defined in \fr{eq:new_probab_errors}. 
Plugging the components into $R_1$ and noting that 
\bes
d_r^{-2} \|\diag(\Xi\, X^T)\|\tinf  \leq 
\tDelxtinf  \leq \tDelvtinf + d_{r+1}  d_r^{-1}  \tDeltinf,
\ees
derive
\be \label{eq:R1_new_fin}
\begin{split}
R_1  & \leq C\, \lkv  \tDelxiutinf  + \tDelvtinf   +  d_{r+1}\, d_r^{-1} (\tDeluo + \tDeltinf) \right. \\
& + \left.  \tDeluuo\, \epsu +  \tilh\, \epsu \, \tepsy) \rkv.
\end{split}
\ee
Now consider 
\be \label{eq:R2new}
\begin{split}
R_2   & =   \|( I - UU^T) \, \tilscrE  (\hU - U W_U)   \hat{\Lambda}^{-1} \|\tinf \\
& \leq R_{21} + R_{22} + R_{23}.
\end{split}
\ee
Denote $\tDelEtinf =     d_r ^{-2}\, \| \tilscrE_1 + \tilscrE_2 \|\tinf$, where $\tilscrE_1$ and $\tilscrE_2$
are defined in \fr{eq:scrE_components}, and observe that 
$$
\tDelEtinf \leq \tDelxitinf + \tDelxtinf.
$$ 
Due to \fr{eq:new_sinTe} and \fr{eq:ineq3},  one has  
\begin{align*}
R_{21} & =   \| U U^T \, (\tilscrE_1 + \tilscrE_2 + \tilscrE_d) \, (\hU - U W_U) \, \hat{\Lambda}^{-1} \|\tinf \\
 & \leq C\, \epsu\, \tDelEo \, \tDeluuo, \\
R_{22} & =   \|(\tilscrE_1 + \tilscrE_2 + \tilscrE_d) \, \, (\hU - U W_U) \, \hat{\Lambda}^{-1} \|\tinf \\
&  \leq C\, \lkv \tDelEtinf + \tilh\ \tepsy \rkv \,  \tDeluuo, \\
R_{23} & =   \| (I - U U^T)\,  X\, \Xi^T\, (\hU - U W_U) \,  \hat{\Lambda}^{-1}\|\tinf \\
& \leq C\,  d_{r+1}\, d_r^{-1} \, \tDelo \,  \tDeluuo.
\end{align*}
Therefore, combining the terms, using \eqref{eq:tDelxtinf_upper} and $\tDeltinf \leq \tDelo$, derive
\be \label{eq:R2_new_fin}
\begin{split}
R_2 & \leq C\,   \tDeluuo\, \lkv  \epsu\, \tDelEo  + \tDelxitinf + \tDelvtinf \right. \\
& + \left.  d_{r+1}\, d_r^{-1}  \tDelo   + \tilh\ \tepsy  \rkv.
\end{split}
\ee
Since the last two terms in \fr{eq:cape-expan} are the same as before, by \fr{eq:R3} 
and \fr{eq:R4},  obtain 
\bes
R_3 \leq C\, d_{r+1}^2\, d_r^{-2}\, \tDeluuo, \quad R_4 \leq C\, \epsu\, \tDeluuo^2.
\ees
Therefore, adding $R_1, R_2,  R_3$  and $R_4$, taking into account that, under assumption \fr{eq:new_cond_sym},
$\tDeluuo$ and $\tDelEo$ are  bounded above by 1/2, and removing smaller order terms, derive 
\begin{align*}
& |\hU - U W_U\|_{2, \infty}  \leq     C\, \lkv  \tDelxiutinf  + \tDelvtinf  \right. \\
& \ \ +    
  \tDeluuo (\epsu + \tDelxitinf) + \tilh\, \tepsy\,  (\tDeluuo + \epsu)\\
& \ \ +   \left.  
    + \frac{d_{r+1}}{d_r}\, (\tDeluo + \tDeltinf + \tDelo  \tDeluuo + \frac{d_{r+1}}{d_r}\,
  \tDeluuo)    \rkv.
\end{align*}

 \medskip 


\noindent
{\bf Proof of Theorem~\ref{thm:symetrized_probab_up_bound}.\  }   
Denote the sets, on which \fr{eq:new_probab_errors} and \fr{eq:assump_A4*} are true, by, respectively, $\tilOmtauo$ and $\tilOmtauo$. 
Denote $\tilOmtau = \tilOmtauo \cap \tilOmtauo$ and observe that $\PP(\tilOmtau) \geq 1 - 2\, n^{-\tau}$.
Use notations \eqref{eq:tDeluuo_tDelxtinf} and note that, by \eqref{eq:tDelxtinf_upper}, one has
$\tDelxtinf \leq \tDelvtinf$.
In order to prove the theorem, we start with expansion \fr{eq:expansion}.
Recall that $d_{r+1} =0$, so that $(I - U U^T) \, X=0$. Therefore,
$\tilscrE = \tilscrE_1 + \tilscrE_2 +   \tilscrE_d$, where components are defined in \fr{eq:new_sym_E}.
Then, with notations in \fr{eq:non-sym-err_more},  under the  conditions of Theorem~\ref{thm:symetrized_up_bound},
derive that $\|(\hU^T \,  U)^{-1}\| \leq C$ and $\|\hLam^{-1}\| \leq C \, d_r^{-2}$. Then, 
\begin{align*}
& \|\hU - U W_U\|\tinf   \leq   C \,  \lfi \epsu\, d_r^{-2}\, \|\tilscrE U\| + d_r^{-2}\, \|\tilscrE U\|\tinf \right.  \\
&\ \ \  + \left. \epsu \, d_r^{-2}\, \|\tilscrE U\| \, \tDeluuo^2 +  \epsu \tDeluuo^2  + d_r^{-2}\, \tilR \rfi 
\end{align*} 
where 
\be \label{eq:tilR}
\tilR = \|\tilscrE\, (\hU\, \hU^T \, U - U) \|\tinf \leq d_r^2\, \tDelEo\, \tDeluuo. 
\ee
Recalling that 
\bes
d_r^{-2}\, \|\tilscrE U\|\tinf \leq \tDelxiutinf + \tDelvtinf + (1-\tilh)\, \tDeltinf^2 + 
\tilh \tepsy  
\ees
and removing   smaller order terms, obtain
\be \label{eq:error_intermed_lem2}
\begin{split}
& \hspace{-2mm} \|\hU - U W_U\|\tinf   \leq  C    \lfi  \epsu\, \tDelEuo + \epsu \tDeluuo^2  + \tilh  \tepsy  \right. \\
 & \ \ +  \tDelxiutinf  +  \tDelvtinf  
  +  (1-\tilh)\, \tDeltinf^2 + \left.
   d_r^{-2}\, \tilR \rfi. 
\end{split}
\ee 
The rest of the proof relies of the following Lemma.

\medskip


\begin{lem}\label{lem:lemma_prob_Symmetized}  
Let conditions of Theorem \ref{thm:symetrized_probab_up_bound} hold. 
Then,  for $\om \in \tilOmtauo$, $\tilR$ defined in \fr{eq:tilR}
satisfies 
\be \label{eq:lemma21}
 d_r^{-2} \, \tilR   \leq \Ctau\, \lkr \tdel_2 + \tdel_{2,U}\, \|\hU - U\, W_U\|\tinf \rkr,
\ee
where $\tdel_{2,U} =  o(1)$  and 
\be \label{eq:tdel2}
\begin{split}
 \tdel_2    & \leq    \Ctau  \lfi  \tepsuuo   \tdelor + \tilh   (\tepstinf \epsu + \tepsy)  \right.   \\
& +  (1 - \tilh) \tepstinf^2 +   \lkr \tepsxiutinf + \tepsvtinf \right.  \\
 & \left. \left. + \epsu\, \tepsEo \rkr  \lkv \tdelo + \tepsEo    
  +  (1 - \tilh) \tepstinf^2 \rkv \rfi,  
\end{split}
\ee 
with 
\be \label{eq:delo_delor}
\begin{split}
\tdelo & = \teps_1 (\tepso +1) + \teps_2 (\tepstinf^T + \epsv), \\
\tdelor & = \sqrt{r}\,  \teps_1 (\tepso +1) + \teps_2 (\tepstinf^T + \epsv).
\end{split}
\ee
\end{lem}


\noindent
Plugging \fr{eq:lemma21}  into \fr{eq:error_intermed_lem2}, adjusting the coefficient for $\|\hU - U W_U\|\tinf$
in a view of $\tdel_{2,U} =  o(1)$, and using Assumption {\bf A3*}, obtain for $n$ large enough and $\om \in \tilOmtau$
\begin{align*} 
& \|\hU - U W_U\|\tinf  \leq    \lfi  \epsu\, \tepsEuo + \epsu (\tepsuuo)^2  +   \tepsxtinf \right.   \\
&\ \  +  \tepsxiutinf + (1-\tilh)\, \tepstinf^2 + \left.
\tilh (\tepsy +   \epsu \, \tepstinf ) + \tdel_2 \rfi 
\end{align*} 
Removing the smaller order terms, we arrive at \fr{eq:symmetrized_up_bound_new}.


\subsection{Proofs of statements in Section~\ref{sec:perfect_clust}} 
\label {sec:proofs_perfect_clust}


\noindent
{\bf Proof of Lemma~\ref{lem:perf_clust}.\  }   
The proof of Lemma~\ref{lem:perf_clust} relies on 
Lemma~D1 of \cite{abbe_fan_AOS2022}. For completeness, we present this lemma 
below, using our notations.

\medskip


\begin{lem}\label{lem:abbe_fan}  
{\bf (Lemma~D1 of \cite{abbe_fan_AOS2022})}. \  Let matrix $B \in \RR^{r \times m}$ with rows $B(k,:)$, $k \in [r]$, 
be the matrix of true means and $z:  [n] \to [r]$ be the true clustering function. 
For a data matrix $\scrX \in \RR^{n \times m}$, any matrix $\tilB \in \RR^{r \times m}$ and any clustering 
function $\tilz: [n] \to  [r]$, define 
\be \label{eq:L_function}
L \lkr \tilB, \tilz \rkr = \sum_{i=1}^n \Big\|\scrX(i,:) -  \tilB(\tilz(i),:) \Big\|^2.
\ee
Let $\hB \in \RR^{r \times m}$ and $\hz [n] \to  [r]$  be solutions to the 
$(1+a)-$approximate k-means problem, i.e.
\bes
L \lkr \hB, \hz \rkr \leq (1 + a)\,  \min_{\tilB, \tilz}\    L \lkr \tilB, \tilz \rkr.
\ees  
Let $s = \displaystyle \min_{i  \neq j} \|B(i,:) - B(j,:)\|$ and 
$\nmin$ be the minimum cluster size.  
 If for some $\del \in (0, s/2)$ one has
\be \label{eq:abbe_del_cond}
\begin{split}
L \lkr B, z \rkr & = \sum_{i=1}^n \Big\|\scrX(i,:) -  B(z(i),:) \Big\|^2 \\
& \leq \frac{\del^2 \, \nmin}{r (1 + \sqrt{1 + a} )^2},
\end{split}
\ee 
then there exists a permutation $\phi: [r] \to [r]$ such that
\begin{align}  \label{eq:cl_err1}
& \lfi i:  \hz(i) \neq \phi(z(i)) \rfi \subseteq \\
& \hspace{9mm}  \Big \{ i:  \|\scrX(i,:) - B(z(i),:)\| \geq s/2 - \del \Big\}, \nonumber \\
%
\label{eq:cl_err2}
& \# \lfi i:\,  \hz(i) \neq \phi(z(i)) \rfi  \leq (s/2 - \del)^{-2}\, L \lkr B, z \rkr. 
\end{align}
\end{lem}

\medskip

\noindent
Recalling \fr{eq:U_rel}, we apply Lemma~\ref{lem:abbe_fan} with $\scrX = \hU$ and $s=\sqrt{2} (n_{\max})^{-1/2}$.
However, since $\hU$ estimates matrix $U$ only up to a rotation, one needs to align matrices $\hU$ and $U$ 
using $W_U$, defined in \fr{eq:W_u}. Specifically,  let matrix $B \in \RR^{r \times m}$
in \fr{eq:abbe_del_cond}   be formed by distinct rows of $U\, W_U$.
Let $D_{sp} (U, \hU)$, $D_{F} (U, \hU)$ and $D_{2,\infty} (U, \hU)$ be defined in 
\fr{eq:dist} and \fr{eq:2inf_dist}, respectively. Then, by \fr{eq:ca-_zhang1}-\fr{eq:D_K_Spectral},
\be \label{eq:rhs}
\begin{split}
L \lkr B, z \rkr & \leq D_{F}^2 (U, \hU) \leq r\, D_{sp}^2 (U, \hU) \\
& \leq 2 r\, \|\sinTeU\|^2.
\end{split}
\ee 
Equating the right hand sides in \fr{eq:abbe_del_cond} and \fr{eq:rhs}, obtain from \fr{eq:clust_assump} and \fr{eq:U_rel},   that
\beqn \label{eq:del_val}
\del   & \leq &  \frac{2r \,   \lkr 1 + \sqrt{1 + a}\rkr^{1/2} \, \|\sinTeU\|  }{\sqrt{2\, \nmin}},\\
s/2 - \del  & \geq &     \frac{1 -  2 r\,    c_0  \,  \lkr 1 + \sqrt{1 + a}\rkr^{-1/2}\, 
   \|\sinTeU\|}{c_0\, \sqrt{2\, \nmin}}. \nonumber
\eeqn
Therefore,  if $r\,  \|\sinTeU\| \to 0$ as $n \to \infty$, then, for $n$ large enough, one has
$s/2 > \del$.
\ignore{
Note that the latter is true in many situations.  
Indeed, under the assumptions \fr{eq:clust_assump}, one obtains that 
$d_r  = \sig_r(X) \geq C_\sig^{-1}\, \sqrt{\nmin} \, \|\Te \|$,
and, therefore,     
\bes
\tDelo \leq \frac{C_\sig \, c_0 \sqrt{r}}{\sqrt{n}} \, \frac{\|\Xi\|}{\|\Te \|}.
\ees
Since $\tDelo$ dominates all random quantities  in \fr{eq:non-sym-err},  $\tepso = o(1)$ as $n \to \infty$ implies that
all  random quantities in \fr{eq:nonsym_up_bound} tend to zero with high probability. Then, if $\hU = \SVD_r(\hX)$, 
one obtains, by Wedin theorem,  that $\sqrt{r}\, \|\sinTeU\| \to 0$ as $n \to \infty$, 
provided $\sqrt{r}\, \tepso = o(1)$. A similar situation occurs when one uses 
$\hU = \SVD_r(\hY)$. If $\sqrt{r}\, \tepsEo = o(1)$ and $\tilh\, \tepsy = o(1)$ as  $n \to \infty$, then, 
when $d_{r+1} =0$, all random quantities in \fr{eq:symetrized_up_bound}  tend to zero 
with high probability, and, moreover, by Davis-Kahan theorem, $\sqrt{r}\, \|\sinTeU\| \to 0$ as $n \to \infty$.
} 
%

%
Under this condition,   due to Lemma~\ref{lem:abbe_fan}, \fr{eq:clust_assump}, \fr{eq:U_rel} and \fr{eq:del_val}, 
 node $i \in [n]$ is certain to be clustered correctly for $n$ large enough, if 
$\|  \hU  (i,:) - (U \, W_U) (i,:)\| \leq  (2\,  c_0\, \sqrt{2\, \nmin})^{-1}$.
Due to $\epsu = (\nmin)^{-1/2}$,  perfect clustering is, therefore,    assured by  
\be \label{eq:perfect_cl}
 \hspace*{-2mm} \|  \hU - U W_U \|\tinf 
 \leq  (2  c_0 \sqrt{2\, \nmin})^{-1} = (2  \sqrt{2} c_0)^{-1}  \epsu.     
\ee 
Since $c_0$ is unknown, the latter  is guaranteed by $ \|  \hU - U \, W_U \|\tinf  = o(\epsu)$
when $n \to \infty$. 

\medskip  \medskip


\noindent
{\bf Proof of Proposition~\ref{prop:perfect_clust}. }\\  
Validity of the first statement \fr{eq:perf_cl1} in Proposition~\ref{prop:perfect_clust} follows directly 
from \fr{eq:probab_nonsym_up_bound} in Theorem~\ref{thm:nonsym_up_bound}. 
Since $d_{r+1} =0$ and, with probability at least $1 - n^{-\tau}$,
one has 
\bes
\epsu^{-1}   \|\hU - U W_U\|\tinf \leq  \Ctau \lkv  \tepsuvo +   \tepso^2  +
\epsu^{-1}  (\tepsvtinf + \tepso   \tepstinf)  \rkv,  
\ees 
where $\tepsuvo \leq \tepso$. Hence, condition \fr{eq:perf_cl1} implies that \fr{eq:perfect_cl} is valid 
and clustering is perfect when $n$ is large enough.  Validity of \fr{eq:perf_cl1_1} follows directly from \fr{eq:nonsym_up_bound_new}
in Theorem~\ref{thm:symetrized_up_bound}.

In order to prove \fr{eq:perf_cl2}, note that it follows from \fr{eq:probab_symetrized_up_bound} that
\begin{align*}
& \epsu^{-1}\,  \|\hU - U W_U\|\tinf  \leq  \Ctau \, \lfi \min(\tepsEo, \sqrt{r}\, \tepsEuo) + \tilh\, \tepsy \right.\\
& \ \  + \epsu^{-1}\, \lkr  \tepsxiutinf +    \tepsvtinf \right. \\
& \ \  + \min(\tepsEo, \sqrt{r}\, \tepsEuo)\, \tepsxitinf   + \left.  \left.
\tilh\ \tepsy\, \tepsEo \rkr \rfi
\end{align*}  
and use the same argument as in the previous case.

Validity of \fr{eq:perf_cl3} follows from \fr{eq:symmetrized_up_bound_new} and \fr{eq:tdel1} of Theorem~\ref{thm:symetrized_probab_up_bound}.

\medskip  \medskip


\noindent
{\bf Proof of Proposition~\ref{prop:perfect_clust_sym}.\  }   
Observe  that, if the second inequality in \fr{eq:clust_assump} holds, relations \fr{eq:U_rel} are valid.
Thus, similarly to the non-symmetric case, perfect clustering is assured by condition \fr{eq:perfect_cl},
which, in turn,  is guaranteed by $ \|  \hU - U \, W_U \|\tinf  = o(\epsu)$
when $n \to \infty$.  Hence, validity of Proposition~\ref{prop:perfect_clust_sym} follows directly from Theorems~\ref{th:sym_up_bound} 
and \ref{th:sym_up_bound_new}. 
 
\medskip \medskip


\noindent
{\bf Proof of Proposition~\ref{prop:didactic}.\  } 
First, consider the case when  one  obtains $\hU = \SVD_r(\hX)$ in Algorithm~\ref{alg:spec_clust}.
Then, for consistency of clustering, one needs $\tepso = o(1)$, hence \fr{eq:tepso_tepstinf} implies that 
the necessary condition for consistent clustering is 
\be \label{eq:necessary_hX}
\frac{\sig\, \sqrt{r}}{\te}\, \lkr \frac{1}{\sqrt{m}} + \frac{1}{\sqrt{n}} \rkr = o(1) \quad \mbox{as}\quad n \to \infty.
\ee
The perfect clustering is guaranteed by conditions in  \fr{eq:perf_cl1}, which,
due to \fr{eq:d_r_epsu_gauss} and \fr{eq:tepso_tepstinf},  are satisfied provided, as $n \to \infty$,
\be \label{eq:gauss_cond_set1}
\begin{split}
& \frac{\sig\, \sqrt{r}}{\te}\,  \frac{(\sqrt{\log n} + \sqrt{r})}{\sqrt{m}}  + 
\frac{\sig\, r}{\te\, \sqrt{n}}\\
& + \frac{\sig^2\, \sqrt{r\, \log n}}{\te^2}\,
\lkr \frac{1}{\sqrt{m}} + \frac{1}{\sqrt{n}} \rkr = o(1).
\end{split}
\ee
Since $r/m = o(1)$, the last condition can be rewritten as condition (S1) in \fr{eq:nes_suf_no_sym}.

Now, consider the case when one applies  symmetrization with hollowing,   
i.e., $\hY = \scrH(\hX\, \hX^T)$.  Then, the necessary condition for consistent clustering becomes $\tepsEo = o(1)$,
which, due to \fr{eq:m_n_te} and \fr{eq:gauss_symmetrized_bounds},  appears as \fr{eq:nes_sym}.
In order to derive sufficient conditions,  we start with the situation when one does not use Assumption~A4* and utilizes only 
conditions \fr{eq:new_probab_errors} in Assumption~A3*.  Then,   Lemma~\ref{lem:err_bounds_gauss} yields
\be  \label{eq:tepsxiinf_gauss_cond}
\begin{split}
 & \epsu^{-1} \, \tepsxiutinf\   \leq  \Ctau \,  \frac{\sig^2\, r}{\te^2}\, \frac{\log n}{\sqrt{m\, n}}, \\
& \epsu^{-1} \, \tepsxtinf \leq \Ctau \, \frac{\sig\, \sqrt{r}}{\te}\, \frac{\log n}{\sqrt{m}},  \\
 & \epsu^{-1} \, \tepsEo\, (\tepsxitinf + \tepsxtinf) \leq   \Ctau \, \lkv \frac{\sig^2\, r}{\te^2}\, \frac{\log n\, \sqrt{r}}{n\, \sqrt{m}} \right.\\
& \left. \ \ \ \ \  \ \  +  \lkr \frac{\sig^2\, r}{\te^2} \rkr^2\, \frac{\log^2 n}{m\, \sqrt{m\, r}} \rkv.  
 \end{split} 
\ee
By checking conditions $\sqrt{r}\,  \tepsEo = o(1)$,   \ $\tepsy = o(1)$, and \fr{eq:perf_cl2} of Proposition~\ref{prop:perfect_clust},
it is easy to see that clustering is perfect,  with probability at least $1 - n^{-\tau}$  for $n$ large enough,  provided, 
as $n \to \infty$,
\begin{align} \label{eq:gauss_cond_set21}
& \frac{\sig^2\, r\, \log n}{\te^2\, \sqrt{m n}} \lkv 1 + \frac{r\, \sqrt{n}}{\sqrt{m}}
+ \frac{\log n\, \sqrt{n}}{\sqrt{m}} \rkv  = o(1),\\ 
 \label{eq:gauss_cond_set22}
&  \frac{\sig^2\, r\, \log n}{\te^2\, \sqrt{m n}} \, \frac{\sqrt{n}}{(m\, r)^{1/4}}  = o(1).
\end{align} 
It is easy to see that combination of \fr{eq:gauss_cond_set21} and \fr{eq:gauss_cond_set22}
is equivalent to combination of conditions in \fr{eq:suf_sym}.

Finally, we consider the situation when  Assumption~A4* holds. In this case, by \fr{eq:perf_cl3}, 
sufficient conditions for perfect clustering are 
\be   \label{eq:gauss_cond_set31}
 \frac{\sig  r \sqrt{\log n}}{\te  \sqrt{m}} \lkv  \frac{\sig  \sqrt{r} }{\te  \min(m,n)} \! +  \!  1 \rkv 
\lkv \frac{\sig^2   r   \log n}{\te^2  m}  \!  +  \!  \frac{r}{n} \rkv  = o(1), 
\ee
\be  \label{eq:gauss_cond_set32}
\begin{split}
& \frac{\sig^2\, r^2\, \log n}{\te^2\, m} = o(1),\quad 
\frac{\sig^2\, r \, \log n}{\te^2\, \sqrt{m n}} = o(1), \\
& \frac{\sig \, \sqrt{r}\, \log n}{\te \, \sqrt{m}} = o(1).    
\end{split}
\ee
Denote 
\be \label{eq:delmnm}
\del^2_{m,n} = \frac{\sig^2\, r\, \log n}{\te^2\, \sqrt{m\, n} }, \quad 
\del^2_m = \frac{\sig^2\, r\, \log n}{\te^2\, m} = \del^2_{m,n}\, \frac{\sqrt{n}}{\sqrt{m}}.
\ee
Then, the three conditions  in \fr{eq:gauss_cond_set32} are guaranteed by \fr{eq:gauss_cond_set21},
which is equivalent to the first condition in \fr{eq:suf_sym}. Now, consider condition \fr{eq:gauss_cond_set31}.
Rewrite it as 
\begin{align}   \label{eq:gauss_cond_set33}
& \del_m^4 \frac{\sqrt{r}}{\sqrt{n}} \lkr 1 + \frac{\sqrt{m}}{\sqrt{n}} \rkr + \del_m^3 \sqrt{r} + \\
& \del^2_m  \frac{\sqrt{r}}{\sqrt{n}} \lkr  1 + \frac{\sqrt{m}}{\sqrt{n}}\rkr \frac{r}{n} + \del_m \frac{r \sqrt{r}}{n} = o(1),
\nonumber
\end{align} 
and observe that \fr{eq:gauss_cond_set21} implies that, as $n, m \to \infty$,
\be \label{eq:gauss_cond_set34}
\del^2_m (r + \log n) = o(1), \quad \del^2_{m,n} = \del^2_m \sqrt{m}/\sqrt{n} = o(1).
\ee
In order to complete the proof, observe that \fr{eq:gauss_cond_set33}  is
guaranteed by \fr{eq:gauss_cond_set34}.

 \medskip 

\ignore{
Now, as an alternative, compare \fr{eq:gauss_cond_set2} with  conditions \fr{eq:teps_conditions} and \fr{eq:perf_cl3} for perfect clustering. 
Since  all calculations above are still valid and $\tilh =1$, we only need to check that, additionally to 
$\epsu^{-1} \, \tepsxiutinf = o(1)$ and $\epsu^{-1} \, \tepsxutinf = o(1)$, one has 
$\sqrt{r} \lkr \tepso\, \teps_1 + \teps_1 \rkr   = o(1)$ and 
$\epsu^{-1}\, \sqrt{r}\, \tepsuuo  \,   \teps_1 (\tepso +1)  = o(1).$

Now we study the cases when $n \ll m$ and $n \geq m$ separately.
Let $n \ll m$. Then, direct  calculations yield  that 
sufficient conditions for perfect clustering are
\begin{align} 
\label{eq:cond_part0}
&   \frac{\sig^2\, r}{\te^2} \lkv \frac{\log n}{\sqrt{m\, n}} +  \frac{\log  n (r + \log n)}{m}  \rkv = o(1),\\
\label{eq:cond_part1}
& \sqrt{r} \lkr \tepso\, \teps_1 + \teps_1 \rkr \leq \Ctau \,   \frac{\sig\, r\, \sqrt{\log n}} {\te\, \sqrt{m\, n}}\, 
\lkv  \frac{\sig\, \sqrt{r}} {\te}\,  \lkr  \frac{1}{\sqrt{m}} +  \frac{1}{\sqrt{n}} \rkr + 1 \rkv = o(1),\\
  & \epsu^{-1}\, \sqrt{r}\, \tepsuuo  \,   \teps_1 (\tepso +1) \leq \Ctau \, \lkv 
  \del^4_{m,n}\,  \frac{\sqrt{r\,n\, \log n}}{\sqrt{m\, \log n}} + 
   \del^{3}_{m,n} \, \frac{r^{1/2} \, n^{3/4}}{m^{3/4}}   \right.  \nonumber \\
 & \left. \hspace{40mm} +\ \del^2_{m,n}  \frac{r \, \sqrt{r}}{n\,  \sqrt{\log n}} + 
  \del_{m,n}\,  \frac{r \, \sqrt{r}} {n^{3/4}\, m^{1/4} } 
  \rkv = o(1),   \label{eq:cond_part2}  
\end{align}
where $\del^2_{m,n} = (\te^2\, \sqrt{m\, n})^{-1} \, \sig^2\, r\, \log n$.
It is easy to verify that conditions \fr{eq:teps_conditions} and \fr{eq:perf_cl3}  hold, 
provided \fr{eq:cond_part0}--\fr{eq:cond_part2}   are valid. Sufficient conditions for the latter are
\be \label{eq:gauss_cond_set3}
 \frac{\sig^2\, r}{\te^2} \, \frac{(r + \log n)}{\sqrt{m\, n}} = o(1), \quad
\frac{n}{m} \lkr \log^2 n + \frac{r}{\log n} \rkr = O(1), 
  \quad \mbox{as}\quad n \to \infty. 
\ee
Here, we would like to draw attention to the fact that 
sufficient conditions \fr{eq:gauss_cond_set2} and \fr{eq:gauss_cond_set3} apply to the same algorithm,
so that clustering is perfect if either  \fr{eq:gauss_cond_set2}, or \fr{eq:gauss_cond_set3}  holds.

Now, we study the case where $n \geq m$. If $\hU = \SVD_r (\hX)$, then conditions in  \fr{eq:perf_cl1} are satisfied provided
\fr{eq:gauss_cond_set1} holds, which, for  $n \geq m$ can be re-written as 
\be \label{eq:gauss_cond_set1_2}
 \frac{\sig \, \sqrt{r}}{\te \, \sqrt{m}}  \lkr \frac{\sig\, \sqrt{\log n}}{\te} + \sqrt{r} + \sqrt{\log n} \rkr
 = o(1)   \quad \mbox{as}\quad n \to \infty. 
\ee 
If  $\hU = \SVD_r (\hY)$, then $\sqrt{r}\, \tepsEo = o(1)$ is guaranteed by 
\be \label{eq:gauss_cond_set2_2}
\frac{\sig^2\, r^2 \, \log n}{\te^2\, m}  = o(1)   \quad \mbox{as}\quad n \to \infty. 
\ee 
Condition \fr{eq:gauss_cond_set2} can be re-written as 
\be \label{eq:gauss_cond_set3_2}
 \frac{\sig^2\, r \log n}{\te^2}\, \frac{1}{m^{3/4}\, r^{1/4}} 
\lkr 1 +    \frac{r^{1/4}\, \log n}{m^{1/4}} \rkr
 = o(1) \quad \mbox{as}\quad n \to \infty.
\ee'/
Note that assumption \fr{eq:gauss_cond_set3_2} is stronger than \fr{eq:gauss_cond_set2_2}.
Now, we make use of Assumption~A4* and check conditions in \fr{eq:teps_conditions} and \fr{eq:perf_cl3}. 
Straightforward calculations yield that, in a view of \fr{eq:m_n_te}, perfect clustering is ensured by 
\be \label{eq:gauss_cond_set4_2}
\frac{\sig \, \sqrt{r\, \log n}}{\te \, \sqrt{m}} \lkv \frac{\sig^2\, r^2 \, \log n}{\te^2\, m} + \frac{r}{n} \rkv
 = o(1) \quad \mbox{as}\quad n \to \infty,
\ee
which, in turn, is guaranteed by condition \fr{eq:gauss_cond_set2_2}.
} 


\medskip \medskip


\noindent
{\bf Proof of Proposition~\ref{prop:SBM_clust}.\  }  
First, we explore the structure of matrix $X$. Denote 
$D_{\calS} =  \ZS \ZS^T = \diag(m_1, ..., m_r)$,  
$D_{\Sc} =  \ZSc \ZSc^T = \diag(N_1, ..., N_r)$,  
$ \US = \ZS\, (D_{\calS})^{-1/2}$ and 
$\USc = \ZSc\, (D_{\Sc})^{-1/2}$.
If $(D_{\calS})^{1/2}\, Q \, (D_{\Sc})^{1/2} = U_Q D_Q V_Q^T$ is the SVD of $(D_{\calS})^{1/2}\, Q \, (D_{\Sc})^{1/2}$, 
where $U_Q, V_Q \in \calO_r$, then the SVD of $X$ is given by 
\begin{align*}    
& X = U D V^T, \quad U = \US U_Q \in \calO_{m,r}, \\
& V  = \USc \, V_Q \in \calO_{n-m,r}, \ \ \ D = D_Q.
\end{align*}
Recall that we are in the environment of Section~\ref{sec:symmetrized}, where  $\tilh=1$ and
$n$ is replaced by  $m$ and $m$ by $n-m$, respectively. Thus,  $X, \hX \in \RR^{m \times (n-m)}$,
and $\hY = \scrH(\hX \hX^T)$.  Note that, \fr{eq:SBM_clust_assump}, $m \to \infty$, $n \to \infty$ and $m=o(n)$ 
guarantee that
\bes
\begin{split}
& \min_k m_k \asymp \max_k m_k \asymp m/r,\\
& \min_k N_k \asymp \max_k N_k \asymp (n-m)/r  \asymp n/r.
\end{split}
\ees
Therefore, one has
\be \label{eq:SBM_ineq1}
\begin{split}
& \epsu \asymp \sqrt{r}/\sqrt{m} = o(1), \quad \epsv \asymp \sqrt{r}/\sqrt{n}, \\
& d_r^2 \asymp r^{-1}  m  n  \rhon^2,
\  \tepsy = d_r^{-2}   n  \rhon^2 \leq C   r/m = o(\epsu).
\end{split}
\ee  
Note that rows of matrix $\Xi = \hX-X$ are independent, hence one can apply \fr{eq:perf_cl3} of 
Proposition~\ref{prop:perfect_clust}. To this end, it is necessary to check that, as $n \to \infty$,
\begin{align} \label{eq:SBM_clust_verify1} 
& \sqrt{r}\, \tepsEo = o(1), \quad 
\epsu^{-1}\, \lkr \tepsxiutinf +  \tepsvtinf  \rkr = o(1), \\
\label{eq:SBM_clust_verify2} 
& \sqrt{r}\, \teps_1 (\tepso +1) = o(1), \quad \teps_2 (\tepstinf^T + \epsv) = o(1),  \\ 
\label{eq:SBM_clust_verify3} 
&  \epsu^{-1}\, \tepsuuo   \lkv \sqrt{r}\, \teps_1 (\tepso +1) + \teps_2 (\tepstinf^T + \epsv)\rkv = o(1).
\end{align} 
where, by \eqref{eq:tDeluuo_tDelxtinf}, $\tDeluuo = \min(\tDelEo, \sqrt{r}\, \tDelEuo)$.

We start with bounding above $\|\tilscrE\|$. Due to \fr{eq:scrE_components}, 
$\|\tilscrE_2\| = \|\tilscrE_3\|$ and $\|\tilscrE_d\| \leq \|\diag(Y)||_{\infty} + \|\tilscrE_2\|$,
it is sufficient to derive upper bounds for  $\|\tilscrE_1\|$ and $\|\tilscrE_2\|$.
By Theorem~3 of \cite{lei2021biasadjusted}, due to $n -m \asymp n$, one has
\be  \label{eq:SBM_ineq2}
\PP \lfi \|\tilscrE_2\| \leq \Ctau \, m \rhon \sqrt{n\, \rhon \log n} \rfi \geq 1 - n^{-\tau}.
\ee 
For $\|\tilscrE_1\|$,  with probability at least $1 - n^{-\tau}$,   Theorem~4  of \cite{lei2021biasadjusted} yields
\be  \label{eq:SBM_ineq3}
\left\|\scrH(\Xi  \Xi ^T) \right\| \leq \Ctau \, \log n\,   \sqrt{m\, n\, \rhon}.
\ee  
Then, \fr{eq:SBM_ineq1}, \fr{eq:SBM_ineq2} and  \fr{eq:SBM_ineq3}  imply that, with probability at least $1 - n^{-\tau}$,  
\be \label{eq:SBM_scrE0}
\hspace*{-2.5mm} \sqrt{r}  \tDelEo \! \leq  \! \sqrt{r}\, \tepsEo = \Ctau \lkr \frac{r   \sqrt{r}   \sqrt{\log n}} {\sqrt{n  \rhon}} 
\! + \!  \frac{r  \sqrt{r}  \log n}{\rhon  \sqrt{m  n}}\rkr.
\ee
Since the first condition in \fr{eq:new_SBM_cond} together with $r^6 \rhon / \log n = o(1)$ guarantees that 
the first term in \fr{eq:SBM_scrE0} tends to zero, 
the first relation in \fr{eq:SBM_clust_verify1} is valid.

Now, we construct an upper bound  for $\tDelxiutinf = d_r^{-2}\, \|\scrH(\Xi \,\Xi^T)\, U \|\tinf$.
For this purpose, for any $\linm$ we construct matrices
$\Xi\upl$ with elements 
\be \label{eq:SBM_Xi_upl}
\Xi\upl(i,j) = \lfi
\begin{array}{ll}
\Xi (i,j), & i \neq l, \\
0, & i=l.
\end{array}\right.
\ee  
Obtain that 
\bes 
\|\scrH(\Xi \Xi^T)\, U\|\tinf = \max_{\linm}  \| \Xi (l,:)\, (\Xi\upl)^T \, U\| 
\ees
Apply Theorem~4  of \cite{lei2021biasadjusted} and observe that, conditioned on $\Xi\upl$, 
with probability at least $1 - n^{-\tau}$,  one has 
\be \label{eq:SBM_ineq11}
\begin{split}
  \max_{\linm}  \| \Xi(l,:)\, (\Xi\upl)^T\,  U\|  &  \leq   \Ctau \lkv \sqrt{\rhon  \log n}
\|\Xi^T U\|_F \right. \\
& \left. + \log n\, \|\Xi^T U\|\tinf \rkv.
\end{split}
\ee 
Here, by Theorem~3 of \cite{lei2021biasadjusted},  with high probability, 
\bes
\begin{split}
& \|\Xi^T U\|_F \leq \sqrt{r}\,  \|\Xi\|  \leq \Ctau\, \sqrt{r\, n \, \rhon\, \log n}, \\
& \|\Xi^T U\|\tinf \leq \Ctau\, \lkr \sqrt{r\, \rhon \, \log n} + m^{-1/2}\, \sqrt{r} \, \log n \rkr.
\end{split}
\ees
Plugging the latter into \fr{eq:SBM_ineq11}, applying the union bound over $\linm$ and adjusting constants, 
obtain that, with probability at least $1 - n^{-\tau}$,  one has
\be \label{eq:SBM_tDelXixi}
\begin{split}
& \max_{\linm} \,  \| \Xi(l,:)\, (\Xi\upl)^T\,  U\|    \leq   \Ctau\,  
\lkr \sqrt{r\, n} \, \rhon\, \log n \right. \\
& + \left.  \log n\, \sqrt{r\, \rhon\, \log n} 
 +  m^{-1/2}\, \sqrt{r} \, \log^2 n \rkr.
 \end{split}
\ee 
Removing the  smaller order terms, derive that $\|\scrH(\Xi \Xi^T)\, U\|\tinf \leq \Ctau \sqrt{r\, n} \, \rhon\, \log n$,
so that, with probability at least $1 - n^{-\tau}$ 
\be \label{eq:SBM_tepsxutinf1}
\tDelxiutinf \leq \tepsxiutinf  = \Ctau\, \frac{\sqrt{r}}{\sqrt{m}}\, \frac{\sqrt{r\, \log n}}{\rhon\, \sqrt{m\, n}} = o(\epsu).
\ee
Now consider $\tDelvtinf = d_r^{-1}\, \displaystyle \max_{\linm} \|\Xi(l,:)\, V\|$. 
Applying Theorem~3 of \cite{lei2021biasadjusted} and the union bound over $\linm$, due to $\|V\|_F^2 = r$, 
$\|V\|\tinf = \epsv$ and \eqref{eq:SBM_ineq1}, obtain  that with probability at least $1 - n^{-\tau}$, one has
\bes
\tDelvtinf \leq \Ctau d_r^{-1}\, \lkr \sqrt{\rhon}\, \sqrt{r\, \log n} + 
 \log n \, \sqrt{r}/\sqrt{n} \rkr.
\ees
Plugging in $d_r$ from \eqref{eq:SBM_ineq1} and removing smaller order terms, derive that 
\be \label{eq:SBM_tepsvtinf2}
\tDelvtinf  \leq \tepsvtinf  = \Ctau\, \frac{\sqrt{r}}{\sqrt{m}}\, \frac{\sqrt{r\, \log n}}{\sqrt{n\, \rhon}} = o(\epsu).
\ee
Therefore, all conditions in  \fr{eq:SBM_clust_verify1}  hold.

In order to check conditions \fr{eq:SBM_clust_verify2}  and \fr{eq:SBM_clust_verify3}, we need 
to obtain the values of $\teps_1$ and $\teps_2$ in \fr{eq:assump_A4*}.  Theorem~3 of \cite{lei2021biasadjusted}
yields that, for any matrix $G \in \RR^{m \times m_0}$, $m_0 \leq m$,  
 with probability at least $1 - n^{-\tau}$,  one has
\bes
\|\Xi\, G\|\tinf \leq \Ctau\, \lkr \sqrt{\rhon\, \log n}\, \|G\|_F + \log n\, \|G\|\tinf \rkr.
\ees
The latter implies that
\be \label{eq:SBM_teps12}
\teps_1 = \Ctau  \frac{\sqrt{r  \log n}}{\sqrt{m  n \rhon}}= o(1), \ 
\teps_2 = \Ctau  \frac{\log n  \sqrt{r}}{\rhon  \sqrt{m  n}} = o(1).
\ee
Now, it is easy to check that, by \cite{lei2015}, $\|\Xi\|\leq \Ctau \sqrt{n\, \rhon}$  with high probability,
so that
\be \label{eq:SBM_tepso}
\tepso \leq \Ctau\, \frac{\sqrt{r}}{\sqrt{m\, \rhon}}.
\ee
Also, $\tDeltinf^T = \displaystyle \max_{l \in [n-m]}\, \|\Xi(:,l)\| \leq \Ctau\, \sqrt{\rhon\, m\, \log n}$ and, therefore,
\be \label{eq:tepstinfT}
\tepstinf^T \leq \Ctau\, \frac{\sqrt{r\, \log n}}{\sqrt{n\, \rhon}}.
\ee
Using \fr{eq:SBM_teps12}, \fr{eq:SBM_tepso}, \fr{eq:tepstinfT} and \fr{eq:SBM_ineq1},
we can  verify validity of  conditions \fr{eq:SBM_clust_verify2}. Obtain
\be \label{eq:SBM_check2}
\begin{split}
& \hspace{-2mm}  \sqrt{r}  \teps_1 (\tepso +1)  \leq \Ctau  \lkr \frac{\sqrt{r  \log n}}{\sqrt{m  n  \rhon}}
\! + \! \frac{r  \sqrt{r  \log n}}{\rhon   m  \sqrt{n}}  \rkr =  o(1), \\ 
& \teps_2 (\tepstinf^T + \epsv)  \leq   \frac{\Ctau r  \log  n}{\rhon  \sqrt{m  n}}  
\frac{\sqrt{r  \log n}}{\sqrt{n  \rhon}} = o(1).
\end{split}
\ee 
Finally,  inequalities \fr{eq:SBM_check2} allows easy checking of conditions in  \fr{eq:SBM_clust_verify3}.
In particular, \fr{eq:SBM_scrE0} and \fr{eq:SBM_check2} yield
\bes
\begin{split}
& \epsu^{-1}  \tepsuuo   \sqrt{r} \teps_1 (\tepso +1) \leq \Ctau  
\lkr \frac{r \, \sqrt{\log n}} {\sqrt{n\, \rhon}} \right. \\
& \left. \ \ \ \ \   + \frac{r\, \log n}{\rhon\, \sqrt{m\, n}}\rkr  
\lkr \frac{ \sqrt{\log n}} {\sqrt{n\, \rhon}} + \frac{r\, \sqrt{\log n}}{\rhon\, \sqrt{m\, n}}\rkr = o(1).
\end{split}
\ees
Also, using \fr{eq:new_SBM_cond}, derive
\bes 
\begin{split}
 \epsu^{-1} \tepsEo     \teps_2 (\tepstinf^T + \epsv) & \leq \Ctau\, 
\lkr \frac{(\log n)^{5/2}\, r^{3/2}}{\rhon^{5/2}\, n^{3/2}\, m^{1/2}} \right. \\
&    \left.  + 
\frac{r^{3/2}\, (\log n)^2}{n^{3/2}\, \rhon^2} \rkr = o(1),
\end{split}
\ees 
which completes the proof.

\medskip \medskip


\noindent
{\bf Proof of Proposition~\ref{prop:error_between_new}.\   }   
Note that, due to \fr{eq:PW_clust_assump}, one has   $\epsu \asymp \sqrt{M}/\sqrt{L}$.
We apply the first part of  Proposition~\ref{prop:perfect_clust} with $r=M$, 
and, therefore, need to show that \fr{eq:perf_cl1} is true. For this purpose, we
 need to upper-bound $\tDelo$, $\tDeltinf$ and $\tDelvtinf$ with high probability.

Similarly to \cite{pensky2021clustering}, we derive  
\bes
\begin{split}
& \|\Xi\|\tinf =   \max_{\linL}\, \left\|\vect(\hU\upl (\hU\upl)^T) - \vect(U\upl (U\upl)^T\right\| \\
&  \hspace{9mm}  \leq 
2 \, \max_{\linL}\, \left\|\sinTe(\hU\upl, U\upl)\right\|_F.
\end{split}
\ees
It  follows from \fr{eq:PW_clust_assump} that 
\be \label{eq:d_M}
d_M = \sig_M (X) \geq C\,  \sqrt{K\, L}/\sqrt{M}.
\ee
Also, it follows from \fr{eq:sparsity} and \fr{eq:minbQl} that, by Davis-Kahan theorem, 
for each $\linL$, with probability at least $1 - n^{- \tau}$,
one has 
\bes
\|\sinTe(\hU\upl, U\upl)\|_F \leq C_\tau \, \frac{K}{\sqrt{n\, \rhon}}  = o(1).
\ees
Therefore, applying  the union bound, obtain that, with probability at least $1 - L\, n^{- \tau}$,
one has simultaneously  
\be \label{eq:Xi_bounds_PW} 
\begin{split}
& \|\Xi\|\tinf \leq \frac{C_\tau \, K\, \log L}{\sqrt{n\, \rhon}}, \\ 
&  \|\Xi\|_F  \leq  \sqrt{L}\, \|\Xi\|\tinf \leq \frac{C_\tau \, K\,\sqrt{L}\,  \log L}{\sqrt{n\, \rhon}}.
\end{split}
\ee
Therefore, the  Wedin theorem, \fr{eq:error_between_PW} and  \fr{eq:d_M} imply that, with probability at least $1 - L\, n^{- \tau}$,
one has 
\be \label{eq:tDel_bounds}
\begin{split}
& \sqrt{M}\, \tDelo \leq \sqrt{M}\,\tepso = \frac{C_\tau \, \sqrt{K}\, M\, \log L}{\sqrt{n\, \rhon}} = o(1),\\
& \tDelvtinf \leq \tDeltinf \leq \tepstinf =  \frac{C_\tau \, \sqrt{K\, M}\, \log L}{\sqrt{n\, L\, \rhon}} = o(1).
\end{split}
\ee
Hence, under the assumption \fr{eq:new_between_cond}, conditions in \fr{eq:perf_cl1} hold, 
and clustering is perfect when $n$ and $L$ large enough.
\\


\subsection{Supplementary inequalities}
\label{sec:Suppl}

\begin{lem}\label{lem:suppl_ineq}  
Let $U, \hU \in \calO_{n,r}$ and $W_U$ be defined in \fr{eq:W_u}. 
Then, the following inequalities hold
\begin{eqnarray}
& \|U^T \hU - W_U \| & \leq \|\sinTeU\|^2, \label{eq:ineq1}\\
& \|\hU - U\, U^T \hU\| &  =  \|\sinTeU\|, \label{eq:ineq2}\\
& \| \hU - U\, W_U \| & \leq \sqrt{2}\, \|\sinTeU\| , \label{eq:ineq3}\\
& \|I - \hU^T U U^T \hU \| & =   \|\sinTeU\|^2.  \label{eq:ineq4}
\end{eqnarray}
\end{lem}

\medskip  

\noindent 
{\bf Proof.} 
Inequalities  \fr{eq:ineq1} and \fr{eq:ineq2} are proved in Lemma~6.7 of \cite{Cape_L2_inf_AOS2019}.
Inequality \fr{eq:ineq3} is established in Lemma~6.8 of \cite{Cape_L2_inf_AOS2019}.
Finally, in order to prove \fr{eq:ineq4}, note that $ U^T \hU = W_1 D_U W_2^T$ where $D_U = \cos(\Te)$
and $\Te$ is the diagonal matrix of the principal angles between the subspaces.
Hence, 
\bes
\begin{split}
& \|I - \hU^T U U^T \hU \|  = \|W_1 \lkv  I - \cos^2(\Te) \rkv W_1^T\| \\
& = \|W_1 \, \sin^2(\Te)\, W_1^T\|,
\end{split}
\ees
which completes the proof.


\bibliographystyle{IEEEtran}
\bibliography{DavisKahan_no_doi.bib}

\begin{thebibliography}{10}
\providecommand{\url}[1]{#1}
\csname url@samestyle\endcsname
\providecommand{\newblock}{\relax}
\providecommand{\bibinfo}[2]{#2}
\providecommand{\BIBentrySTDinterwordspacing}{\spaceskip=0pt\relax}
\providecommand{\BIBentryALTinterwordstretchfactor}{4}
\providecommand{\BIBentryALTinterwordspacing}{\spaceskip=\fontdimen2\font plus
\BIBentryALTinterwordstretchfactor\fontdimen3\font minus \fontdimen4\font\relax}
\providecommand{\BIBforeignlanguage}[2]{{%
\expandafter\ifx\csname l@#1\endcsname\relax
\typeout{** WARNING: IEEEtran.bst: No hyphenation pattern has been}%
\typeout{** loaded for the language `#1'. Using the pattern for}%
\typeout{** the default language instead.}%
\else
\language=\csname l@#1\endcsname
\fi
#2}}
\providecommand{\BIBdecl}{\relax}
\BIBdecl

\bibitem{Davis_Kahan_1970}
C.~Davis and W.~M. Kahan, ``The rotation of eigenvectors by a perturbation. iii,'' \emph{SIAM Journal on Numerical Analysis}, vol.~7, no.~1, pp. 1--46, 1970.

\bibitem{10.1093/biomet/asv008}
Y.~Yu, T.~Wang, and R.~J. Samworth, ``A useful variant of the davis-kahan theorem for statisticians,'' \emph{Biometrika}, vol. 102, no.~2, pp. 315--323, 2014.

\bibitem{10.1214/17-AOS1541}
T.~T. Cai and A.~Zhang, ``{Rate-optimal perturbation bounds for singular subspaces with applications to high-dimensional statistics},'' \emph{The Annals of Statistics}, vol.~46, no.~1, pp. 60 -- 89, 2018.

\bibitem{procrustes2736}
J.~C. Gower and G.~B. Dijksterhuis, \emph{Procrustes problems}, ser. Oxford Statistical Science Series.\hskip 1em plus 0.5em minus 0.4em\relax Oxford, UK: Oxford University Press, January 2004, vol.~30.

\bibitem{Cape_L2_inf_AOS2019}
J.~Cape, M.~Tang, and C.~E. Priebe, ``{The two-to-infinity norm and singular subspace geometry with applications to high-dimensional statistics},'' \emph{The Annals of Statistics}, vol.~47, no.~5, pp. 2405 -- 2439, 2019.

\bibitem{abbe_fan_AOS2022}
E.~Abbe, J.~Fan, and K.~Wang, ``{An ${\ell _{p}}$ theory of PCA and spectral clustering},'' \emph{The Annals of Statistics}, vol.~50, no.~4, pp. 2359 -- 2385, 2022.

\bibitem{abbe_fan_aos2020}
E.~Abbe, J.~Fan, K.~Wang, and Y.~Zhong, ``{Entrywise eigenvector analysis of random matrices with low expected rank},'' \emph{The Annals of Statistics}, vol.~48, no.~3, pp. 1452 -- 1474, 2020.

\bibitem{cai_AOS2021_unbalanced_incomplete}
C.~Cai, G.~Li, Y.~Chi, H.~V. Poor, and Y.~Chen, ``{Subspace estimation from unbalanced and incomplete data matrices: ${\ell _{2,\infty }}$ statistical guarantees},'' \emph{The Annals of Statistics}, vol.~49, no.~2, pp. 944 -- 967, 2021.

\bibitem{chen_Assymetry_helps_AOS2021}
Y.~Chen, C.~Cheng, and J.~Fan, ``{Asymmetry helps: Eigenvalue and eigenvector analyses of asymmetrically perturbed low-rank matrices},'' \emph{The Annals of Statistics}, vol.~49, no.~1, pp. 435 -- 458, 2021.

\bibitem{Chen_2021}
Y.~Chen, Y.~Chi, J.~Fan, and C.~Ma, ``Spectral methods for data science: A statistical perspective,'' \emph{Foundations and Trends in Machine Learning}, vol.~14, no.~5, pp. 566--806, 2021.

\bibitem{lihua_lei_2020_generic_symmetric}
L.~Lei, ``Unified $\ell_{2\rightarrow\infty}$ eigenspace perturbation theory for symmetric random matrices,'' \emph{ArXiv: 1909.04798}, 2020.

\bibitem{tsyganov2026_matrixfree_2-to-infty}
A.~Tsyganov, E.~Frolov, S.~Samsonov, and M.~Rakhuba, ``Matrix-free two-to-infinity and one-to-two norms estimation,'' \emph{ArXiv: 2508.04444}, 2026.

\bibitem{wang2024_singular_subspaces_random}
K.~Wang, ``Analysis of singular subspaces under random perturbations,'' \emph{The Annals of Statistics}, vol.~54, no.~2, pp. 667--691, 2026.

\bibitem{Xie-Bernoulli2024}
F.~Xie, ``{Entrywise limit theorems for eigenvectors of signal-plus-noise matrix models with weak signals},'' \emph{Bernoulli}, vol.~30, no.~1, pp. 388 -- 418, 2024.

\bibitem{xie2025AOS}
F.~Xie and Y.~Zhang, ``Higher-order entrywise eigenvectors analysis of low-rank random matrices: Bias correction, edgeworth expansion, and bootstrap,'' \emph{The Annals of Statistics}, vol.~53, no.~4, pp. 1667--1693, 2025.

\bibitem{Yan_AOS2024_MissingData}
Y.~Yan, Y.~Chen, and J.~Fan, ``{Inference for heteroskedastic PCA with missing data},'' \emph{The Annals of Statistics}, vol.~52, no.~2, pp. 729 -- 756, 2024.

\bibitem{zhou2024_illposedPCA}
Y.~Zhou and Y.~Chen, ``Deflated heteropca: Overcoming the curse of ill-conditioning in heteroskedastic pca,'' \emph{ArXiv: 2303.06198}, 2024.

\bibitem{Chen_Fan_2019_PNAS_Matrix_Completion}
Y.~Chen, J.~Fan, C.~Ma, and Y.~Yan, ``Inference and uncertainty quantification for noisy matrix completion,'' \emph{Proceedings of the National Academy of Sciences}, vol. 116, no.~46, pp. 22\,931--22\,937, 2019.

\bibitem{Jedra2024_LowRank_Bandits}
Y.~Jedra, W.~R'eveillard, S.~Stojanovic, and A.~Prouti{\`e}re, ``Low-rank bandits via tight two-to-infinity singular subspace recovery,'' in \emph{International Conference on Machine Learning}, 2024.

\bibitem{pensky2021clustering}
M.~Pensky and Y.~Wang, ``Clustering of diverse multiplex networks,'' \emph{IEEE Transactions on Network Science and Engineering}, vol.~11, no.~4, pp. 3441--3454, 2024.

\bibitem{lei2021biasadjusted}
J.~Lei and K.~Z. Lin, ``Bias-adjusted spectral clustering in multi-layer stochastic block models,'' \emph{Journal of the American Statistical Association}, vol. 118, no. 544, pp. 2433--2445, 2023.

\bibitem{ndaoud_aos2022}
M.~Ndaoud, ``{Sharp optimal recovery in the two component Gaussian mixture model},'' \emph{The Annals of Statistics}, vol.~50, no.~4, pp. 2096 -- 2126, 2022.

\bibitem{tropp2015introduction}
J.~A. Tropp, ``An introduction to matrix concentration inequalities,'' \emph{Foundations and Trends in Machine Learning}, vol.~8, no. 1–2, pp. 1--–230, 2015.

\bibitem{giraud2018RelaxedKmeans}
C.~Giraud and N.~Verzelen, ``Partial recovery bounds for clustering with the relaxed k-means,'' \emph{Mathematical Statistics and Learning}, vol.~1, no. 3-4, pp. 317--374, 2018.

\bibitem{H_Zhou_AOS2021_Spec_Clust}
M.~L{\"o}ffler, A.~Y. Zhang, and H.~H. Zhou, ``{Optimality of spectral clustering in the Gaussian mixture model},'' \emph{The Annals of Statistics}, vol.~49, no.~5, pp. 2506 -- 2530, 2021.

\bibitem{NIPS2017_6776}
M.~Royer, ``Adaptive clustering through semidefinite programming,'' in \emph{Advances in Neural Information Processing Systems 30}, I.~Guyon, U.~V. Luxburg, S.~Bengio, H.~Wallach, R.~Fergus, S.~Vishwanathan, and R.~Garnett, Eds.\hskip 1em plus 0.5em minus 0.4em\relax Curran Associates, Inc., 2017, pp. 1795--1803.

\bibitem{giraud_2024_comp_gap}
B.~Even, C.~Giraud, and N.~Verzelen, ``Computation-information gap in high-dimensional clustering,'' in \emph{Proceedings of the 37th Annual Conference on Learning Theory}, ser. Proceedings of Machine Learning Research, S.~Agrawal and A.~Roth, Eds., vol. 247.\hskip 1em plus 0.5em minus 0.4em\relax PMLR, 2024, pp. 1--67.

\bibitem{lei2015}
J.~Lei and A.~Rinaldo, ``Consistency of spectral clustering in stochastic block models,'' \emph{Ann. Statist.}, vol.~43, no.~1, pp. 215--237, 2015.

\bibitem{1366265}
A.~Kumar, Y.~Sabharwal, and S.~Sen, ``A simple linear time (1 + epsiv;)-approximation algorithm for k-means clustering in any dimensions,'' in \emph{45th Annual IEEE Symposium on Foundations of Computer Science}, Oct 2004, pp. 454--462.

\bibitem{JMLR:v18:16-480}
E.~Abbe, ``Community detection and stochastic block models: Recent developments,'' \emph{J. Mach. Learn. Res.}, vol.~18, no. 177, pp. 1--86, 2018.

\bibitem{abbe_bandeira_2016}
E.~Abbe, A.~Bandeira, and G.~Hall, ``\BIBforeignlanguage{English (US)}{Exact recovery in the stochastic block model},'' \emph{\BIBforeignlanguage{English (US)}{IEEE Transactions on Information Theory}}, vol.~62, no.~1, pp. 471--487, 2016.

\bibitem{DBLP:journals/corr/AminiL14}
A.~A. Amini and E.~Levina, ``On semidefinite relaxations for the block model,'' \emph{Ann. Statist.}, vol.~46, no.~1, pp. 149--179, 2018.

\bibitem{rohe2011spectral}
K.~Rohe, S.~Chatterjee, and B.~Yu, ``Spectral clustering and the high-dimensional stochastic blockmodel,'' \emph{Ann. Statist.}, vol.~39, no.~4, pp. 1878--1915, 2011.

\bibitem{Ander_Zhang_IEEE2024_SBM}
A.~Y. Zhang, ``Fundamental limits of spectral clustering in stochastic block models,'' \emph{IEEE Transactions on Information Theory}, vol.~70, no.~10, pp. 7320--7348, 2024.

\bibitem{chakrabarty2023sonnet}
S.~Chakrabarty, S.~Sengupta, and Y.~Chen, ``Subsampling based community detection for large networks,'' \emph{Statistica Sinica}, vol. in press, 2023.

\bibitem{mukherjee2021two}
S.~S. Mukherjee, P.~Sarkar, and P.~J. Bickel, ``Two provably consistent divide-and-conquer clustering algorithms for large networks,'' \emph{Proceedings of the National Academy of Sciences}, vol. 118, no.~44, p. e2100482118, 2021.

\bibitem{bhadra2025scalablecomm_detect}
S.~Bhadra, M.~Pensky, and S.~Sengupta, ``Scalable community detection in massive networks via predictive assignment,'' \emph{ArXiv:2503.16730}, 2025.

\bibitem{GDPG}
P.~Rubin-Delanchy, J.~Cape, M.~Tang, and C.~E. Priebe, ``{A Statistical Interpretation of Spectral Embedding: The Generalised Random Dot Product Graph},'' \emph{Journal of the Royal Statistical Society Series B: Statistical Methodology}, vol.~84, no.~4, pp. 1446--1473, 2022.

\bibitem{pensky2024signed}
M.~Pensky, ``Signed diverse multiplex networks: Clustering and inference,'' \emph{IEEE Transactions on Information Theory}, vol.~71, no.~9, pp. 7076--7096, 2025.

\bibitem{AgterbergLubbertsPriebe2022}
J.~Agterberg, Z.~Lubberts, and C.~E. Priebe, ``Entrywise estimation of singular vectors of low-rank matrices with heteroskedasticity and dependence,'' \emph{IEEE Transactions on Information Theory}, vol.~68, no.~7, pp. 4618--4650, 2022.

\bibitem{ChengWeiChen2021}
C.~Cheng, Y.~Wei, and Y.~Chen, ``Tackling small eigen-gaps: Fine-grained eigenvector estimation and inference under heteroscedastic noise,'' \emph{IEEE Transactions on Information Theory}, vol.~67, no.~11, pp. 7380--7419, 2021.

\bibitem{ZhangCaiWu2022}
A.~R. Zhang, T.~T. Cai, and Y.~Wu, ``Heteroskedastic pca: Algorithm, optimality, and applications,'' \emph{The Annals of Statistics}, vol.~50, no.~1, pp. 53--80, 2022.

\bibitem{bandeira2016}
A.~S. Bandeira and R.~van Handel, ``Sharp nonasymptotic bounds on the norm of random matrices with independent entries,'' \emph{Ann. Probab.}, vol.~44, no.~4, pp. 2479--2506, 07 2016.

\bibitem{Latala2005}
R.~Lata{\l}a, ``Some estimates of norms of random matrices,'' \emph{Proceedings of the American Mathematical Society}, vol. 133, no.~5, pp. 1273--1282, 2005.

\bibitem{Seginer2000}
Y.~Seginer, ``The expected norm of random matrices,'' \emph{Combinatorics, Probability and Computing}, vol.~9, no.~2, pp. 149--166, 2000.

\bibitem{Wedin1972PerturbationBI}
P.-{\AA}. Wedin, ``Perturbation bounds in connection with singular value decomposition,'' \emph{BIT Numerical Mathematics}, vol.~12, pp. 99--111, 1972.

\bibitem{vershynin2018_book}
R.~Vershynin, \emph{High-Dimensional Probability}, ser. Cambridge Series in Statistical and Probabilistic Mathematics.\hskip 1em plus 0.5em minus 0.4em\relax Cambridge University Press, 2018, vol.~47.

\end{thebibliography}

\ifCLASSOPTIONcaptionsoff
  \newpage
\fi

%

\begin{IEEEbiographynophoto}{Marianna Pensky}
 is a Professor with the Department of Mathematics, University of Central Florida. 
She is a Fellow of the Institute of Mathematical Statistics and a Fellow of the American Statisical Association. 
She is also  an executive editor of the Journal of the Statistical Planning and Inference. 
\end{IEEEbiographynophoto}



 \newpage

 \clearpage
\onecolumn

\setlength{\topmargin}{-1.8cm} \setlength{\evensidemargin}{-5mm}
\setlength{\oddsidemargin}{-5mm} \setlength{\textheight}{22.7cm}
\setlength{\textwidth}{17.8cm} 
\setlength{\parindent}{0.5cm}

\setcounter{page}{1}

\section{Supplementary Material:\ Proofs of supplementary lemmas}
\label{sec:supplementary}

\renewcommand{\theequation}{S.\arabic{equation}}
 \setcounter{equation}{0}




\noindent
{\bf Proof of Lemmas~\ref{lem:lemma_Assump2}  and \ref{lem:lemma_Assump4}}.\\ 
Validity of statements a) and b) in Lemmas~\ref{lem:lemma_Assump2}  and \ref{lem:lemma_Assump4} 
follow from \cite{vershynin2018_book}. Validity of statements c) follow from  Theorem~3 of \cite{lei2021biasadjusted}. 
\\


\medskip

\noindent
{\bf Proof of Lemma~\ref{lem:err_bounds_gauss}.  }\\  
First, consider the case where $\hU = \SVD_r(\hX)$. Then, it is well known  (see, e.g.,  \cite{vershynin2018_book})  
that, due to expansion \fr{eq:X_SVD} of $X$, asymptotic relations in \fr{eq:tepso_tepstinf} are valid.

Now, consider the case, where $\hU = \scrH(\hX \hX^T)$. Then, $\tilscrE$ is given by \fr{eq:new_sym_E}--\fr{eq:barXixi}
with $\tilh =1$. We first find $\tepsEo$, which requires evaluation of $\|\tilscrE\|$.
It is easy to see that, by   \fr{eq:X_SVD}
\bes
\|\tilscrE_2\| = \|\tilscrE_3 \| = \|\Xi X^T \| \leq d_1 \|\Xi V\| \asymp d_1\, \sig \, \lkr \sqrt{n} + \sqrt{r} \rkr,
\ees
where $d_1 = \|X\|$. In order to obtain an upper bound for $\|\tilscrE_1\|$, apply Theorem~7 of  \cite{lei2021biasadjusted},
which yields
\bes
\|\tilscrE_1\| \leq \Ctau \, \sig^2\, \lkv n \log n + \sqrt{n}\, (\log n)^{3/2} + \sqrt{n}\, \log n + (\log n)^2 \rkv \leq 
\Ctau \, \sig^2 \, n\, \log n.
\ees
Finally, $\tilscrE_d \leq m\, \te^2 + \Ctau\, d_1\, \sig \, \lkr \sqrt{n} + \sqrt{r} \rkr$.
Therefore, using \fr{eq:d_r_epsu_gauss}, derive
\be \label{eq:gauss_tepsEo}
\tepsEo \leq \Ctau\, \lkr \frac{\sig^2\, r\, \log n}{\te^2\, m} + \frac{r}{n} \rkr.
\ee
 
The next objective is to bound above $\|\scrH(\Xi\, \Xi^T)\, A\|\tinf = \displaystyle \max_l  \|\Xi(l,:)  \, (\Xi\upl)^T  A\|$
with $A=U$ and $A = I_n$, where  $\Xi\upl$ is defined in \fr{eq:Xiupl}.  Since $\|\Xi(l,:)$ and $\Xi\upl$ are independent  for any $l \in [n]$, 
 using Bernstein's inequality and conditioning on  $\Xi\upl$, derive, for any $l$ and any $t_1>0$
\bes 
\PP \lfi \left\|\Xi(l,:)  \, (\Xi\upl)^T  U \right\| \geq t_1 \rfi \leq 2\, (n+r) \, \exp \lkr -\frac{t_1^2}{2\, (\sig^2 \, a_1^2 + \sig\, b_1\, t_1)} \rkr,
\ees
where 
\beqns
a_1^2 & = &   \|(\Xi\upl)^T\, U\|^2_F \leq \Ctau\, \sig^2 r\, m\, \log n,\\
b_1 & = &   \|(\Xi\upl)^T\, U\|\tinf \leq \Ctau\, \sig \, \sqrt{r}\, \sqrt{\log n}
\eeqns
with high probability. Set $t_1 = \Ctau \, \sig^2 \, \lkr \sqrt{r\, m} \, \log n + \sqrt{r}\,  \log n\, \sqrt{\log n} \rkr$.
Since taking the union bound over $l \in [n]$ just leads to changing the constant $\Ctau$, obtain, that with probability at least $1 - n^{-\tau}$,
\be \label{eq:gauss_up1}
\|\scrH(\Xi\, \Xi^T)\, U\|\tinf \leq \Ctau\, \sig^2 \, \log n \lkr \sqrt{r\, m} + \sqrt{r\, \log n} \rkr.
\ee  
Then, combination of \fr{eq:d_r_epsu_gauss} and  \fr{eq:gauss_up1} yields the expression for $\tepsxiutinf$.

Similarly, using Bernstein inequality, derive that, for any $t_2>0$ 
\bes 
\PP \lfi \lnor \Xi(l,:)  \, (\Xi\upl)^T   \rnor \geq t_2 \rfi \leq 4\, n \, \exp \lkr -\frac{t_2^2}{2\, (\sig^2 \, a_2^2 + \sig\, b_2\, t_2)} \rkr,
\ees
where $a_2^2 = \|\Xi\upl\|^2_F \leq \Ctau\, \sig^2\, m\, n\, \log n$ and $b_2 = \|(\Xi\upl)^T\|\tinf \leq \Ctau\, \sig\, \sqrt{n\, \log n}$ with high probability. 
Therefore, obtain that, with probability at least $1 - n^{-\tau}$, 
\be \label{eq:gauss_up2}
\|\tilscrE_1\|\tinf = \|\scrH(\Xi\, \Xi^T) \|\tinf \leq \Ctau\, \sig^2 \, \log n \, \sqrt{m\,n}.
\ee   
We shall use the inequality above later, for obtaining an upper bound for $\tepsEtinf$.

Now, consider $\| \Xi X^T U\|\tinf = \displaystyle \max_l \|\|\Xi(l,:) \, X^T U\|$. 
Since $\|X^T U\|^2_F \leq r\, d_1^2$ and $\|X^T U\|\tinf \leq d_1\,\sqrt{r}$,
obtain that, with high probability, $\|\|\Xi(l,:) \, X^T U\| \leq \Ctau\, d_1\, \sig \sqrt{r}\,  \log n$. Then,  
\fr{eq:d_r_epsu_gauss} yields the expression for $\tepsxutinf$.

It remains to obtain an upper bound for $\tepsEtinf$. For this purpose, it is necessary to 
bound  above $\|\tilscrE_1\|\tinf + \|\tilscrE_2\|\tinf$. Note that 
\be \label{eq:gauss_up3}
\|\tilscrE_2\|\tinf = \max_{l \in [n]}\, \|\Xi(l,:)\, X^T \| \leq  \Ctau\, d_1\, \sig  \,  \sqrt{r\, \log n}.
\ee   
Then, combination of \fr{eq:d_r_epsu_gauss}, \fr{eq:gauss_up2} and   \fr{eq:gauss_up3}, leads to the upper bound for 
$\tepsEtinf$.

Finally,  \fr{eq:assump_A4*} holds with $\teps_1$ and $\teps_2$ given in \fr{eq:teps12},  by Hanson-Wright inequality 
(Theorem 6.2.1 of \cite{vershynin2018_book}).
\\


\medskip

\noindent
{\bf Proof of Lemma~\ref{lem:lemma1}.  }\\   
Recall that, by \fr{eq:inv_norm}, $\|(U^T \, \hU)^{-1}\|\leq 2$.  Then,
\bes
\| \hU\, \hU^T \, U - U\|\tinf \leq 2\,  \| U U^T \hU - \hU \|\tinf + 2 \|\hU\, \hU^T \, U U^T \, \hU - \hU\|\tinf.
\ees
Here, 
\bes 
\| U U^T \hU - \hU \|\tinf \leq \|\hU - U W_U  \|\tinf + \|U\|\tinf\, \|U^T \hU - W_U\|,
\ees
\bes
\|\hU\, \hU^T \, U U^T \, \hU - \hU\|\tinf = \|\hU\|\tinf\, \|\hU^T \, U U^T \, \hU - I\|.
\ees   
Note that,  by \fr{eq:ineq3} and \fr{eq:ineq4}, for $\om \in \Omtauo$, one has  
$\|\hU^T \, U U^T \, \hU - I\| = \|\sinTeU\|^2 \leq \Ctau \, \epso^2$ 
and $\|U^T \hU - W_U\| \leq \Ctau \,  \epso^2$. Also, 
\bes
\|\hU\|\tinf \leq  \|\hU - U W_U\|\tinf + \|U\|\tinf.
\ees
Combining all inequalities above and recalling that $\epso = o(1)$, immediately obtain \fr{eq:lemma11}.
\\

In order to prove  \fr{eq:lemma12}, we use the ``leave one out'' method.
Specifically, fix $l \in [n]$ and let 
$\hY\upl = \hU\upl \hLam\upl  (\hU\upl)^T + \hU_{\perp}\upl \hLam_{\perp}\upl (\hU_{\perp}\upl)^T$
be the SVD of $\hY\upl$, where $\hU\upl \in \calO_{n,r}$ and $\hU_{\perp}\upl \in \calO_{n,n-r}$.
Since $\|\scrE\upl\| \leq \|\scrE\|$, one has 
\be \label{eq:Upl_rel}
 \|\hLam\upl - \Lam\| \leq \| \hLam - \Lam\|, \quad 
\| \sinTe(\hU\upl, U)\| \leq \|\sinTeU\|.
\ee
Note that 
\be \label{eq:R1_R2}
\|\scrE (\hU \hU^T U - U) \|\tinf \leq R_1 + R_2,
\ee
 where 
\be \label{eq:R1_R2-for}
R_1 = \max_{l \in [n]} \lnor \scrE(l,:) \lkv \hU\upl (\hU\upl)^T U - U\rkv \rnor,
\quad
R_2 = \|\scrE\| \, \lnor \lkv \hU \hU^T - \hU\upl (\hU\upl)^T \rkv \, U \rnor_F
\ee 

Start with the second term. Note that, by Davis-Kahan theorem (\cite{Davis_Kahan_1970}),
\be  \label{eq:lem2_for11}
\| [\hU \hU^T - \hU\upl (\hU\upl)^T]\, U \|_F \leq C\, |\lam_r|^{-1} \lnor (\hY - \hY\upl) \hU\upl\rnor_F
\ee
Here, 
\bes
(\hY - \hY\upl) \hU\upl = e_l \scrE(l,:)\hU\upl + \lkv \scrE(:,l) - \scrE(l,l)  e_l\rkv\,  e_l^T \hU\upl,
\ees
where $e_l$ is the $l$-th  canonical vector in $\RR^n$. Since both components above have ranks one, derive that
\be \label{eq:lem2_for12}
\lnor  (\hY - \hY\upl) \hU\upl  \rnor_F \leq \| \scrE(l,:)\hU\upl\| + \|\scrE(:,l) - \scrE(l,l)  e_l \| \, \|e_l^T \hU\upl \|.
\ee
Denote $H = \hU^T U$, $H\upl = (\hU\upl)^T\, U$. 
Then, by \fr{eq:inv_norm} and   \fr{eq:Upl_rel},   for $n$ large enough, $\|H^{-1}\| \leq 2$ and   $\|(H\upl)^{-1}\| \leq 2$.
Hence,
\be  \label{eq:lem2_for15}
 \lnor  \scrE(:,l) - \scrE(l,l)  e_l \rnor \, \|e_l^T \hU\upl \|  \leq
2\, \|\scrE\| \, \lnor \hU\upl (\hU\upl)^T \, U \rnor\tinf.
\ee
Plugging  \fr{eq:lem2_for15} into \fr{eq:lem2_for12}, obtain
\be \label{eq:lem2_for16}
\lnor (\hY - \hY\upl) \hU\upl \rnor_F \leq \| \scrE(l,:)\hU\upl\| + 2\, \|\scrE\| \, \| \hU\upl (\hU\upl)^T \, U - \hU \hU^T\, U \|\tinf 
+  2\, \|\scrE\| \, \| \hU \hU^T\, U \|\tinf.
\ee
Now, combine \fr{eq:lem2_for16}  and \fr{eq:lem2_for11}:
\begin{align*}
\| [\hU \hU^T - \hU\upl (\hU\upl)^T]\, U \|_F & \leq C\, |\lam_r|^{-1}\, \lkr 
\| \scrE(l,:)\hU\upl\| +  \|\scrE\| \, \| \hU \hU^T\, U \|\tinf  \right. \\
& + \left.  \|\scrE\| \, \lnor \hU\upl (\hU\upl)^T \, U - \hU \hU^T\, U \rnor\tinf  \rkr.
\end{align*}
Note that, for $\om \in \Omtau$, the coefficient of the last term is bounded above by $\Ctau   \epso$, and,
by assumption \fr{eq:eps_conditions}, it is below 1/2 when $n$ is large enough. Therefore,
the last inequality can be rewritten as
\beqn 
\lnor [\hU \hU^T - \hU\upl (\hU\upl)^T]  U \rnor_F & \leq &  C\, |\lam_r|^{-1}\, \lkr 
\| \scrE(l,:)\hU\upl\| +  \|\scrE\| \, \| \hU \hU^T  U - U \|\tinf \right. \nonumber \\
& + & \left.  \|\scrE\| \, \| U \|\tinf  \rkr.  \label{eq:lem2_for17}
\eeqn
Consider the first term in \fr{eq:lem2_for17}: 
\beqn 
\| \scrE(l,:)\hU\upl\| & = &  \lnor \scrE(l,:) \hU\upl H\upl (H\upl)^{-1}\rnor  \leq 
2 \, \lnor \scrE(l,:) \hU\upl (\hU\upl)^T\, U \rnor \nonumber \\
& \leq & 2\, \|\scrE(l,:) \, U \| + 2\, \lnor \scrE(l,:) [\hU\upl (\hU\upl)^T\, U - U] \rnor.
\label{eq:lem2_for18}
\eeqn
Now observe that, due to the conditions of the theorem, $\scrE(l,:)$ and $\hU\upl (\hU\upl)^T\, U - U$
are independent, so, conditioned on  $\hY\upl$, by assumption \fr{eq:assump_A2}, 
obtain that, for $\om \in \Omtaut$,    one has
\bes
\|\scrE(l,:) [\hU\upl (\hU\upl)^T\, U - U] \| \leq \Ctau |\lam_r| \lkr \eps_1 \|\hU\upl (\hU\upl)^T\, U - U\|_F
+ \eps_2\, \|\hU\upl (\hU\upl)^T\, U - U\|\tinf \rkr.
\ees
Now, rewrite the last inequality as 
\beqn
\lnor  \scrE(l,:) [\hU\upl (\hU\upl)^T\, U - U] \rnor  &  \leq & \Ctau\,  |\lam_r| 
\lfi \eps_1 \,  \| [\hU\upl (\hU\upl)^T  - \hU \, \hU^T]\, U \|_F   
 \right. \nonumber \\
& + &   \eps_1 \,  \|\hU \, \hU^T \, U - U\|_F  + \eps_2 \,  \|\hU \, \hU^T  - U\|\tinf 
\label{eq:lem2_for19} \\
& + & \left. \eps_2 \,  \| [\hU\upl (\hU\upl)^T  - \hU \, \hU^T]\, U - U\|\tinf 
 \rfi  \nonumber 
\eeqn
Plugging \fr{eq:lem2_for19} into \fr{eq:lem2_for18}  and  \fr{eq:lem2_for18} into \fr{eq:lem2_for17},
due to $\|\scrE(l,:)\, U \| \leq \|\scrE \, U \|\tinf$ for any $l \in [n]$,  and $\|\cdot\|\tinf \leq  \|\cdot\|_F$,
obtain that, for $\om \in \Omtau$ 
\begin{align*}
\| [\hU\upl (\hU\upl)^T  - \hU \, \hU^T]\, U \|_F & \leq \Ctau\, \lfi \Delo\,  \|\hU \, \hU^T\, U  - U\|\tinf
 + \Delo\,  \|U\|\tinf + \DelEu  \right.  \\
& + \eps_1\, \|\hU\upl (\hU\upl)^T\, U - \hU \, \hU^T\, U \|_F 
+ \eps_1 \, \|\hU \, \hU^T\, U  - U\|_F
\\
& \left.  + \eps_2\, \|\hU\upl (\hU\upl)^T\, U - \hU \, \hU^T\, U \|\tinf
+ \eps_2 \, \|\hU \, \hU^T\, U  - U\|\tinf \rfi
\end{align*}
Combine the terms under the assumptions $ \Ctau (\eps_1 + \eps_2) \leq 1/2$,
which is true for $\om \in \Omtau$ if $n$ is large enough. Obtain
\beqn 
\| [\hU\upl (\hU\upl)^T  - \hU \, \hU^T]\, U \|_F & \leq &   \Ctau\, \lkv  \epsEu +   \epso\, \epsu
+  \eps_1 \, \|\hU \, \hU^T\, U  - U\|_F \right. \nonumber \\
& + &  \left.
(\epso + \eps_2) \, \|\hU \, \hU^T\, U  - U\|\tinf  \rkv. \label{eq:lem2_for20}
\eeqn
Plugging  \fr{eq:lem2_for20} into \fr{eq:lem2_for19},  combining the terms  and 
removing the smaller order terms, derive an upper bound for $R_1$ in \fr{eq:R1_R2}:
\beqn 
R_1 & \leq & \Ctau\, |\lam_r|\, \lfi   (\eps_1 + \eps_2) (\epsEu + \epso \epsu) +
\eps_1 \, \|\hU \, \hU^T\, U  - U\|_F +   \right. \nonumber \\
& + & \left.
(\epso\, \eps_1 + \eps_2)  \, \|\hU \, \hU^T\, U  - U\|\tinf   \rfi
\label{eq:lem2_for21}
\eeqn 
Now, combine \fr{eq:R1_R2-for} and \fr{eq:lem2_for20} to obtain an upper bound for $R_2$, when $\om \in \Omtau$:
\beqn \label{eq:lem2_for22}
R_2 & \leq & 8\, |\lam_r|\, \epso\, \lfi \epsEu + \epso\, \epsu + \Ctau\, \eps_1 \, \|\hU \, \hU^T\, U  - U\|_F
+ (\Ctau \eps_2 + \epso) \, \|\hU \, \hU^T\, U  - U\|\tinf \rfi. \ \ \  
\eeqn
Plugging \fr{eq:lem2_for21} and \fr{eq:lem2_for22} into \fr{eq:R1_R2} and adjusting coefficients,
due to $\|\hU \, \hU^T\, U  - U\|_F \leq \sqrt{r}\, \|\hU \, \hU^T\, U  - U\| \leq \sqrt{r}\, \epso$, 
for $\om \in \Omtau$,  infer that
\beqns 
 \|\scrE\, (\hU\, \hU^T \, U - U)\|\tinf & \leq &   \Ctau\, \|\lam_r|^{-1}\, \lfi  (\epsEu + \epso\, \epsu) (\epso + \eps_1 + \eps_2)
+ \sqrt{r}\, \epso\, \eps_1 \right. \nonumber \\
& + & \left. (\epso^2 + \epso\, \eps_1 + \eps_2)\,  \|\hU\, \hU^T\, U - U \|\tinf \rfi.
\eeqns 
Eliminating smaller order terms, we arrive at \fr{eq:lemma12}.
\\

\medskip


\noindent
{\bf Proof of Lemma~\ref{lem:lemma_prob_Symmetized}.  }\\   
Fix $l \in [n]$, and decompose 
\be \label{eq:Xiupl}
\Xi = \Xi\upl + e_l \Xi(l,:), \quad \mbox{where} \quad \Xi\upl(i,:) = \lfi 
\begin{array}{ll}
\Xi(i,:), & \mbox{if}\ \ i\neq l, \\
0, & \mbox{if}\ \ i= l,
\end{array} \right.
\ee 
and $e_l$ is the $l$-th canonical  vector in $\RR^n$. Observe that $\Xi\upl$ and $\Xi(l,:)$ are independent from each other.
Define $\tilscrE\upl = \tilscrE\upl_1 +  \tilscrE\upl_2 +  \tilscrE\upl_d$, where
\beqns  
\tilscrE_1\upl & = & \barXixil, \quad 
\tilscrE_2\upl = \Xi\upl\, X^T,   \quad
\tilscrE_d = -   \tilh\,  [\diag(Y) +  2 \, \diag(\Xi\upl\, X^T)]. 
\eeqns
Also, denote $\hY\upl = Y + \tilscrE\upl$ and consider its eigenvalue decomposition
\bes
\hY\upl  = \hU\upl \hLam\upl (\hU\upl)^T + \hU\upl_\perp \hLam_\perp\upl (\hU_\perp\upl)^T.
\ees
Similarly to the symmetric case, $\|\tilscrE\upl\| \leq \|\tilscrE\|$, and \fr{eq:Upl_rel} holds.
Also, \fr{eq:R1_R2} and  \fr{eq:R1_R2-for} are valid.
In order to simplify the presentation, denote
\be \label{eq:R_UU}
\Ruu = \hU \hU^T U - U, \quad \Ruul = \lkv \hU\upl (\hU\upl)^T   - \hU \hU^T \rkv \, U,
\ee
so that, for $\tilR$ defined in \fr{eq:tilR}, one has   $\tilR  \leq  R_1 + R_2$ where  
\be \label{eq:R1_R2-new}
R_1 = \max_{l \in [n]} \lnor \tilscrE(l,:) \lkv \hU\upl (\hU\upl)^T U - U\rkv \rnor,
\quad
R_2 = \|\tilscrE\| \, \|\Ruul \|_F
\ee 
Observe that, by Davis-Kahan theorem 
\be \label{eq:L20}
\|\Ruul \|_F \leq \| \hU\upl (\hU\upl)^T   - \hU \hU^T \|_F \leq C \, d_r^{-2}  \|(\hY - \hY\upl)\hU\upl\|_F.
\ee 
Decompose $\hY - \hY\upl$ as 
\be \label{eq:delscrE}
\hY - \hY\upl = \tilscrE - \tilscrE\upl=  \dscrEl_1 +  \dscrEl_2 + \dscrEl_d, 
\ee
where $\dscrEl_1 = \barXixi - \barXixil$, $\dscrEl_2 = (\Xi - \Xi\upl)\, X^T$ and   
$\dscrEl_d = - 2\, \tilh\, \diag \lkr (\Xi - \Xi\upl)\, X^T \rkr$.
Due to \fr{eq:Xiupl}, one has
\beqn \label{eq:dscrE}
\dscrEl_1 & = & e_l \, \Xi(l,:)  \, (\Xi\upl)^T + \Xi\upl (\Xi(l,:))^T\, e_l^T + (1-\tilh)\, \|\Xi(l,:)\|^2 \, e_l e_l^T,\\
\dscrEl_2 & = & e_l \Xi(l,:)\, X^T, \quad \quad \dscrEl_d =  2\, \tilh\, \diag(e_l\, \Xi(l,:) \, X^T). \nonumber
\eeqn
Plugging \fr{eq:delscrE} and \fr{eq:dscrE}  into the r.h.s. of \fr{eq:L20}, obtain
\beqn \label{eq:L28}
\|(\hY  & - & \hY\upl)\hU\upl\|_F   \leq    \|\Xi(l,:)  \, (\Xi\upl)^T \, \hU\upl\| +  \|\Xi(l,:)\, X^T  \hU\upl\| \\
& + &  \| e_l^T \hU\upl\| \, \lkv \|\Xi\upl (\Xi(l,:))^T\| + (1 - \tilh) \,  \|\Xi(l,:)\|^2 + 
2\tilh \, | \Xi(l,:)\, (X(l,:))^T | \rkv. \nonumber
\eeqn
Denote $H\upl = (\hU\upl)^T \, U$, $H = \hU^T U$, and observe that, if $\tDelEo$ is small enough (which is true 
for $\om \in \tilOmtau$), then $\|H^{-1}\| \leq 2$ and $\|(H\upl)^{-1}\| \leq 2$.
In this proof, we shall use the following two representations of $\hU\upl$:
\beqn \label{eq:L31}
\hU\upl & = &   \Ruu \, \lkr H\upl \rkr^{-1}  + \Ruul \, \lkr H\upl \rkr^{-1}  + U  \, \lkr H\upl \rkr^{-1} , \\
\hU\upl & = &    (\hU\upl (\hU\upl)^T U - U) \, \lkr H\upl \rkr^{-1}   + U \, \lkr H\upl \rkr^{-1} ,
\label{eq:L32}
\eeqn
where $\Ruu$ and $\Ruul$ are defined in \fr{eq:R_UU}. 
Note that, by \fr{eq:L31}, for $\om \in \tilOmtau$, one has  
\bes
  \| e_l^T \hU\upl\| \leq 2\, \|\Ruu\|\tinf +  2\, \|\Ruul\|\tinf + 2 \epsu.
\ees 
Hence, combination of \fr{eq:L20}, \fr{eq:L28} and the last inequality yields
\begin{align} \label{eq:L33}
&  \|\Ruul \|_F     \leq    d_r^{-2}\, \|\Xi(l,:)  \, (\Xi\upl)^T \, \hU\upl\| +  \|\Xi(l,:)\, X^T  \hU\upl\| \\
& +    2\, d_r^{-2}\, \lkr  \|\Ruu\|\tinf +    \|\Ruul\|\tinf +  \epsu \rkr \, \breve{R}, 
\nonumber
\end{align}
where 
\be \label{eq:breveR}
\breve{R} =  \|\Xi\upl (\Xi(l,:))^T\|  + (1 - \tilh) \,  \|\Xi(l,:)\|^2 + 2\tilh \, |\Xi(l,:)\, (X(l,:))^T|.
\ee

Observe that, in the first two terms in \fr{eq:L33}, one has
\beqn  \label{eq:L34}
  \|\Xi(l,:)  \, (\Xi\upl)^T \, \hU\upl\| & \leq & 2\,  \|\Xi(l,:)  \, (\Xi\upl)^T \,  \lkv  \hU\upl (\hU\upl)^T U - U \rkv \| + 
2\,   \|\Xi(l,:)  \, (\Xi\upl)^T \,  U\|,\ \ \ \   \\
  \|\Xi(l,:)  \, X^T \, \hU\upl\| & \leq & 2\,  \|\Xi(l,:)  \, X^T \,  \lkv  \hU\upl (\hU\upl)^T U - U \rkv \| + 
2\,  \|\Xi(l,:)  \, X^T \,  U\|.  
\label{eq:L36}
\eeqn  
Note that $\Xi(l,:)$ and $\Xi\upl$ are independent, so that $\Xi(l,:)$ and $\hU\upl$ are independent also. 
Therefore,  conditioned on $\Xi\upl$, by Assumption {\bf A4*}, for  $\om \in \tilOmtau$,  
derive
\beqn   
\|\Xi(l,:)  \, (\Xi\upl)^T \,  \lkv  \hU\upl (\hU\upl)^T U - U \rkv \| 
& \leq &
\Ctau \, d_r \lfi \teps_1 \|(\Xi\upl)^T \, \lkv \Ruul + \Ruu   \rkv \|_F  \right. \ \ \nonumber \\
& + & \left. \teps_2 \|(\Xi\upl)^T \, \lkv \Ruul + \Ruu   \rkv \|\tinf \rfi,  
\label{eq:L35} 
\eeqn
\beqn   
\|\Xi(l,:)  \, X^T \,  \lkv  \hU\upl (\hU\upl)^T U - U \rkv \| 
& \leq &
\Ctau \, d_r \lfi \teps_1 \|X^T \, \lkv \Ruul + \Ruu   \rkv \|_F  \right. \ \ \nonumber \\
& + & \left. \teps_2 \|X^T \, \lkv \Ruul + \Ruu   \rkv \|\tinf \rfi.  
\label{eq:L37} 
\eeqn
Plug \fr{eq:L35} into \fr{eq:L34},  \fr{eq:L37} into \fr{eq:L36} and then both 
\fr{eq:L34} into \fr{eq:L36}  into \fr{eq:L33}. Observing  that 
$d_r^{-2}\, \|\Xi(l,:)  \, (\Xi\upl)^T \,  U\|  = \tDelxiutinf$,\ \  
$d_r^{-2}\,  \|\Xi(l,:)  \, X^T \,  U\|   = \tDelxtinf$,  derive
\beqns 
\|\Ruul \|_F &\leq &  2\, \tDelxiutinf + 2 \, \tDelxtinf + 2\, \Ctau\,  \lfi 
\teps_1 (\tDelo + C_d)  \, \lkv \Ruulf + \Ruuf \rkv \right. \\
& +  & \left. 
\teps_2 (\tDeltinf + C_d\, \epsv) \, \lkv \Ruulop + \Ruuop \rkv \rfi \\
& +  & 2\,  \lkv \Ruultinf + \Ruutinf + \epsu \rkv\,   \breve{R}
\eeqns 
where $C_d$ and $\breve{R}$ are defined in \fr{eq:singval_cond} and \fr{eq:breveR}, respectively. Then, 
\be \label{eq:L38}
\breve{R} = d_r^{-2}\,  \lkv \|\Xi(l,:)  \, (\Xi\upl)^T\| + (1 - \tilh)\, \|\Xi\|\tinf^2 + 
2\, \tilh \|\Xi(l,:)  \, X^T\| \rkv \leq  \tDelEtinf + (1 - \tilh)\, \tDeltinf^2,
\ee  
and, due to $\tDelEtinf \leq \tDelEo$ and  \fr{eq:teps_conditions}, for $\om \in \Omtau$, one has $\breve{R} = o(1)$
as $n \to \infty$ with high probability. 
Hence, adjusting the coefficient in front of $\|\Ruul \|_F$, due to $\Ruuf \leq \sqrt{r}\, \Ruuop$ and  \fr{eq:teps_conditions}, 
and using \eqref{eq:tDelxtinf_upper}, derive  that 
\beqn  
\max_{l \in [n]}\, \Ruulf & \leq & \Ctau \lfi \tDelxiutinf + \tDelvtinf + 
\lkv \Ruutinf + \epsu \rkv \, \breve{R} 
 \right. \nonumber \\
& + & \left.  
 \Ruuop \, \lkv \sqrt{r}\,    \teps_1 (1 + \tDelo) + \teps_2\, (\tDeltinf^T +  \epsv] \rkv \rfi.
\label{eq:L40}
\eeqn

Now, we return to $R_1$ and $R_2$ in \fr{eq:R1_R2-new}. Note that, due to the structure of $\tilscrE$, one can 
define $R_1  (l) = \lnor \tilscrE(l,:) [\hU\upl (\hU\upl)^T U - U] \rnor$  and bound above $R_1$ as 
\bes 
R_1 \leq \max_{l \in [n]} \lkv R_{11}(l) + R_{12}(l) + (1 - \tilh) R_{13}(l) + \tilh\,  R_{14}(l)  \rkv,
\ees 
where 
\begin{align*}
 R_{11}(l) & =   \lnor \Xi(l,:)  \, (\Xi\upl)^T \, [\hU\upl (\hU\upl)^T U - U]\rnor \leq 
\Ctau \, d_r^2 \lfi \teps_1 \tDelo \sqrt{r}\, \Ruuop \right. \\
& \left.  + \teps_2\, \tDeltinf \Ruuop  +  (\tDelo\,  \teps_1  + \tDeltinf^T\, \teps_2)\, \Ruulf \rfi;\\
 R_{12}(l) & =   \left\|\Xi(l,:)  \, X^T \, [\hU\upl (\hU\upl)^T U - U]\right\| \leq 
\Ctau \, d_r^2 \,  \lfi (\teps_1  \sqrt{r} +  \teps_2  \epsv)\, \Ruuop \right. \\
& \left. + (\teps_1   +  \teps_2  \epsv)\, \Ruulf \rfi; \\
 R_{13}(l) & =  \|\Xi(l,:)\|^2 \, \lkr \Ruultinf + \Ruutinf \rkr \leq 
  d_r^2 \, (\tDeltinf)^2 \, \lkr \Ruultinf + \Ruutinf \rkr; \\
 R_{14}(l) & =  d_r^2   \tepsy + 2\,   \|\Xi\|\tinf \|X\|\tinf \leq d_r^2 \, \lkr \tepsy + 2 \, C_d\, \tDeltinf \, \epsu \rkr.
\end{align*}
Also, it follows from  \fr{eq:R1_R2-new} that 
\bes 
R_2 \leq d_r^2 \, \tDelEo\, \max_{l \in [n]}\,  \Ruulf.
\ees
Taking the union bound over $\linL$, and combining all components of $R_1(l)$ and $R_2$, derive,
for $\om \in \tilOmtau$:
\beqn 
\tilR & \leq & \Ctau \, d_r^2 \lfi \Ruuop \,  \tdelor + \max_{l \in [n]}\, \Ruulf \lkv 
\tdelo + (1 - \tilh)\, \tDeltinf^2  + \tDelEo \rkv \right.  \nonumber\\
& + & \left. (1 - \tilh)\, \tDeltinf^2   + \tilh (\tepsy +   \tDeltinf \epsu) \rfi, 
\label{eq:L47}
\eeqn
where $\tdelo$ and $\tdelor$ are defined in \fr{eq:delo_delor}.
\ignore{
\beqn 
\tilR & \leq & \Ctau \, d_r^2 \lfi \Ruuop \, \lkv \tDelo \, (\teps_1    + \teps_2)  + 
C_d\, (\teps_1 \sqrt{r} + \teps_2 \epsv) \rkv \right. \nonumber\\
& + & \max_{l \in [n]}\, \Ruulf \lkv \tDelo \, (\teps_1    + \teps_2)  + 
C_d\, (\teps_1   + \teps_2 \epsv) + (1 - \tilh)\, \tDeltinf^2  + \tDelEo \rkv  \nonumber\\
& + & \left. (1 - \tilh)\, \tDeltinf^2 \, \Ruutinf + \tilh \tepsy + 2\, C_d\, \tilh\, \tDeltinf \epsu \rfi 
\label{eq:L47}
\eeqn
}
Recall that $\Ruuop = \|\sinTeU\| \leq 2 \, \tDeluuo$.
In addition, by Lemma~\ref{lem:lemma1}, one has
\be \label{eq:L48}
\Ruutinf \leq 4\, \|\hU - U W_U\|\tinf + C \epsu\,  \|\sinTeU\|^2 \leq 4\, \|\hU - U W_U\|\tinf + \Ctau\, \epsu\, \tepsuuo^2.
\ee 
Plugging  the latter into \fr{eq:L40} and removing the smaller order terms, obtain
\beqn 
\max_{l \in [n]}\, \Ruulf & \leq & \Ctau \, \lfi \tDelxiutinf + \tDelvtinf +   \tDeluuo \, \lkr \tdelor +   \epsu\, 
\breve{R}\rkr \right. \nonumber\\
& + & \left. 4\, \breve{R}\,  \|\hU - U W_U\|\tinf \rfi, 
\label{eq:L49}
\eeqn 
where, due to \fr{eq:L38}, for $\om \in \Omtau$, one has
\bes
\breve{R} \leq \tepsEtinf + (1 - \tilh)\, \tepstinf^2 = o(1)\quad \mbox{as}  \quad n \to \infty.
\ees
Now, substituting   \fr{eq:L48}  and \fr{eq:L49}  into \fr{eq:L47},
obtain that  $d_r^{-2}\, \tilR$ satisfies \fr{eq:lemma21}, for $\om \in \Omtau$, with $\tdel_2$ defined in \fr{eq:tdel2} 
and 
\bes
\tdel_{2,U} = \tepsEuo + \tepsEo^2 + \tepso\, \tdelo = o(1),
\ees
which, together with \fr{eq:teps_conditions} and \fr{eq:delo_delor}, completes the proof.
\\


\medskip

\ignore{

\subsection{Supplementary inequalities}
\label{sec:Suppl}

\begin{lem}\label{lem:suppl_ineq}  
Let $U, \hU \in \calO_{n,r}$ and $W_U$ be defined in \fr{eq:W_u}. 
Then, the following inequalities hold
\begin{eqnarray}
& \|U^T \hU - W_U \| & \leq \|\sinTeU\|^2, \label{eq:ineq1}\\
& \|\hU - U\, U^T \hU\| &  =  \|\sinTeU\|, \label{eq:ineq2}\\
& \| \hU - U\, W_U \| & \leq \sqrt{2}\, \|\sinTeU\| , \label{eq:ineq3}\\
& \|I - \hU^T U U^T \hU \| & =   \|\sinTeU\|^2.  \label{eq:ineq4}
\end{eqnarray}
\end{lem}

\medskip \medskip 

\noindent 
{\bf Proof.} 
Inequalities  \fr{eq:ineq1} and \fr{eq:ineq2} are proved in Lemma~6.7 of \cite{Cape_L2_inf_AOS2019}.
Inequality \fr{eq:ineq3} is established in Lemma~6.8 of \cite{Cape_L2_inf_AOS2019}.
Finally, in order to prove \fr{eq:ineq4}, note that $ U^T \hU = W_1 D_U W_2^T$ where $D_U = \cos(\Te)$
and $\Te$ is the diagonal matrix of the principal angles between the subspaces.
Hence, 
\bes
\|I - \hU^T U U^T \hU \|  = \|W_1 \lkv  I - \cos^2(\Te) \rkv W_1^T\| = \|W_1 \, \sin^2(\Te)\, W_1^T\|,
\ees
which completes the proof. 
} 

\end{document}